\UseRawInputEncoding
\documentclass[aps,nofootinbib,pra,twocolumn,showpacs,superscriptaddress,notitlepage,superscriptaddress,amsmath]{revtex4-2}
\usepackage{listings}
\usepackage{inconsolata}
\usepackage{siunitx}
\usepackage{xcolor}

\usepackage[colorlinks=true,citecolor=blue,linkcolor=blue,urlcolor=blue]{hyperref}
\usepackage{xcolor}
\usepackage{amsmath,amssymb,bm,mathtools,amsthm, amssymb}
\usepackage[autostyle, english = american]{csquotes}
\MakeOuterQuote{"}
\usepackage{tikz}
\usepackage{physics}
\usepackage[compat=0.6]{yquant}
\usepackage{dsfont}
\usepackage{tabularx}
\usepackage[normalem]{ulem}
\usepackage[T1]{fontenc}

\newcommand{\nocontentsline}[3]{}
\newcommand{\tocless}[2]{\bgroup\let\addcontentsline=\nocontentsline#1{#2}\egroup}

\let\oldaddcontentsline\addcontentsline
\newcommand{\stoptocentries}{\renewcommand{\addcontentsline}[3]{}}
\newcommand{\starttocentries}{\let\addcontentsline\oldaddcontentsline}

\usepackage[normalem]{ulem}

\usepackage{lineno}

\begin{document}

\begin{abstract}
The Fermi-Hubbard model is the starting point for the simulation of many strongly correlated materials, including high-temperature superconductors, whose modelling is a key motivation for the construction of quantum simulation and computing devices~\cite{arovas_hubbard_2022, bloch_quantum_2012}. However, the detection of superconducting pairing correlations has so far remained out of reach, both because of their off-diagonal character---which makes them inaccessible to local density measurements---and because of the difficulty of preparing superconducting states~\cite{schlomer_local_2024}.
Here, we report measurement of significant pairing correlations in three different regimes of Fermi-Hubbard models simulated on Quantinuum's Helios trapped-ion quantum computer. Specifically, we measure non-equilibrium pairing induced by an electromagnetic field in the half-filled square lattice model, $d$-wave pairing in an approximate ground state of the checkerboard Hubbard model at $1/6$-doping, and $s$-wave pairing in a bilayer model relevant to nickelate superconductors.
These results show that a quantum computer can reliably create and probe physically relevant states with superconducting pairing correlations, opening a path to the exploration of superconductivity with quantum computers.

\end{abstract}

\title{Superconducting pairing correlations on a trapped-ion quantum computer}

\newcommand{\QTN}{Quantinuum, 303 S Technology Ct, Broomfield, CO 80021, USA}

\newcommand{\QTNC}{Quantinuum, Terrington House, 13-15 Hills Road, Cambridge CB2 1NL, UK}

\newcommand{\QTNBP}{ Quantinuum, 8799 Brooklyn Boulevard, Brooklyn Park, MN 55428, USA}

\newcommand{\QTNUK}{Quantinuum, Partnership House, Carlisle Place, London SW1P 1BX, UK}

\newcommand{\QTNGER}{Quantinuum, Leopoldstrasse 180, 80804 Munich, Germany}

\author{Etienne Granet}
\affiliation{\QTNGER}
\author{Sheng-Hsuan Lin}
\affiliation{\QTNGER}
\author{Kevin H\'emery}
\affiliation{\QTNGER}
\author{Reza Haghshenas}
\affiliation{\QTN}
\author{Pablo Andres-Martinez}
\affiliation{\QTNC}
\author{David T. Stephen}
\affiliation{\QTN}
\author{Anthony Ransford}
\affiliation{\QTN}
\author{Jake Arkinstall}
\affiliation{\QTNC}
\author{M.S. Allman}
\affiliation{\QTN}
\author{Pete Campora}
\affiliation{\QTN}
\author{Samuel F. Cooper}
\affiliation{\QTN}
\author{Robert D. Delaney}
\affiliation{\QTN}
\author{Joan M. Dreiling}
\affiliation{\QTN}
\author{Brian Estey}
\affiliation{\QTN}
\author{Caroline Figgatt}
\affiliation{\QTN}
\author{Cameron Foltz}
\affiliation{\QTN}
\author{John P. Gaebler}
\affiliation{\QTN}
\author{Alex Hall}
\affiliation{\QTN}
\author{Ali Husain}
\affiliation{\QTNBP}
\author{Akhil Isanaka}
\affiliation{\QTN}
\author{Colin J. Kennedy}
\affiliation{\QTN}
\author{Nikhil Kotibhaskar}
\affiliation{\QTNUK}
\author{Ivaylo S. Madjarov}
\affiliation{\QTN}
\author{Michael Mills}
\affiliation{\QTN}
\author{Alistair R. Milne}
\affiliation{\QTNUK}
\author{Annie J. Park}
\affiliation{\QTN}
\author{Adam P. Reed}
\affiliation{\QTN}
\author{Brian Neyenhuis}
\affiliation{\QTN}
\author{Justin G. Bohnet}
\affiliation{\QTN}
\author{Michael Foss-Feig}
\affiliation{\QTN}
\author{Andrew C. Potter}
\affiliation{\QTN}
\author{Ramil Nigmatullin}
\affiliation{\QTNC}
\author{Mohsin Iqbal}
\affiliation{\QTNGER}
\author{Henrik Dreyer}
\email{henrik.dreyer@quantinuum.com}
\affiliation{\QTNGER}

\date{\today}

\maketitle

\tocless\section{Introduction}
Despite decades of study and increasing technological relevance~\cite{coombs_high-temperature_2024}, the mechanism underlying unconventional superconductivity remains unresolved. Theoretical modelling can be invaluable in understanding these materials and informing  experimental and technological progress. The starting point for modelling the electronic correlations in many of these systems is the Fermi-Hubbard model
\begin{equation}
\label{eq_hubbard}
    H_\mathrm{Hubbard} = -\sum_{ij \sigma} t_{ij} c^\dagger_{i \sigma} c_{j \sigma} + U \sum_i n_{i \uparrow} n_{i \downarrow}
\end{equation}
where $t_{ij}$ describes spinful fermions hopping between proximate orbitals on a lattice and $U$ models an on-site Coulomb repulsion. The ability to compute arbitrary observables in this model, including superconducting pairing correlations, would be highly valuable for elucidating the basic mechanisms of superconductivity and connecting with experiments~\cite{qin2022hubbard}.

In the pursuit of this goal, remarkable progress has been made in the development of classical computer simulations, to the point where different methods are starting to agree on ground state predictions for certain experimentally relevant parameters~\cite{xu_coexistence_2024,simons_collaboration_on_the_many-electron_problem_absence_2020}. Yet, extending these capabilities to simulate thermal and non-equilibrium properties essential to phenomena like strange metallicity~\cite{bruin_linear_resistivity} and light-induced superconductivity~\cite{fava_magnetic_2024} remains a formidable challenge.
Indeed, the difficulty of classically simulating quantum many-body systems of this kind was one of the initial motivations to develop quantum simulation platforms, which promise to capture their dynamics in an unbiased way using only polynomially scaling resources. Such devices have become experimental reality, with purpose-built analog simulators now challenging classical simulation techniques in low-but-finite temperature regimes of the Hubbard model~\cite{xu_neutral-atom_2025, bourgund_formation_2025, kendrick2025pseudogapfermihubbardquantumsimulator, chalopin2024probingmagneticoriginpseudogap}, and digital quantum computers making fast progress on their path towards fault-tolerance~\cite{dasu_breaking_2025, lacroix_scaling_2025, sales_rodriguez_experimental_2025}. In the context of superconducitivity, using these platforms to produce and measure a state with superconducting pairing correlations is an essential step towards realising their potential. However, despite the existence of experimental proposals~\cite{mark2025efficiently, schlomer_local_2024}, to the best of our knowledge, no quantum simulation platform has observed superconducting pairing correlations in the repulsive Fermi-Hubbard model to date.

\begin{figure*}[t]
    \centering
\includegraphics[width=0.99\linewidth]{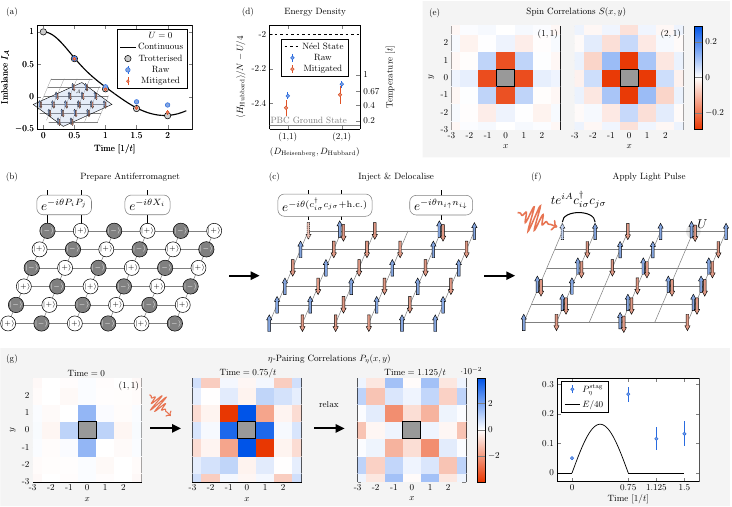}
    \caption{\textbf{The Hubbard model at half-filling and light-induced $\eta$-pairing.} (a) To benchmark the quantum computer,
    a region $\mathcal{A}$ of a periodic $N=6 \times 6$ lattice is densely packed with 36 non-interacting fermions. The system is evolved and the imbalance $I_\mathcal{A}=n_\mathcal{A}-n_{\overline{\mathcal{A}}}$ measured. (b) To prepare low-energy states at $U=8$, an approximate ground state of the $6 \times 6$ Heisenberg model is prepared using a classically optimised circuit with $D_\mathrm{Heisenberg}$ layers. (c) The qubit state is then injected into the fermionic Fock space and the fermions delocalised using a Trotterised adiabatic ansatz circuit with $D_\mathrm{Hubbard}$ layers. (d) Energy and temperature corresponding to a thermal state with the same energy. (e) Spin-spin correlations $S(x,y) = 1/N\sum_i \langle S^z_i S^z_{i+(x,y)} \rangle - \langle S^z_i \rangle \langle S^z_{i+(x,y)} \rangle$ for ansatz circuits with ($D_\mathrm{Heisenberg}, D_\mathrm{Hubbard})$ layers. Avg. (max.) standard error on the mean is 0.011 (0.014).
    (f) The low-energy state is subjected to a light pulse modelled by a time-dependent electric field shown in (g), leading to an increase in $\eta$-pairing correlations~\eqref{eq_Peta}, especially for the staggered average $P_\eta^\mathrm{stag}$. Avg. (max.) standard error on the mean of $P_\eta(x,y)$  is 0.004 (0.009). Note that $S(x,y) = S(-x,-y)$ and $P_\eta(x,y)=P_\eta(-x,-y)$ by definition but we choose to show all sites for visual completeness.\label{fig1}}
\end{figure*}

Here, we report measurement of non-zero pairing correlations in three regimes of the Hubbard model, simulated on a trapped-ion quantum computer. 
First, we show that electromagnetic radiation can induce non-equilibrium $\eta$-type~\cite{yang_ensurematheta_1989} pairing correlations in a half-filled square lattice, whose ground-state is a non-superconducting antiferromagnet. Second, we prepare an approximate ground state of the $1/6$-doped checkerboard model in which $2 \times 2$ Hubbard plaquettes are weakly coupled~\cite{yao_myriad_2007} and observe singlet pairing correlations with $d$-wave symmetry. Finally, we prepare low-energy states of a bilayer Hubbard model relevant to the recently discovered nickelate superconductor $\mathrm{La}_3\mathrm{Ni}_2\mathrm{O}$~\cite{sun_signatures_2023}, and observe $s$-wave pairing in the limit of strong interlayer spin-exchange coupling. 

All of our experiments are carried out on Quantinuum's Helios quantum computer~\cite{to_be_published_helios_nodate}, which operates on 98 effectively all-to-all connected $^{137}\mathrm{Ba}^{+}$ hyperfine qubits. Our results are enabled by several theoretical advancements: (i) we develop ground state preparation techniques based on the injection of states with locally fixed fermionic parity, (ii) we use new time evolution circuits for hopping and exchange terms that balance Trotter error and gate count, and (iii) we adapt and extend local fermionic encodings ~\cite{derby_compact_2021,derby2021compactfermionqubitmapping} and develop new error mitigation strategies.
\\

\tocless\section{Half-Filling}
To benchmark the performance of quantum hardware and software, we begin our experiments with the $6 \times 6$ square lattice Hubbard model at half-filling and nearest-neighbour hopping, which offers opportunities to compare with exact classical simulations. To this end, we first set $t=1$ and $U=0$, initialise a state with 36 non-interacting fermions in a region $\mathcal{A}$ and track the population imbalance between $\mathcal{A}$ and its complement $\overline{\mathcal{A}}$, denoted $I_\mathcal{A} = n_\mathcal{A} - n_{\overline{\mathcal{A}}}$, over time. To carry out the evolution we use a Trotter circuit $U_\mathrm{hop}(\tau t)$ with $\tau$ some Trotter step, which approximates $e^{i \tau t \sum_{\langle ij \rangle \sigma} c^\dagger_{i \sigma} c_{j \sigma} + \mathrm{h.c.}}$ (see Section ~\ref{sec_supplement_trotter_circuits}). Here, and in all other experiments, we use mixed periodic boundary conditions (mPBC), i.e. we average our results over periodic and anti-periodic boundary conditions in the $x$- and $y$-direction, which minimises finite-size effects~\cite{lin_twist-averaged_2001} (see Section~\ref{sec:boundaryconditions}).

In order to capture the fermionic statistics, we use a local fermionic encoding~\cite{derby_compact_2021,derby2021compactfermionqubitmapping, nigmatullin_experimental_2025}, which requires us to prepare a toric code state on 18 ancillae prior to evolution (see Sections~\ref{sec_supplement_fermionic_encoding} and~\ref{sec_supplement_toric_code_preparation}). The 18 stabilisers of the toric code commute with all fermionic bilinears and we choose to measure them simultaneously with $I_\mathcal{A}$. As such, each shot comes with diagnostic information about how much noise has occurred. A simple strategy to exploit that information is to group shots into two sets: those with lower-than-median number of violated stabilisers ("good" shots) and the complement ("bad" shots). Extrapolating the corresponding expectation values yields an error mitigation strategy that avoids the need to run additional, noise-amplified circuits (see Section~\ref{sec_supplement_error_mitigation} for more details).

The results of this method, as well as the raw data are shown in Fig.~\ref{fig1}a. We observe that Trotter errors up to time $2/t$ are small, despite choosing $\tau=0.5 /t$. The raw data tracks the exact values up to time $1/t$. Deviations occur for deeper circuits, although there is still sufficient signal in the raw data for the mitigation to recover the exact result within 20\% relative error (see also section~\ref{supp:moredataeta_imbalance}). While the data supports the utility of the mitigation method, we stress that any error mitigation is only viable in the presence of high-quality raw data and all data presented in this work is raw unless otherwise stated (note that the Helios quantum computer is capable of heralding leakage errors---for a discussion of what "raw" means on such a device see Section~\ref{sec_supplement_leakage_heralding}).

\begin{figure*}[t]
    \centering
\includegraphics[width=\linewidth]{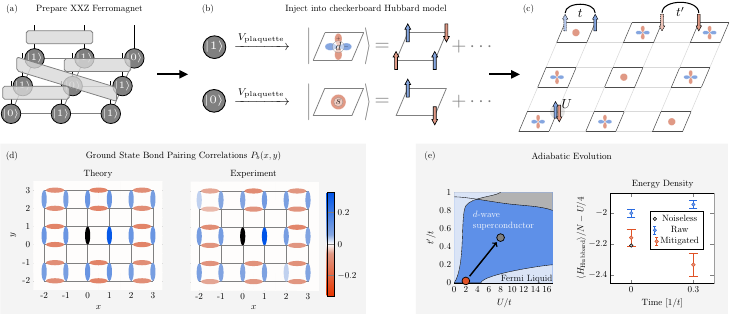}
    \caption{\textbf{$D$-wave pairing in the doped checkerboard model.}
    (a) Using a brickwall circuit, an approximate ground state of the ferromagnetic $3 \times 3$ lattice XXZ model is prepared.
    (b) The state is injected into the fermionic Fock space via a local isometry $V_\mathrm{plaquette}$ which maps the qubit states to the ground states of the $2 \times 2$ Hubbard model in the sector with two $(s)$ and four $(d)$ fermions, respectively.
    (c) The resulting wave function is a superposition of states with three $s$- and six $d$-plaquettes (one configuration shown). It approximates the ground state of the $6 \times 6$ square lattice with inhomogeneous hopping $t \gg t'$, $U/t=2$ and $1/6$ doping.
    (d) Pairing correlations~\eqref{eq_Pb} of the approximate ground state which show $d$-wave symmetry. For bonds on the left of a strong plaquette $P_b(x,y)=P_b(-x,-y)$ by definition, but all bonds are shown for visual completeness. Avg. (max.) standard error on the mean is 0.016 (0.017).
    (e) Sketch of the phase diagram of the checkerboard model at low doping, as conjectured in~\cite{tsai_optimal_2008}, with the prepared state indicated by the red circle (left). We evolve the prepared state by a one-step adiabatic evolution towards $U/t=8,\ t'/t=1/2$ and measure the energy density with respect to the target parameters (right).
    \label{fig2}}
\end{figure*}

After using the exactly solvable limit for validation, we now turn to the interacting case. Specifically, we aim to prepare a low-energy state in the half-filled model at $U/t=8$, which is experimentally relevant to cuprate materials and a standard benchmark for numerical simulations and analog simulator platforms \cite{bohrdt2021exploration,simons_collaboration_on_the_many-electron_problem_solutions_2015}. To this end, we approach the model from its $J = 4t^2/U \rightarrow 0$ limit, at which point the low-energy physics is captured by the isotropic ($\delta=1$) Heisenberg limit of the XXZ model
\begin{equation}
\label{eq_heisenberg}
    H_\mathrm{XXZ} = J \sum_{\langle ij \rangle} X_i X_j + Y_i Y_j + \delta Z_i Z_j.
\end{equation}
For a qubit-based quantum computer, ground state preparation for the Heisenberg model is much simpler than for the Hubbard model, first due to the smaller two-qubit gate count per Trotter step (e.g. 216 vs. 612 for the periodic $6 \times 6$ lattice), and second because one Trotter step in the Heisenberg model evolves the system by a time step $\tau\sim 1/J \gg 1/t$. Operationally, we initialise the quantum computer in a N\'eel state $\ket{-+ \dots -+}$ which is the ground state of $H_x = \sum_i (-1)^\mathrm{sublattice} X_i$ and apply a sequence of unitaries (Fig.~\ref{fig1}b)
\begin{equation}
    \prod_{j=1}^{D_\mathrm{Heisenberg}} e^{-i \theta^j_{xx} H_{xx}} e^{-i\theta^j_{x_2} H_{x}} e^{-i \theta^j_{yy} H_{yy}} e^{-i \theta^j_{x_1} H_{x}} e^{-i \theta^j_{zz} H_{zz}}
\end{equation}
where $H_{xx} = \sum_{\langle ij \rangle} X_i X_j$ and similarly for Pauli-$Y$ and $Z$. The approximate Heisenberg model ground state is now injected into the fermionic Fock space using a depth-1 quantum circuit denoted $V_\mathrm{site}$. We write symbolically
\begin{equation}
\begin{aligned}
    V_\mathrm{site} |0 \rangle_i & \rightarrow c^\dagger_{i \uparrow}|\mathrm{vac} \rangle \\
    V_\mathrm{site} |1 \rangle_i & \rightarrow c^\dagger_{i \downarrow}|\mathrm{vac} \rangle.
\end{aligned}
\end{equation}
Subsequently, fermions are delocalised through the application of the unitary shown in Fig.~\ref{fig1}c:
\begin{equation}
    \prod_{j=1}^{D_\mathrm{Hubbard}} e^{i \theta^j_{S^x} \sum_{i} (-1)^\mathrm{sublattice} S^x_i } U_\mathrm{int}(\theta^j_\mathrm{int}) U_\mathrm{hop}(\theta^j_\mathrm{hop}),
\end{equation}
where $S^x_i = (c^\dagger_{i \uparrow } c_{i \downarrow } + \mathrm{h.c.})/2$ and $U_\mathrm{int}(\theta^j_\mathrm{int}) = e^{-i \theta^j_\mathrm{int} \sum_i n_{i \uparrow} n_{i \downarrow}}$. While we envision the parameter $\theta$ to be obtained from an adiabatic schedule in the general case, here we use parameters that minimise the energy with respect to the Heisenberg (Hubbard) model for the first $D_\mathrm{Heisenberg}$ (last $D_\mathrm{Hubbard}$) layers on a classically tractable $4 \times 4$ lattice.

The results of this strategy are shown in Fig.~\ref{fig1}d. The lowest raw (mitigated) energy density is $\langle H \rangle/N - U/4 = -2.36 \pm 0.02$ $(-2.43 \pm 0.04)$ and is achieved for $(D_\mathrm{Heisenberg}, D_\mathrm{Hubbard}) = (1,1)$, indicating that increasing noise may overcompensate for the increase in variational power of deeper circuits. The ground state energy for periodic boundary conditions is approximately $-2.5278$~\cite{qin_qmc_benchmark_2016}. Note that, throughout this work, we set the energy scale such that the fully depolarised state ($\rho \propto \mathds{1}$) has energy zero.

The ability to separately measure energy density and correlations allows us to characterise the state. On the one hand, despite the relatively shallow preparation circuit, the raw spin-spin correlations in Fig.~\ref{fig1}e show clear antiferromagnetism and their spatial decay is in reasonable agreement with that of a thermal state at temperatures between $t/3$ and $t/2$ 
(see Section~\ref{sec_supplement_qmc}). On the other hand, the lowest achieved raw energy density corresponds to temperatures in the range $\sim t/2 - t$. Thus, the flexibility to measure various observables can be a valuable tool for thermometry and, in this case, allows us to conclude that local correlations in the raw state deviate from those of a perfect thermal state.

Nevertheless, these low-energy states can be used as a resource to investigate non-equilibrium phenomena, including light-induced superconductivity. Specifically,  Ref.~\cite{kaneko_photoinduced_2019} showed that the antiferromagnetic ground states of a 14-site Fermi-Hubbard chain develop doublon pairing correlations
\begin{equation}
\label{eq_Peta}
    P_\eta(x,y) = \frac{1}{N} \sum_i \Delta^\dagger_{i} \Delta_{i+(x,y)}+\mathrm{h.c.}
\end{equation} when exposed to radiation in a range of frequencies around $\sim U=8t$, where $\Delta^\dagger_{i}  = c^\dagger_{i \uparrow} c^\dagger_{i \downarrow}$~\cite{yang_ensurematheta_1989}. Moreover, the sign of $P_\eta$ was found to follow a staggered pattern which stabilises when turning off the radiation and even increases for larger distances.

On Helios, we can investigate this phenomenon at a scale that is likely difficult to access with classical simulations (see Section~\ref{sec:classicaldifficulty}). We start with the state prepared using the $(D_\mathrm{Heisenberg}, D_\mathrm{Hubbard}) = (1,1)$-strategy and apply a short electromagnetic pulse, as shown in Fig.~\ref{fig1}g. The pulse is modelled by a Peierls substitution on all bonds $c^\dagger_{i \sigma} c_{j \sigma} \rightarrow e^{iA(s)} c^\dagger_{i \sigma} c_{j \sigma}$, where the time-dependent vector potential $A(s) = \pi(1-\cos(\omega s))/2$ is aligned with the lattice diagonal and $\omega = 4\pi/3 \approx U/2$. The state is evolved up to time $\pi/ \omega$, using two Trotter steps $U_\mathrm{int}(U\tau/2) U'_\mathrm{hop}(t\tau) U_\mathrm{int}(U\tau) U''_\mathrm{hop}(t\tau) U_\mathrm{int}(U\tau/2)$, where $\tau=\pi/2\omega=0.375$ and $U'_\mathrm{hop}$, $U''_\mathrm{hop}$ differ from the aforementioned $U_\mathrm{hop}$ by single-qubit gates due to the Peierls substitution (see Sections~\ref{sec_supplement_trotter_circuits} and \ref{sec_etasupplement_details}). Finally, we measure $\eta$-pairing correlations both in the initial state and after the pulse has been applied.

Fig.~\ref{fig1}g shows a significant increase of overall pairing correlations induced by the pulse. These correlations are especially strong for nearest neighbours and occur predominantly in a staggered pattern. The staggered average $P^\mathrm{stag}_\eta = \sum_{x,y \neq 0,0} (-1)^\mathrm{sublattice} P_\eta(x,y)$ decreases during a subsequent relaxation implemented via two more Trotter steps of size $\tau = 0.375$ with the field turned off, although it remains significantly larger than the pre-pulse value. Contributions to the staggered average from nearest neighbours decrease during the relaxation, while those from larger distances increase (see Section~\ref{supp:moredataeta}). These results provide evidence that the phenomena observed in Ref.~\cite{kaneko_photoinduced_2019} can carry over to larger lattices, and may even be robust to initial states with non-zero energy density above the ground state and to Hamiltonian imperfections caused by finite Trotter step size.
\\

\tocless\section{The doped checkerboard model}
While we have found pairing correlations in a half-filled model in the previous section, modelling light-matter interactions in cuprates almost certainly requires accounting for lattice distortions as well as dopants~\cite{tsai_superconductivity_2006, zhang_photoinduced_2023, fava_magnetic_2024}. To study both of these concepts, we now introduce 1/6 hole dopants into the $6 \times 6$ checkerboard model~\cite{tsai_superconductivity_2006,yao_myriad_2007}, in which the lattice is partitioned into non-overlapping $2 \times 2$ plaquettes and $t_{ij} = t$ ($t'$) if $\langle ij \rangle$ is a strong (weak) bond within (between) the plaquettes (Fig.~\ref{fig2}c). Lattice inhomogeneities of this kind can either be a natural part of the material structure or exist transiently due to lattice vibrations. An exact diagonalisation study on a $4 \times 4$-lattice showed the checkerboard model to potentially host a $d$-wave superconducting phase in a large region of parameter space (Fig.~\ref{fig2}e)~\cite{tsai_optimal_2008}.

To access this phase experimentally, we follow a strategy inspired by Ref.~\cite{rey_controlled_2009}. Specifically, we consider the limit $t'=0, U/t=2$ at which the plaquettes decouple. The two lowest-energy states of the plaquette are given by a state $|s \rangle$ with $s$-wave symmetry in the quarter-filled sector and a state $|d \rangle$ with $d$-wave symmetry in the half-filled sector~\cite{scalapino_local_1996}.
At 1/6-doping on the $6 \times 6$ lattice, all states with three $|s \rangle$ and six $|d \rangle$ plaquettes are degenerate ground states in this limit. The degeneracy is lifted when turning on hopping $t' > 0$ between the plaquettes. The unique perturbative ground state that emerges is precisely the ground state of the ferromagnetic XXZ model $-H_\mathrm{XXZ}$ at $\delta \approx -1$ and $\sum_i Z_i = -3$ on a $3 \times 3$ lattice (see Section~\ref{sec_supplement_perturbation_theory}).

We exploit this drastic reduction in Hilbert space size to find a brickwall circuit consisting of ten layers of general $U(1)$-symmetry-preserving two-qubit unitaries to prepare the XXZ ground state on the $3 \times 3$ lattice (Fig.~\ref{fig2}a, see also Section~\ref{sec_supplement_state_preparation_circuits}).
Subsequently, we approximate an isometry $V_\mathrm{plaquette}$ which coherently maps the qubit degrees of freedom to the ground states of the $2 \times 2$ plaquettes (Fig.~\ref{fig2}b)
\begin{equation}
\begin{aligned}
        V_\mathrm{plaquette} |0 \rangle &\rightarrow |s \rangle \\
        V_\mathrm{plaquette} |1 \rangle &\rightarrow |d \rangle.
\end{aligned}
\end{equation}
The exact ground state energy density   is $\langle H_\mathrm{Hubbard} \rangle/N - U/4 = -1.27367$, whereas the raw (mitigated) quantum measurement yields $-1.12 \pm 0.01$ ($-1.24 \pm 0.03$).

Finally, we measure singlet pairing correlations between strong bonds
\begin{equation}
\label{eq_Pb}
    P_b(x,y) = \frac{4}{N} \sum_{\langle ij \rangle} \Delta_{ij} \Delta^\dagger_{ij+(x,y)} + \mathrm{h.c.},
\end{equation}
where $\Delta^\dagger_{ij} = \frac{1}{\sqrt{2}} ( c^\dagger_{i \uparrow} c^\dagger_{j \downarrow} - c^\dagger_{i \downarrow} c^\dagger_{j \uparrow} )$ creates a singlet and $\langle ij \rangle$ denotes strong vertical bonds.
The resulting pairing correlations are shown in Fig.~\ref{fig2}d. We observe significant pairing correlations across the lattice. Moreover, the alternating sign between vertical-vertical and vertical-horizontal bond correlations indicates that a state with $d$-wave pairing correlations has been prepared.
To quantify the $d$-wave symmetry, we compute the average over all strong bond-bond correlators relative to a fixed strong vertical bond $\sum_{x,y \neq 0,0} (-1)^\mathrm{orientation} P_b(x,y)/(N-3)$. The experimental (theoretical) value for this $d$-wave average is $0.079 \pm 0.005$ ($0.108$).

The approximate ground state can be used to explore the surrounding superconducting phase. Specifically, we simulate a linear ramp of the parameters towards $U/t=8$, $t'/t= 1/2$, using one second-order Trotter step $U_\mathrm{int} U_\mathrm{hop} U_\mathrm{int}$ of size $\tau=0.3$, where, in this instance, $U_\mathrm{hop}$ implements inhomogeneous hopping by modifying selected rotation angles in the original hopping circuit (see Section~\ref{supplement_fig_plaquette_hopping}). The decrease in mitigated energy density shown in Fig.~\ref{fig2}e is consistent with a cooling effect for the noiseless circuit, although this trend is not visible in the raw data, potentially due to being offset by gate errors.
\\
\tocless\section{A Bilayer Hubbard Model}
\begin{figure*}[t]
    \centering
\includegraphics[width=\linewidth]{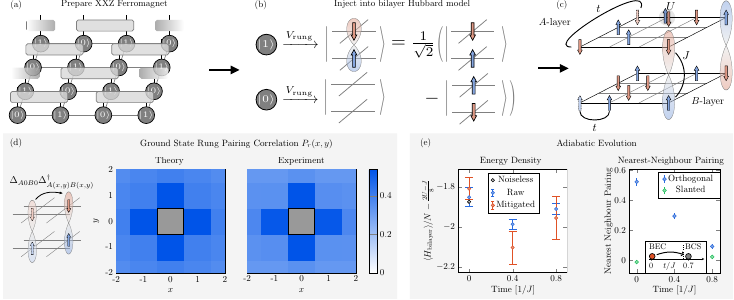}
\caption{\textbf{A Bilayer Hubbard model.} (a) The approximate ground state of a ferromagnetic XXZ model at $\delta=-2/3$ on a periodic $4 \times 4$ lattice is prepared using a brickwall circuit. (b) The state is injected into the Fock space of a $4 \times 4 \times 2$ bilayer Hubbard model by means of an isometry $V_\mathrm{rung}$ which maps the qubit states to either a hole-pair or a singlet on a given rung. (c) The state thus prepared is an approximate ground state of a bilayer Hubbard model in the limit of strong interlayer exchange coupling $J \gg t$.
(d) Rung-rung pairing correlations~\eqref{eq_Pr}. Note that $P_r(x,y)=P_r(-x,-y)$ by definition, although we show all rungs for visual completeness. Avg. (max.) standard error on the mean is 0.0266 (0.0271). (e) The initial state is used to test an adiabatic evolution using $M=1,2$ Trotter steps of size $\tau=0.4$, using a linear ramp towards $t/J=0.7$, while keeping $U=10t$. A lowering of the energy density is observed for $M=1$ before noise heats the state. Pairing correlations between singlets orthogonal to the $A$- and $B$-layers decrease rapidly, but correlations between slanted pairs are likely to increase. Both adiabatic evolutions are complete. 
\label{fig3}}
\end{figure*}

One advantage of an effectively all-to-all connected quantum computer is the ability to simulate arbitrary geometries without changes to the experimental setup. We utilize this capability to simulate a model of the recently discovered bilayer nickelate superconductor $\mathrm{La}_3\mathrm{Ni}_2\mathrm{O}$~\cite{sun_signatures_2023}. This material has been modelled by two quarter-filled square lattice Hubbard models~\eqref{eq_hubbard} with nearest-neighbour hopping and strong inter-layer exchange coupling 
\begin{equation}
    H_\mathrm{bilayer} = H_\mathrm{Hubbard}^A + H_\mathrm{Hubbard}^B + H_\mathrm{exchange},
\end{equation}
where $H_\mathrm{exchange} = J \sum_i \mathbf{S}_{iA} \cdot \mathbf{S}_{iB} - n_{iA} n_{iB}/4$ and $A$ and $B$ denote the two layers of the system~\cite{lu_interlayer-coupling-driven_2024, qu_bilayer_2024, schlomer_superconductivity_2024, oh_type-ii_2023}. In the limit $t/J \rightarrow 0$ (and $U \geq 0$), the ground state subspace is spanned by tightly bound singlets between the layers (Fig.~\ref{fig3}c). Second-order degenerate perturbation theory again leads to a single-layer effective ferromagnetic XXZ model~\eqref{eq_heisenberg}, in this instance with anisotropy $\delta = -2/3$.

Similar to the doped plaquette case, a brickwall circuit which prepares an approximate ground state of the XXZ model on the periodic $4 \times 4$ lattice can be obtained classically (Fig~\ref{fig3}a, cf. also Section~\ref{sec_supplement_state_preparation_circuits}). Subsequently, we apply isometries on rungs (Fig~\ref{fig3}b)
\begin{equation}
\begin{aligned}
    V_\mathrm{rung} |0 \rangle_i &\rightarrow |\mathrm{vac}\rangle_{Ai Bi} \\
    V_\mathrm{rung} |1 \rangle_i &\rightarrow \Delta^\dagger_{Ai Bi}|\mathrm{vac} \rangle_{Ai Bi}.
\end{aligned}
\end{equation}
and measure rung-rung singlet pairing correlations

\begin{equation}
\label{eq_Pr}
    P_r(x,y) = \frac{1}{N} \sum_{i} \Delta_{AiBi} \Delta^\dagger_{Ai+(x,y)Bi+(x,y)} + \mathrm{h.c.}
\end{equation}
Fig.~\ref{fig3}d shows large positive pairing correlations for all rung-rung pairs on the lattice, compatible with theoretical predictions.

In reality, the relevant parameters describing $\mathrm{La}_3\mathrm{Ni}_2\mathrm{O}$ deviate from the perturbative limit,  likely being closer to $t/J=0.7, U/t = 10$~\cite{lu_interlayer-coupling-driven_2024, qu_bilayer_2024} and the tightly coupled singlet pairs were proposed to undergo a BEC-BCS crossover at $t/J \sim 0.6$ in the related $t_\parallel - J_\perp - J_\parallel$ model~\cite{schlomer_superconductivity_2024}. In our final set of experiments, we study the feasibility of an adiabatic evolution from the perturbative ground state towards those parameters. To that end, we consider a short linear ramp of the parameters $(t/J,U/J) = (0,0) \rightarrow (0.7, 7)$ while keeping $J=1$, simulated via $M=1,2$ Trotter steps 
\begin{equation}
    \prod_{m=1}^M U_\mathrm{int}(\tau U_m/2) U_\mathrm{exchange}(\tau J) U_\mathrm{hop}(\tau t_m) U_\mathrm{int}(\tau U_m/2),
\end{equation} where $\tau=0.4$. The relatively large Trotter step size is afforded by the use of novel circuit gadgets (shown in Section~\ref{sec_supplement_trotter_circuits}) to implement the exchange coupling $U_\mathrm{exchange}(\tau J) = e^{-i\tau J \sum_i \mathbf{S}_{iA} \cdot \mathbf{S}_{iB} - n_{iA} n_{iB}/4}$ directly, rather than via a high-frequency expansion~\cite{schlomer_local_2024, bourgund_formation_2025}. Fig.~\ref{fig3}e shows that, although the $M=1$ strategy succeeds at lowering the energy with respect to the target Hamiltonian at $t/J=0.7$, $U/t=10$, heating from hardware errors eventually outcompetes adiabatic cooling for longer evolution times. Nearest-neighbour pairing correlations involving rungs orthogonal to the $A$- and $B$- layers decrease rapidly during the adiabatic evolution, although correlations between "slanted" singlets that involve coordinates $(x,y)$ in the $A$ layer and $(x+1,y)$ in the $B$ layer are likely to increase, indicating that the coupling between the pairs becomes less tight.
\\
\tocless\section{Conclusion}
We have used a trapped-ion quantum computer to simulate the Fermi-Hubbard model on periodic square lattices of up to 72 orbitals, using circuits with up to 90 qubits and 3439 two-qubit gates. These circuits were used to prepare and measure states with non-equilibrium $\eta$-pairing correlations at half-filling, $d$-wave correlations in a doped model of weakly coupled plaquettes, and $s$-wave correlations in a bilayer system. While these results are encouraging for future explorations of superconductivity with quantum computers, many questions remain.

First and foremost are the limits placed on adiabatic algorithms caused by entropy accumulation due to hardware errors. One option that appears sufficient (though maybe not necessary) to reach low-temperature states is to adiabatically evolve large systems for hundreds of hopping times~\cite{xu_neutral-atom_2025}. This is likely achievable with a factor 1000 more Trotter steps than we explored here~\cite{granet_dilution_2025, heyl_quantum_2019, mizuta_trotterization_2025, kovalsky_self-healing_2023}, which translate to per-two-qubit-gate errors on the order of $10^{-6}$. While this requirement is mild compared to other quantum computing applications, understanding the most effective combination of (fermionic) error corrections codes~\cite{algaba_fermion--qubit_2025, chen_error-correcting_2024} and algorithmic techniques like bath-assisted cooling~\cite{matthies_programmable_2024,kishony_efficiently_2025} is essential to reach this goal. Reductions in gate error can also be used to simulate more complex Hamiltonians or longer evolution times which may be used to gain a better microscopic understanding of light-induced superconductivity in experimental setups~\cite{fausti_light-induced_2011, baykusheva_ultrafast_2022, fava_magnetic_2024}.

On the technical side, we have introduced techniques to coherently inject states with locally fixed fermionic parity into fermionic encodings. These circuits can be used to access and explore superconducting phases and they show that ground state preparation for spin models can be useful beyond the study of magnetism. The techniques we have demonstrated can be adapted to other geometries, next-nearest neighbour hopping with a negative sign~\cite{xu_coexistence_2024}, and other Hamiltonian terms, imparting, for example, the ability to introduce dopants coherently.
\stoptocentries
\bibliography{ref_editable}
\starttocentries

\clearpage
\onecolumngrid

\begin{center}
    \textbf{\large Supplemental Material}
\end{center}
\vspace{0.5cm}
\setcounter{section}{0}
\renewcommand{\thesection}{S\arabic{section}} 
\renewcommand{\theequation}{S\arabic{equation}} 
\setcounter{equation}{0}
\renewcommand{\thefigure}{S\arabic{figure}} 
\setcounter{figure}{0}
\renewcommand{\thetable}{S\arabic{table}} 

\begingroup
  \def\contentsname{Table of Contents for the Supplementary Material}
  \setcounter{tocdepth}{1}
  \tableofcontents
\endgroup

\section{Leakage Heralding}
\label{sec_supplement_leakage_heralding}
As is the case on virtually all quantum computing platforms, the ionic qubits used in the Helios quantum computer can be in quantum states that are different from the computational subspace that is spanned by the hyperfine states $\ket{F=1, m_F=0}$ and $\ket{F=2, m_F=0}$, most notably other Zeeman states in the electronic ground state manifold. When reading out qubits at the end of the computation, we utilize a built-in trinary measurement capability \cite{to_be_published_helios_nodate} that returns whether the qubit was 0, 1, or L (the latter indicating it has leaked out of the qubit subspace). The measurement works by first shelving both $\ket{0}$ and $\ket{1}$ to different sets of states in $D_{5/2}$ and fluorescing on a cycling transition ($S\leftrightarrow P_{1/2}\leftrightarrow D_{3/2}$), with fluorescence indicating leakage. Then, the $\ket{1}$ state is deshelved to a combination of $S$ and $D_{3/2}$ states and fluorescence repeated. The four possibilities of that double fluorescence sequence are assigned readout values
\begin{equation*}
\begin{aligned}
(\mathrm{fluorescence,\,fluorescence}) &\rightarrow \textrm{L} \\
(\mathrm{fluorescence,\, no \, fluorescence}) &\rightarrow \textrm{L} \\
(\mathrm{no \, fluorescence,\,fluorescence}) &\rightarrow 1 \\
(\mathrm{no\, fluorescence,\,no \, fluorescence}) &\rightarrow 0 \\
\end{aligned}
\end{equation*}
An exemplary shot on $N=6$ qubits might return the result
\begin{equation*}
    0 \quad 0\quad 0\quad 1\quad 1\quad \textrm{L}.
\end{equation*}
Suppose we want to measure the observable $1/N\sum_{i=1}^N Z_i$ on this shot. How should the leaked qubit be counted in the computation of such an observable? What is the correct value of the "raw" data in this case? One possible strategy is to extend the definition of Pauli-$Z$ to take the value 0 on the leaked subspace, leading to the
\begin{equation*}
\begin{aligned}
    \textrm{L}=0-\mathrm{Strategy}&: (1+1+1-1-1+0)/6 = 1/6.
\end{aligned}
\end{equation*}
A second and third option is to assign $Z$ to $\pm 1$ on the leaked subspace:
\begin{equation*}
\begin{aligned}
    \textrm{L}=+1-\mathrm{Strategy}&: (1+1+1-1-1+1)/6 = 2/6 \\
    \textrm{L}=-1-\mathrm{Strategy}&: (1+1+1-1-1-1)/6 = 0.
\end{aligned}
\end{equation*}
However, all of these three strategies are somewhat arbitrary in that a particular value of $\langle \textrm{L} | Z | \textrm{L} \rangle$ has to be chosen by the user. One desideratum of "raw" data is that minimal assumptions are required to obtain it. In this case, we argue that $Z$ should not be assigned an arbitrary numerical value on the leaked subspace, similar to the use of NaN (Not a Number) in classical computing. This leads to the
\begin{equation*}
\begin{aligned}
    \textrm{L}=\mathrm{NaN-Strategy}&: (1+1+1-1-1)/5 = 1/5.
\end{aligned}
\end{equation*}
It is the NaN-Strategy that we use to report data labeled "Raw" in this work.

\begin{figure}[t!]
    \centering
    \includegraphics[scale=0.8]{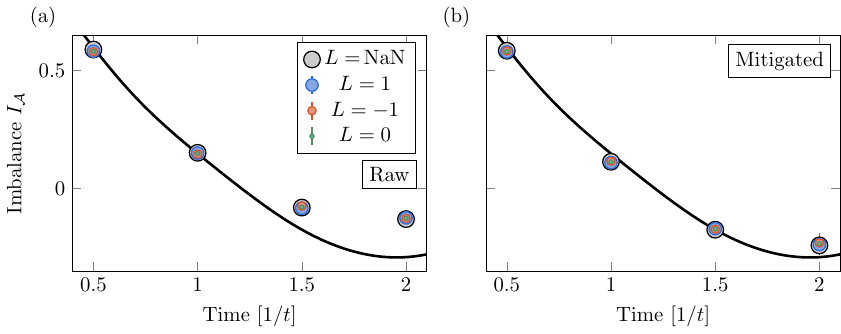}
    \caption{\textbf{Comparison of the} $\textrm{L}=0$, $\textrm{L}=\pm 1$ \textbf{and} $\textrm{L}=$NaN \textbf{strategies.} Expectation values obtained for the imbalance benchmark setup. (a) shows unmitigated data, and (b) mitigated data with the noise mitigation described in section \ref{sec_supplement_error_mitigation}. The exact value (without Trotter error) is shown with the black continuous line as indication.}
    \label{fig:leakage}
\end{figure}

In practice, as shown in Fig.~\ref{fig:leakage} all of the four strategies lead to quantitatively similar data. Of course, in the limit of zero leakage, all strategies must yield the same values and in fact, the good agreement between all strategies may be due to the overall small proportion of leaked qubits across the different experiments, as shown in Table~\ref{table_leakage}: From the imbalance experiments, we observe that around $1\%$ of the qubits leak per Hubbard-Trotter step (that is, $612$ two-qubit gates and depth $18$). Although the proportion of leaked qubits is small, because the experiments use $90$ qubits, it is rather unlikely that a shot contains no leaked qubit, except for the shallowest circuits. For this reason, post-selecting the shots onto no leakage events is not a viable approach.

\begin{table}[b!]
\label{table_leakage}
\begin{tabular}{ |c||c|c||c|c||c| } 
 \hline
 experiment & \begin{tabular}{@{}c@{}}proportion  \\ of leaked qubits\end{tabular}  &\begin{tabular}{@{}c@{}}proportion of shots  \\ without leaked qubits\end{tabular}    &number of shots & \begin{tabular}{@{}c@{}}number of  \\ two-qubit gates\end{tabular}   &date \\ 
 \hline
   \hline
 imbalance $1$ step & $0.015$ & $0.255$ & $200$&$631$& 29 - 30 Aug\\ 
   \hline
 imbalance $2$ steps & $0.021$ & $0.135$ &$200$&$1243$& 29 - 30 Aug\\ 
   \hline
 imbalance $3$ steps & $0.028$ & $0.09$  &$200$&$1855$& 4 Sept\\ 
   \hline
 imbalance $4$ steps & $0.039$ & $0.025$ &$200$&$2467$& 6 - 21 Sept\\ 
 \hline
 energy $(1,1)$ & $0.017$ & $0.2$ &$1500$&$919-1027$& 6 - 8 Sept \\ 
 \hline
 energy $(2,1)$ & $0.021$ & $0.13$ &$1693$&$1207-1315$& 29 Aug - 7 Sept\\ 
 \hline
 eta before pulse & $0.018$ & $0.22$&$400$& $955$& 3 - 4 Sept\\ 
   \hline
 eta after pulse & $0.032$ & $0.07$ &$400$&$2215$& 6 - 7 Sept\\ 
   \hline
 eta after pulse + 1 step  & $0.043$ & $0.022$ &$225$&$2827$& 6 - 10 Oct\\ 
   \hline
 eta after pulse + 2 steps  & $0.053 $ & $0.022$&$225$&$3439$& 6 - 10 Oct\\
   \hline
   \hline
doped pairing correlation & $0.020 $ & $0.17$ & $2000$ & $1057$  & 26 Aug - 18 Sept\\
   \hline
doped energy (step-0)  & $0.020$ & $0.16$ & $1100$ & $1021-1165$  & 29 - 30 Aug\\
   \hline
doped energy (step-1)  & $ 0.024$ & $ 0.11$ & $1100$ & $1669-1813$  & 20 - 27 Oct\\
   \hline
   \hline
Bilayer state prep. &$0.0095$ & $0.5$ &$1300$&$368-498$& 26 Aug - 6 Sept\\
\hline 
Bilayer 1 step & $0.019$ &$0.19$ &$1400$ &$1228-1324$&31 Aug - 6 Sept \\
\hline
Bilayer 2 steps & $0.02901$&$0.102$ &$1300$ & $2060-2150$& 8 - 18 Sept \\
\hline
Bilayer state prep., slanted & $0.0130$ &$0.34$ &$400$ &$498$& 18 Sept - 15 Oct \\
\hline
Bilayer 2 steps, slanted&$0.028$ &$0.095$&$400$ &$2092$&18 Sept - 15 Oct \\
\hline
\end{tabular}

\caption{\textbf{Qubit leakage statistics}. Per Hubbard-Trotter steps we observe a roughly $1 \%$ increase in the number of leaked qubits.
}
\end{table}



Let us also note that the NaN-strategy and the $\textrm{L}=0$ - strategy are related by a computation subspace projector in the following way. Let us consider an observable $\mathcal{O}$ defined in the computational space that is a sum of Pauli strings with coefficients that are equal in absolute value. We normalize the observable by the number of such Pauli strings, so that the measured value is between $-1$ and $1$
\begin{equation}
    \mathcal{O}=\frac{1}{N_P}\sum_n P_n=\frac{\sum_n P_n}{\sum_n P_n^2}\,,
\end{equation}
with $N_P$ the number of Pauli strings in the sum. The second equality holds true because every Pauli string squares to $1$. We extend this observable to the space $0$, $1$, L by
\begin{equation}
    \mathcal{O}=\frac{\sum_n \tilde{P}_n}{\sum_n \tilde{P}_n^2}\,,
\end{equation}
where $\tilde{P}_n$ acts as $P_n$ on the computational space $0,1$ where it can take values among $\pm 1$, and takes value $0$ whenever one of the qubits in the support of $P_n$ has leaked. $\tilde{P}_n^2$ is not the identity anymore, but takes value $1$ in the computational space, and value $0$ if one of the qubits in the support of $P_n$ has leaked. This rule exactly corresponds to discarding the Pauli strings that involve leaked qubits, and to normalize the overall sum by the number of Pauli strings taken into account.

The NaN strategy may have a larger variance of the estimator with respect to the other two strategies, but, for a given leakage rate, this difference is constant with system size $N$ and thus all strategies are equally scalable.

\newpage
\section{Fermionic Encodings}
\label{sec_supplement_fermionic_encoding}

One difficulty in measuring pairing correlations is that their local fermionic parity is odd when restricted to a single spin species. Previous Hubbard model simulations on quantum computers have used encodings in which the up- and down-fermions are encoded separately~\cite{arute_observation_2020, stanisic_observing_2022, hemery_measuring_2024, nigmatullin_experimental_2025}. On the one hand, this makes it relatively easy to find efficient circuits for the time evolution under the hopping terms of the Hamiltonian $-t \sum_{\langle ij \rangle, \sigma} c^\dagger_{i \sigma} c_{j \sigma}$. On the other hand, in such encodings, measuring distant pairing correlations requires extended Jordan-Wigner strings, which makes it difficult to simultaneously measure many correlators, and leads to high-weight operators, which are highly susceptible to noise~\cite{granet_dilution_2025}. 

\begin{figure*}[t!]
    \centering
\includegraphics[width=1\linewidth]{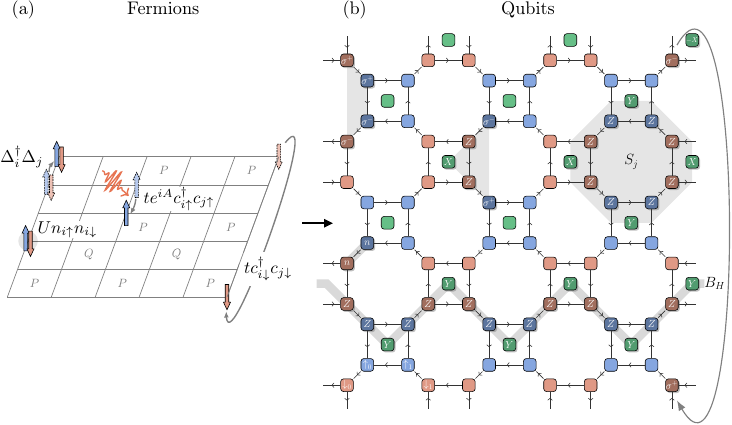}
\caption{\textbf{Octagon Fermion-to-Qubit Encoding for the operators and observables in the single-layer Hubbard Model.} Each fermionic site is represented by an up-qubit (blue) and a down-qubit (red). Ancilla qubits (green) are added into all odd faces of the lattice. Nearest-neighbour hopping operators $c^\dagger_{i \sigma} c_{j \sigma}$ are mapped to three-qubit operators (boundary-crossing-term shown) and five-body operators (triangular shape) which act on one ancilla each. On-site repulsion $U n_{i \uparrow} n_{i \downarrow}$ is implemented using to the qubit $n=(1-Z)/2$-operator. Including both up- and down-fermions in the same encoding allows us to simultaneously measure many $\eta$-pairing correlation observables $\Delta^\dagger_i \Delta_j = c^\dagger_{i \uparrow} c^\dagger_{i \downarrow} c_{j \uparrow} c_{j \downarrow}$ (trapezoidal shape). To avoid the presence of magnetic fluxes through even plaquettes, we initialise the stabilisers $S_j = +1$, by preparing a toric code state on the green qubits and a fixed fermionic parity state on the blue and red qubits. The logical operator $B_H$ sets the magnetic flux through one handle of the torus (+1 for anti-periodic, -1 for periodic boundary conditions). Not shown is another logical operator $B_V$, which is obtained by rotating $B_H$ by 90 degrees and swapping $Y \rightarrow X$ on the ancilla qubits.
\label{supplement_fig_mapping_half_filled}}
\end{figure*}

To overcome these difficulties, in Figure~\ref{supplement_fig_mapping_half_filled}, we show an "Octagon" fermion-to-qubit encoding. The encoding rules are closely related to the Compact Encoding~\cite{derby_compact_2021, derby2021compactfermionqubitmapping} and are as follows: Each site $i$ of the square-lattice Hubbard model is encoded by two qubits, one representing the "up" (blue, $\sigma = \uparrow$) and one representing the "down" (red, $\sigma = \downarrow$) fermionic modes. A \textit{diagonal} edge is placed between each such pair. The square lattice is partitioned into even and odd plaquettes, corresponding to a checkerboard pattern in Fig \ref{supplement_fig_mapping_half_filled}a, with the plaquette at the south-west corner being an odd plaquette. Then, the odd plaquettes are further partitioned into $P$-plaquettes and $Q$-plaquettes, as illustrated in Fig \ref{supplement_fig_mapping_half_filled}a, so that plaquettes of one type contain disjoint sets of qubits. Edges are now placed between all nearest-neighbour "up"-qubits in the $P$-plaquettes and between all nearest-neighbour "down"-qubits in the $Q$-plaquettes. Depending on their geometry, these edges are \textit{horizontal} or \textit{vertical} edges. Within each odd plaquette, an ancilla qubits is placed. Thus each horizontal and vertical edge $(i \sigma ,j \sigma)$ has a unique ancilla $a(i,j)$ associated to it. Edges are assigned an orientation, as shown via the arrows in Figure~\ref{supplement_fig_mapping_half_filled}. Finally, identification between fermionic modes and qubit operators is made by associating to each edge $(i \sigma,j \sigma')$ the Majorana bilinear $A_{i \sigma, j \sigma'} = -i \gamma_{i \sigma} \gamma_{j \sigma'}$, where $\gamma_{i \sigma} = c^\dagger_{i \sigma} + c_{i \sigma}$ and the qubit operator $A_{i \sigma, j \sigma'}  = X_{i \sigma} Y_{j \sigma'} P$, where $P$ is $Y_{a(i,j)}$ for horizontal edges, $-X_{a(i,j)}$ for down-pointing vertical edges, $X_{a(i,j)}$ for up-pointing vertical edges, and $\mathds{1}$ for diagonal edges. Note that the order of the indices of $A_{i \sigma, j \sigma'}$ follows the orientation of arrows on the lattice (in this case $i \sigma \rightarrow j \sigma')$, and reversed operators are defined as $A_{j \sigma', i \sigma} = -A_{i \sigma, j \sigma'}$. To complete the algebra of fermionic bilinears, we associate to each mode $i \sigma$ an on-site term $B_{i \sigma} = -i \gamma_{i \sigma} \overline{\gamma_{i \sigma}} \rightarrow Z_{i \sigma}$, where $\overline{\gamma_{i \sigma}} = i(c^\dagger_{i \sigma} - c_{i \sigma})$. Edge operators between non-adjacent modes $(i \sigma,k \sigma')$ can be defined if there exists a path of edges $\{(j_1, j_2), \dots (j_{l-1}, j_l) \}$, via $A_{i,k} = i^{l-1} \prod_{m=1}^{l-1} A_{j_m, j_{m+1}}$, where $j_1 = i \sigma$ and $j_l = k \sigma'$. For paths $j_1, \dots j_l$ that form closed loops $j_l = j_1$, one associates an operator $S = i A_{j_1, j_l}$. The eigenvalues of $S=\pm 1$ indicate whether or not there is a $\pi$-magnetic flux through the enclosed area. If the flux-free sector is desired, one needs to ensure $S=1$ for all closed loops (this is the case for all topologically trivial loops of the experiments in this work, except for the closed loops that wrap around the torus, which are set to $\langle S\rangle=0$ to obtain mixed boundary conditions, see Section \ref{sec:boundaryconditions}). Note that, in the qubit representation, the constraints on minimal loops that are topologically trivial are given by equations of the form $Z^{\otimes 8} XYXY = +1$, whereas if periodic boundary conditions were desired then one would need to set the qubit operators for minimal topologically non-trivial loops on the $6 \times 6$ lattice (Figure~\ref{supplement_fig_mapping_half_filled}) $Z^{\otimes 12} Y^{\otimes 6} = -1$ and $Z^{\otimes 12} X^{\otimes 6} = -1$, i.e., there is a  negative sign for the non-trivial operators. The difference in sign arises due to the fact that the topologically trivial elementary loops touch an odd number of up-pointing vertical edges (namely, one), which are associated with a $-X$ operator on the ancilla.

Using these rules, we can take products and linear combinations of Majorana bilinears in order to write down the qubit expressions for the operators and observables used in the experiments on the single-layer Hubbard model described in the main text:
\begin{itemize}
    \item The on-site number operator $n_{i \sigma} \rightarrow (1-Z_{i \sigma})/2$. 
    \item The on-site repulsion $n_{i \uparrow} n_{i \downarrow} \rightarrow (1-Z_{i \uparrow})(1-Z_{i \downarrow})/4$. 
    \item The spin-spin correlator $S^z_i S^z_j = (n_{i \uparrow} - n_{i \downarrow}) (n_{j \uparrow} - n_{j \downarrow})/4 \rightarrow (Z_{i \downarrow} - Z_{i \uparrow}) (Z_{j \downarrow} - Z_{j \uparrow})/16$. Note that in the Octagon Encoding of the single layer Hubbard model it would also be straightforward to measure spin correlators along other axes, e.g. $S^x_i = (c^\dagger_{i \uparrow} c_{i \downarrow} + \mathrm{h.c.})/2 \rightarrow (X_{i \uparrow} X_{i \downarrow} + Y_{i \uparrow} Y_{i \downarrow})/4$ and $S^y_i = 2i S^x_i S^z_i \rightarrow ( X_{i \uparrow} Y_{i \downarrow}-Y_{i \uparrow} X_{i \downarrow})/4$.
    \item The nearest-neighbour hopping operator $c^\dagger_{i \sigma} c_{i \sigma} + \mathrm{h.c.} \rightarrow (X_{i \sigma} X_{j \sigma} + Y_{i \sigma} Y_{j \sigma}) P/2$, for nearest-neighbour bonds associated with an edge $\langle ij \rangle$ within a $P$-plaquette ($Q$-plaquette) for $\sigma = \uparrow$ ($\sigma = \downarrow$) and the operator $P$ is $Y_{a(i,j)}$ for horizontal edges, $-X_{a(i,j)}$ for down-pointing vertical edges, $X_{a(i,j)}$ for up-pointing vertical edges. Half of all nearest-neighbour hopping terms are of this form.
    
    \item For the other half of nearest-neighbour hopping terms on bonds associated with nearest-neighbour edges $\langle ij \rangle$ within a $P$-plaquette ($Q$-plaquette) for $\sigma = \downarrow$ ($\sigma = \uparrow$), we obtain a five-body operator that arises from multiplication of edge operators that go through the other species $\bar{\sigma}$, i.e. $c^\dagger_{i \sigma} c_{i \sigma} + \mathrm{h.c.} \rightarrow (X_{i \sigma} X_{j \sigma} + Y_{i \sigma} Y_{j \sigma}) Z_{i \bar{\sigma}} Z_{j \bar{\sigma}} P/2$. Again, $P$ is $Y_{a(i,j)}$ for horizontal edges, $-X_{a(i,j)}$ for down-pointing vertical edges, $X_{a(i,j)}$ for up-pointing vertical edges.
    \item Pairing correlators of the $\eta$-kind $\Delta^\dagger_i \Delta_j = c^\dagger_{i \uparrow} c^\dagger_{i \downarrow} c_{j \uparrow} c_{j \downarrow}$. Since we work in a fixed particle number sector, we have $\langle c_{i \uparrow} c_{i \downarrow} c_{j \uparrow} c_{j \downarrow}\rangle = \langle c^\dagger_{i \uparrow} c^\dagger_{i \downarrow} c^\dagger_{j \uparrow} c^\dagger_{j \downarrow}\rangle = 0$, and so for $i\neq j$
    \begin{equation}
    \begin{aligned}
        \langle \Delta^\dagger_i \Delta_j + \mathrm{h.c.} \rangle &= \langle (c^\dagger_{i \uparrow} c^\dagger_{i \downarrow} c_{j \downarrow} c_{j \uparrow} + c_{i \uparrow} c_{i \downarrow} c_{j \uparrow} c_{j \downarrow}) + \mathrm{h.c.}\rangle \\ &= \langle (c^\dagger_{i \uparrow} c^\dagger_{i \downarrow} + c_{i \downarrow} c_{i \uparrow}) (c^\dagger_{j \uparrow} c^\dagger_{j \downarrow} + c_{j \downarrow} c_{j \uparrow})\rangle \\
        &=(-1)^{i+j} \langle(X_{i \uparrow} X_{i \downarrow} - Y_{i \uparrow} Y_{i \downarrow}) (X_{j \uparrow} X_{j \downarrow} - Y_{j \uparrow} Y_{j \downarrow}) \rangle/4
    \end{aligned}
    \end{equation}
    which is a sum of weight-4 operators independent of the distance between $i$ and $j$. The term $(-1)^{i}$ and $(-1)^{j}$ takes values $\pm 1$ if site $i,j$ is even or odd in a checkerboard pattern. This sign factor comes from the fact that with our fermionic encoding we have
    \begin{equation}
\begin{aligned}
X_{j\uparrow}X_{j\downarrow}&=c^\dagger_{j\uparrow}c_{j\downarrow}+c^\dagger_{j\downarrow}c_{j\uparrow}\mp (c^\dagger_{j\uparrow}c^\dagger_{j\downarrow}+c_{j\downarrow}c_{j\uparrow})\\
Y_{j\uparrow}Y_{j\downarrow}&=c^\dagger_{j\uparrow}c_{j\downarrow}+c^\dagger_{j\downarrow}c_{j\uparrow}\pm (c^\dagger_{j\uparrow}c^\dagger_{j\downarrow}+c_{j\downarrow}c_{j\uparrow})\,,
\end{aligned}
\end{equation}
with $\pm$ alternating signs in a checkerboard pattern, because of the alternating direction of the arrows linking qubits $i\uparrow$ and $i\downarrow$ in Fig.~\ref{supplement_fig_mapping_half_filled}(b).

    \item Singlet-pairing correlations $\Delta^\dagger_{ij} \Delta_{kl} + \mathrm{h.c.}$, where  $\Delta^\dagger_{ij} = \frac{1}{\sqrt{2}} ( c^\dagger_{i \uparrow} c^\dagger_{j \downarrow} - c^\dagger_{i \downarrow} c^\dagger_{j \uparrow} )$ creates a singlet on a vertical or horizontal bond $\langle ij \rangle$. In a fixed particle sector, we have $\langle \Delta_{ij} \Delta_{kl}\rangle = \langle \Delta^\dagger_{ij} \Delta^\dagger_{kl}\rangle = 0$, so that we can write 
    \begin{equation}
    \begin{aligned}
        \langle \Delta^\dagger_{ij} \Delta_{kl} + \mathrm{h.c.} \rangle &= \langle (\Delta_{ij} + \Delta^\dagger_{ij})( \Delta_{kl} + \Delta^\dagger_{kl} )\rangle \\
        &\rightarrow \langle [(X_{i \uparrow} X_{j \downarrow} - Y_{i \uparrow} Y_{j \downarrow})Z_{i \uparrow \rightarrow j \downarrow} + (X_{i \downarrow} X_{j \uparrow} - Y_{i \downarrow} Y_{j \uparrow})Z_{i \downarrow \rightarrow j \uparrow} ]P_{ij} \\
        &\times [(X_{k \uparrow} X_{l \downarrow} - Y_{k \uparrow} Y_{l \downarrow})Z_{k \uparrow \rightarrow l \downarrow} + (X_{k \downarrow} X_{l \uparrow} - Y_{k \downarrow} Y_{l \uparrow})Z_{k \downarrow \rightarrow l \uparrow} ]P_{kl}\rangle/8
    \end{aligned}
    \end{equation}
    where $Z_{i \sigma \rightarrow j \bar{\sigma}}$ is either $Z_{i \bar{\sigma}}$ or $Z_{j \sigma}$, whichever qubit is the one connecting $i \sigma$ and  $j \bar{\sigma}$ and $P_{ij}$ is $Y_{a(i,j)}$ for horizontal edges, $-X_{a(i,j)}$ for down-pointing vertical edges, $X_{a(i,j)}$ for up-pointing vertical edges. This observable is thus a sum of weight-8 Pauli observables for any pair of bonds $\langle ij \rangle, \langle kl \rangle$.
\end{itemize}

The same techniques can be adapted to simulate the bilayer Hubbard model, as shown in Figure~\ref{supplement_fig_mapping_bilaer}. One qubit is again assigned to each individual fermionic mode $i \sigma$. The key difference to the single-layer case is that not every pair of qubits is connected via edge operators. Instead, we define two disconnected qubit subsystems. For clarity, let us point out that the two subsystems are \textit{not} encoding the $A$- and $B$-layers separately. Instead, we place all "up"-modes of layer $B$ and all "down"-modes of layer $A$ in subsystem 1 and all "down"-modes of layer $B$ and all "up"-modes of layer $A$ in subsystem 2. By doing so, we sacrifice the ability to efficiently implement terms like single-site $S^x$ which now would have odd fermionic parity within each subsystem.
However, we can still use the same strategy to map hopping operators as in the single-layer case, which will turn out to be much more efficient than had we attempted to place all qubits into a single connected component. What's more, the following bilayer-specific operators can still be implemented and measured with system-size independent cost:

\begin{figure*}[t!]
    \centering
\includegraphics[width=1\linewidth]{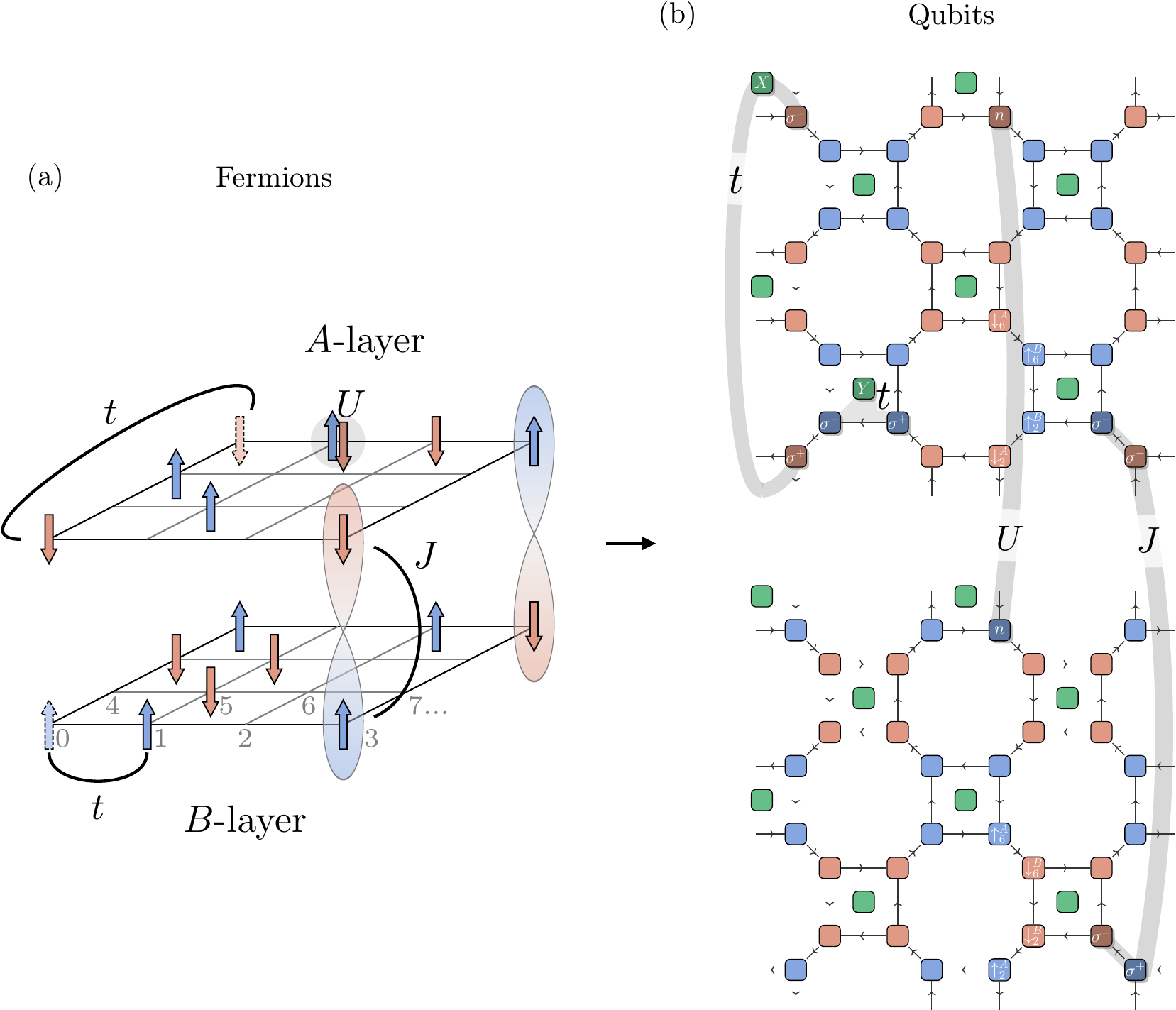}
\caption{\textbf{Octagon Fermion-to-Qubit Encoding for the operators and observables in the bilayer Hubbard Model.} (a) Hamiltonian terms in the bilayer Hubbard model that is used to model bilayer nickelates~\cite{lu_interlayer-coupling-driven_2024, qu_bilayer_2024, schlomer_superconductivity_2024, oh_type-ii_2023}. The terms within each layer are the same as for in the single-layer case, namely a nearest-neighbour hopping $-t \sum_{\langle ij \rangle \sigma} c^\dagger_{i \sigma} c_{j \sigma}$ and an on-site interaction $U \sum_i n_{i \uparrow} n_{i \downarrow}$. The layers are coupled via an exchange coupling $J \sum_i \mathbf{S}_{iA} \cdot \mathbf{S}_{iB} - n_{iA}n_{iB}/4$. (b) The system is mapped to two separate Octagon Encodings, one containing all "up"-modes of layer $B$ and all "down"-modes of layer $A$ (top) and one containing all "down"-modes of layer $B$ and all "up"-modes of layer $A$ (bottom). This is possible because all three Hamiltonian terms, as well as the rung-singlet-pairing observable $\Delta^\dagger_{AiBi} \Delta_{AjBj}$ have even fermionic parity in each subsystem.
\label{supplement_fig_mapping_bilaer}}
\end{figure*}

\begin{itemize}
    \item The exchange interaction
    \begin{equation}
\label{eq_exchange_mapping}
    \begin{aligned}
        \mathbf{S}_{iA} \cdot \mathbf{S}_{iB} - \frac{n_{iA}n_{iB}}{4} &= S^x_{iA} S^x_{iB} + S^y_{iA} S^y_{iB} + S^z_{iA} S^z_{iB} - \frac{n_{iA}n_{iB}}{4} \\
        & = S^x_{iA} S^x_{iB} + S^y_{iA} S^y_{iB} -\frac{n_{iA \uparrow }n_{iB \downarrow} + n_{iA \downarrow }n_{iB \uparrow}}{2} \\
        & = \frac{c^\dagger_{iA \uparrow} c_{iA \downarrow} c^\dagger_{iB \downarrow} c_{iB \uparrow} + \mathrm{h.c.}}{8} -\frac{n_{iA \uparrow }n_{iB \downarrow} + n_{iA \downarrow }n_{iB \uparrow}}{2} \\
        &= \frac{ (c^\dagger_{iA \uparrow} c^\dagger_{iB \downarrow})(c_{iB \uparrow} c_{iA \downarrow}) + \mathrm{h.c.}}{8} -\frac{n_{iA \uparrow }n_{iB \downarrow} + n_{iA \downarrow }n_{iB \uparrow}}{2} \\
        & \rightarrow \frac{ (\sigma^+_{iA \uparrow} \sigma^+_{iB \downarrow})(\sigma^-_{iB \uparrow} \sigma^-_{iA \downarrow}) + \mathrm{h.c.}}{8} -\frac{n_{iA \uparrow }n_{iB \downarrow} + n_{iA \downarrow }n_{iB \uparrow}}{2} \\
    \end{aligned}
    \end{equation}
    with qubit operators $\sigma^{\pm} = (X \mp iY)/2$ and $n=(1-Z)/2$. While circuits for \textit{simulating} the exchange term are found in section~\ref{sec_supplement_trotter_circuits}, for the \textit{measurement} of the exchange energy in a fixed particle sector, one may use the fact that $\langle c^\dagger_{iA \uparrow} c^\dagger_{iB \downarrow} c^\dagger_{iB \uparrow} c^\dagger_{iA \downarrow}\rangle = \langle c_{iA \uparrow} c_{iB \downarrow} c_{iB \uparrow} c_{iA \downarrow}\rangle = 0$ to simplify the above operator expression
    \begin{equation}
        \begin{aligned}
        \left \langle \mathbf{S}_{iA} \cdot \mathbf{S}_{iB} - \frac{n_{iA}n_{iB}}{4} \right \rangle &= \frac{ (c^\dagger_{iA \uparrow} c^\dagger_{iB \downarrow} + \mathrm{h.c.})(c_{iB \uparrow} c_{iA \downarrow} + \mathrm{h.c.})}{8} -\frac{n_{iA \uparrow }n_{iB \downarrow} + n_{iA \downarrow }n_{iB \uparrow}}{2} \\
        &= \left \langle\frac{(X_{iA \uparrow} X_{iB \downarrow}-Y_{iA \uparrow}Y_{iB \downarrow})(X_{iB \uparrow}X_{iA \downarrow}-Y_{iB \uparrow}Y_{iA \downarrow})}{32} - \frac{n_{iA \uparrow }n_{iB \downarrow} + n_{iA \downarrow }n_{iB \uparrow}}{2} \right \rangle.
    \end{aligned}
    \end{equation}
    Measurement of the exchange energy is thus reduced to measuring sums of weight-2 and weight-4 Pauli operators. Note that the overall sign is positive for all $i$, due to the fact that in our arrow convention we have $\{(iA \uparrow \rightarrow iB \downarrow ), (iB \uparrow \rightarrow iA \downarrow \}$ or $\{(iA \uparrow \leftarrow iB \downarrow ), (iB \uparrow \leftarrow iA \downarrow) \}$ for each site $i$.
    \item The singlet pairing correlation between rungs $i$ and $j$, $\Delta^\dagger_{AiBi} \Delta_{AjBj}$, where  $\Delta^\dagger_{AiBi} = \frac{1}{\sqrt{2}} ( c^\dagger_{Ai \uparrow} c^\dagger_{Bi \downarrow} - c^\dagger_{Ai \downarrow} c^\dagger_{Bi \uparrow} )$ creates a singlet on rung $i$. Since we have placed $Ai \downarrow$ and $Bi \uparrow$ next to each other on a diagonal bond, no $Z$-strings or fSWAPs are required to map
    \begin{equation}
    \begin{aligned}
        \langle \Delta^\dagger_{AiBi} \Delta_{AjBj} + \mathrm{h.c} \rangle &= \langle (\Delta_{AiBi} + \Delta^\dagger _{AiBi}) (\Delta_{AjBj} + \Delta^\dagger _{AjBj}) \rangle \\
        & \rightarrow \frac{(-1)^\mathrm{orientation}}{8} \left \langle (X_{iA \uparrow} X_{iB \downarrow}-Y_{iA \uparrow}Y_{iB \downarrow}) (X_{jA \uparrow} X_{jB \downarrow}-Y_{jA \uparrow}Y_{jB \downarrow}) \right \rangle
    \end{aligned}
    \end{equation}
    where we have again used that $ \langle \Delta^\dagger_{AiBi} \Delta^\dagger_{AjBj} \rangle = \langle \Delta_{AiBi} \Delta_{AjBj} \rangle = 0$ in a fixed particle number sector.  The orientation-dependent sign originates from the fact that on some sites we have oriented edge operators pointing from $iA \uparrow$ to $iB \downarrow$, whereas for other sites the orientation is from $iB \downarrow$ to $iA\uparrow$. Whenever the orientation at \textit{either} $i$ \textit{or} $j$ is from $B \downarrow$ to $A \uparrow$, the expectation value has to be multiplied with $-1$.
    For the computation of slanted pair-pair correlators, we apply fSWAPs on one layer to swap e.g., $j \leftrightarrow k$.
\end{itemize}

\begin{figure*}[t]
    \centering
\includegraphics[width=\linewidth]{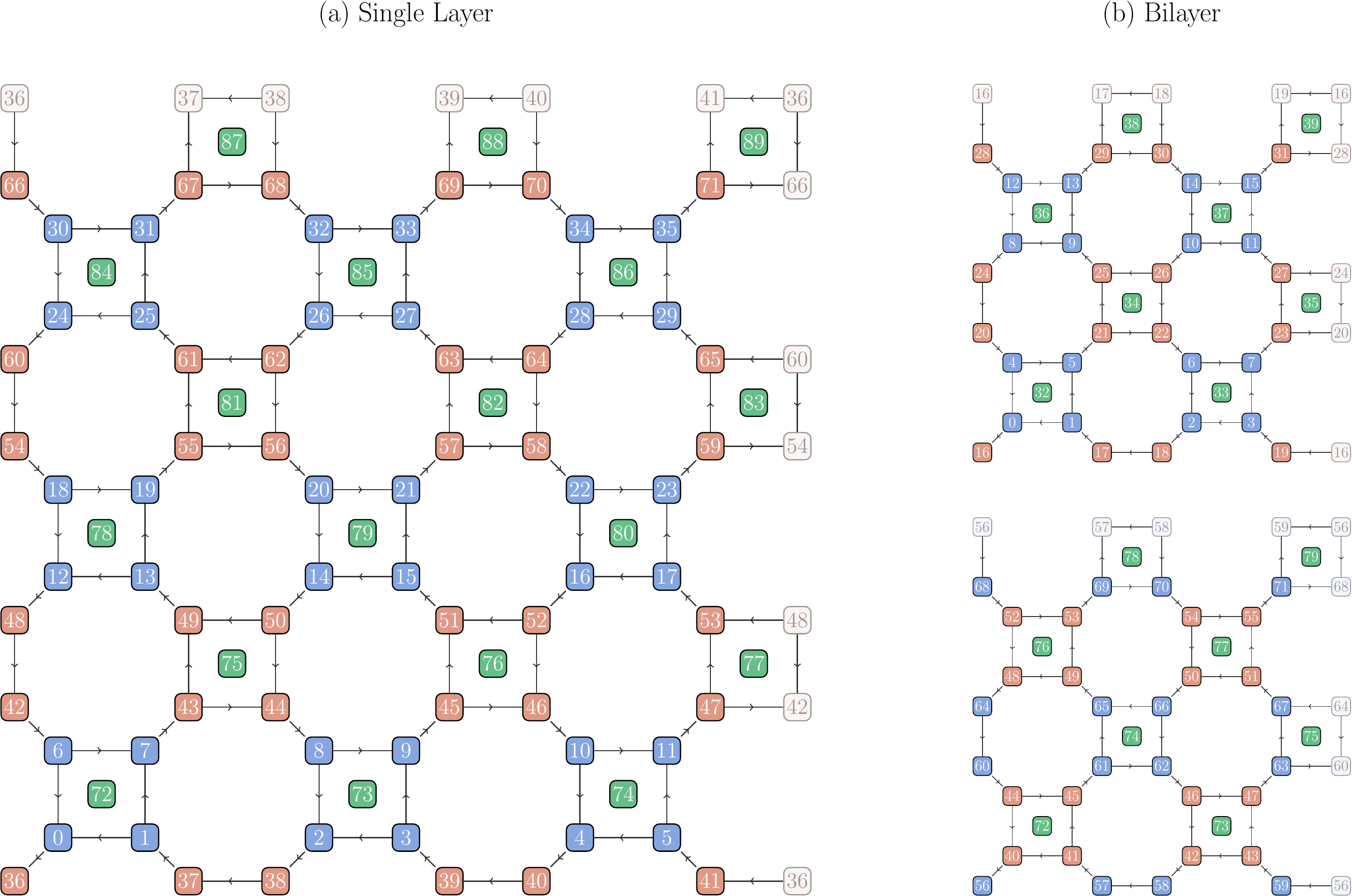}
    \caption{\textbf{Qubit Labeling.} (a) Qubit labeling for the single layer Hubbard model simulation circuits. (b) Qubit labeling for the bilayer Hubbard model simulation circuits.\label{supplement_fig_qubit_labeling}}
\end{figure*}

For reference, the qubit labeling that is used in the actual circuits and data analysis is presented in Figure~\ref{supplement_fig_qubit_labeling}, for both the single layer and bilayer settings.

\newpage
\section{Error Mitigation}
\label{sec_supplement_error_mitigation}
The objective of this section is to present the stabiliser-based noise mitigation technique used in the main text.

\subsection{Stabilisers \label{sec:stabilizers}}
We briefly recall the expression of the stabilisers presented in section \ref{sec_supplement_fermionic_encoding}. The fermionic encoding that we use involves $N/2$ "local" stabilisers $S$ that are associated to every even plaquette of the lattice, where $N$ is the number of sites. They read
\begin{equation}
    S=Y_{a\uparrow} Y_{a\downarrow} X_{a\rightarrow}X_{a\leftarrow} \prod_{j\in {\rm face}}Z_j\,,
\end{equation}
where $a\uparrow,a\downarrow,a\rightarrow,a\leftarrow$ are the four ancillas around the plaquette, and where the product of $Z_j$ runs over the $8$ sites of the plaquette, $4$ of which belong to the down-spin lattice and the other $4$ to the up-spin lattice. So in total, these stabilisers are $12$-body operators. We note that, among these $N/2$ stabilisers, only $N/2-2$ are independent, because of the operator identities $\prod_{\rm all} S=1$ and $\prod_{\rm even}S=\prod_j Z_j$. On top of these "local" stabilisers, there also exist $2$ "winding" stabilisers. These two "winding" stabilisers correspond to the logical qubits of the toric code when used as a quantum error code. In our case, they will play a role in the boundary conditions, as explained in section \ref{sec:boundaryconditions}. We will not take them into account for the noise mitigation technique.

All these stabilisers (local or winding) commute with the hopping terms and interaction terms, so their value is conserved in all quantum circuits implemented in this work. The local stabilisers are initialized to $S=1$ by preparing the ground state of the toric code on the ancillas. Let us denote $U$ the unitary operator that corresponds to this toric code state preparation. This operator maps the stabilisers onto just $Z$ strings on the ancillas (the Paulis on the system qubits are untouched)
\begin{equation}
    U S U^\dagger= \prod_{k \subset {\rm ancillas}} Z_k \prod_{j\in {\rm face}}Z_j\,.
\end{equation}
The precise form of the stabilisers after the application of this toric code state preparation is given in section \ref{sec_supplement_toric_code_preparation}. Hence, by applying the inverse toric code preparation at the end of the circuit, and measuring all the qubits in the $Z$ basis, one can measure the value of the stabilisers. More generally, this holds true whenever the measurement setting on the system (face) qubits allows us to access the global fermionic parity $\prod_{j\in {\rm face}}Z_j$, which is the case for all the circuits run on hardware. From every shot on the quantum computer, we thus get two pieces of information: the value of an observable $\mathcal{O}$ computed on this shot, and the stabiliser syndrome, i.e. the list of stabilisers measurements. We are now going to explain how to use this additional information to perform noise mitigation on the observable.

\subsection{Computing stabilisers in the presence of leakage}
As explained in section \ref{sec_supplement_leakage_heralding}, every qubit can be measured as $0$, $1$ or L when it is leaked. There, we explained that when computing expectation values of observables, leaked qubits are interpreted as NaN and the corresponding Pauli strings are discarded in the average. A similar approach can be implemented for stabilisers, by assigning to stabilisers the values $+1$ or $-1$ (if none of their $12$ qubits have leaked) or L (if at least one of their $12$ qubits has leaked). However, for the purpose of performing noise mitigation based on stabiliser information, this approach combined with the NaN strategy of section \ref{sec_supplement_leakage_heralding} would lead to over-mitigate leakage. Indeed, the NaN strategy of section \ref{sec_supplement_leakage_heralding} already mitigates the effect of leakage errors (although not completely, because leakage can have an effect on other qubits than the leaked qubit).

To avoid mitigating leakage twice, we would thus like to not take into account any information about leakage in the computation of the stabilisers. Of course, we cannot know what the stabilisers would have been if a qubit had not leaked. Any choice to interpret leaked qubits will thus be imperfect. However, the most reasonable strategy is to always interpret leaked qubits to be in the $|1\rangle$ state when computing stabilisers, i.e., the $\textrm{L}=-1$ strategy. Indeed, if we did not implement the leakage detection gadgets in the circuit, for hardware reasons any leaked qubit would have been measured in the $|1\rangle$ basis. Adopting the $\textrm{L}=-1$ strategy for the stabilisers only corresponds thus to discarding the additional information brought by leakage measurement gadgets. Moreover, half of the time (more precisely, whenever the qubit would have been measured in the $|1\rangle$ state if it has not leaked) this assignment does not introduce errors. In the other cases, this assignment does introduce a bit flip error. This error is  milder than setting the stabiliser to be leaked, because a leaked stabiliser cannot be un-leaked by other errors, contrary to bit-flips, and so has statistically more weight than a bit-flip. For the purpose of the mitigation, we will thus adopt the $\textrm{L}=-1$ strategy when computing stabilisers. We note that this argument is not used to derive a rigorously optimal (or even unbiased) error mitigation strategy.

\subsection{Effect of noise on stabilisers}
\begin{figure}
    \centering
    \includegraphics[scale=0.7]{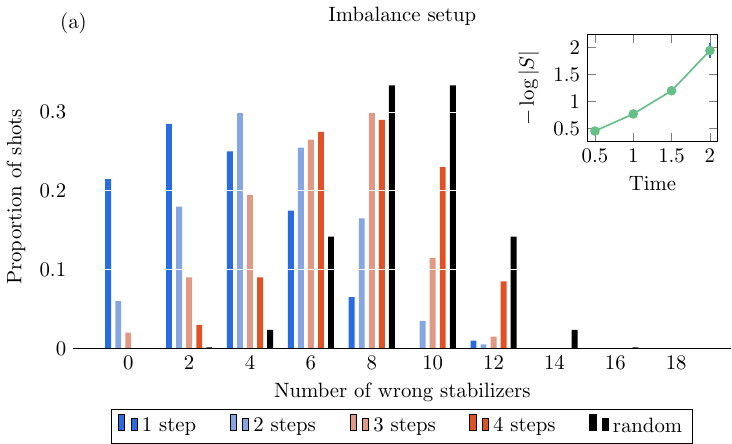}
    \includegraphics[scale=0.7]{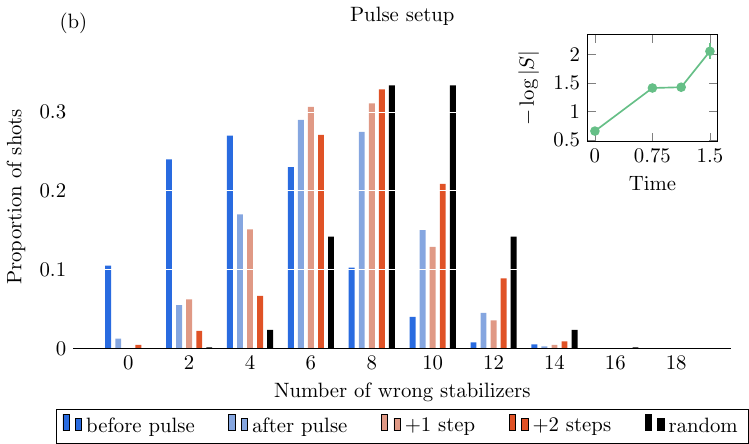}
    \caption{\textbf{Proportion of shots as a function of the number of wrong local stabilisers for different experiments.} The inset shows the quantity $-\log |S|$ as a function of time, with $S=\frac{1}{18}\sum_{j=1}^{18}S_j$ with $S_j$ the $18$ different local stabilisers.}
    \label{fig:stabilizers}
\end{figure}

Given a list of shots, with the value of an observable and the stabiliser syndrome for each shot, the simplest noise mitigation is to post-select the shots on $0$ wrong stabilisers. This noise mitigation has two shortcomings. The first one is that it does not remove all errors, because (i) some errors are undetectable as they do not flip any stabilisers (for a depolarizing noise channel, in our circuits around one error out of five is undetectable) and because (ii) errors can flip a stabiliser twice making those combinations of errors undetectable. However, this postselection still removes an appreciable part of errors in circuits, and is expected to significantly decrease the bias due to noise in the expectation values. The second and more important shortcoming of this noise mitigation is that it is not scalable: the proportion of shots with zero flipped stabilisers decreases exponentially with both the number of qubits and the depth of the circuit. For many of the circuits we ran on hardware, the proportion of shots with $0$ wrong stabilisers is very small, which would lead to high variance of the post-selected expectation value. The objective of the noise mitigation that we introduced is to alleviate the second shortcoming of post-selecting onto $0$ wrong stabilisers, without improving the first shortcoming. Namely, the noise mitigation will, under certain assumptions, have the same expectation value as post-selection, but will have much smaller variance. To illustrate our point, we display in Table~\ref{tab:stabilizers} the proportion of shots with a certain number of wrong stabilisers, for different experiments run on hardware. We also show in Fig.~\ref{fig:stabilizers} histograms of these quantities, as well as their values for completely random strings of bits.

\begin{table}[b!]
\begin{tabular}{ |c|c|c|c|c|c|c|c|c|c|c| } 
 \hline
 number of violated stabilisers & $0$ & $2$ & $4$ & $6$ & $8$ & $10$ & $12$ & $14$ & $16$ & $18$ \\ 
 \hline
   \hline
 imbalance $1$ step & $0.215$ & $0.285$ & $0.25$ & $0.175$ & $0.065$ & $0$ & $0.01$ & $0$ & $0$ & $0$ \\ 
   \hline
 imbalance $2$ steps & $0.06$ & $0.18$ & $0.3$ & $0.255$ & $0.165$ & $0.035$ & $0.005$ & $0$ & $0$ & $0$ \\ 
   \hline
 imbalance $3$ steps & $0.02$ & $0.09$ & $0.195$ & $0.265$ & $0.3$ & $0.115$ & $0.015$ & $0$ & $0$ & $0$ \\ 
   \hline
 imbalance $4$ steps & $0$ & $0.03$ & $0.09$ & $0.275$ & $0.29$ & $0.23$ & $0.085$ & $0$ & $0$ & $0$ \\ 
   \hline
 energy $(1,1)$ & $0.11$ & $0.23$ & $0.28$ & $0.24$ & $0.10$ & $0.027$ & $0.085$ & $0.003$ & $0$ & $0$ \\ 
   \hline
 energy $(2,1)$ & $0.067$ & $0.18$ & $0.27$ & $0.27$ & $0.14$ & $0.057$ & $0.015$ & $0.001$ & $0$ & $0$ \\ 
   \hline
 eta before pulse & $0.105$ & $0.24$ & $0.27$ & $0.23$ & $0.1025$ & $0.04$ & $0.0075$ & $0.005$ & $0$ & $0$ \\ 
   \hline
 eta after pulse & $0.0125$ & $0.055$ & $0.17$ & $0.29$ & $0.275$ & $0.15$ & $0.045$ & $0.0025$ & $0$ & $0$ \\ 
   \hline
 eta after pulse + 1 step  & $0.0622$ & $0.151$ & $0.307$ & $0.311$ & $0.129$ & $0.0356$ & $0.004$ & $0$ & $0$ & $0$ \\ 
   \hline
 eta after pulse + 2 steps  & $0.0044$ & $0.022$ & $0.067$ & $0.271$ & $0.329$ & $0.209$ & $0.089$ & $0.0089$ & $0$ & $0$ \\ 
   \hline
   \hline
 doped pairing correlation & $0.081$ & $0.2435$ & $0.316$&$0.231$&$0.099$& $0.025$& $0.004$ &    $0.0005$ & $0 $& $0$ \\ 
   \hline
 doped energy (step-0) & $0.103$ & $0.268$ & $0.301$ &  $0.202$ & $0.09$ & $0.0309$ & $0.005$ & $0$ & $0$ & $0$\\ 
   \hline
doped energy (step-1) & $0.0282$ &  $0.107$ &  $0.245$ & $0.295$ &  $0.215$ &  $0.092$ &
 $0.0155$ &  $0.0018$ &  $0.0009$ & $0$\\ 
\hline
\hline
(random, single layer) & $0$ & $0.0012$ & $0.0233$ & $0.142$ & $0.333$ & $0.333$ & $0.142$ & $0.0233$ & $0.0012$ & $0$ \\ 
\hline
\hline
Bilayer state prep. & $0.303$ & $0.35625$ & $0.1975$ & $0.1125$ & $0.02625$ & $0.0025$ & $0$ & $0.00125$& $0$ &\\
   \hline
Bilayer 1 step &$0.0325$ & $0.1244$ & $0.2321$ & $0.3153$ & $0.2134$ & $0.0732$ & $0.0091$ & $0$ & $0$ &\\
    \hline
Bilayer 2 steps& $0.0043$ & $0.0405$ & $0.1448$ & $0.3195$ & $0.3238$ & $0.1438$ & $0.0224$ & $0.0010$ & $0$& \\
   \hline
Bilayer state prep., slanted & $0.0059$ & $0.1500$ & $0.3735$ & $0.2794$ & $0.1441$ & $0.0412$ & $0.0059$ & $0$ & $0$ &\\
   \hline
Bilayer 2 steps, slanted & $0.0025$ & $0.0300$ & $0.1225$ & $0.3275$ & $0.3350$ & $0.1525$ & $0.0300$ & $0$ & $0 $& \\
\hline
\hline
(random, bilayer) & $0$ & $0.0034$ & $0.0568$ & $0.243$ & $0.394$ & $0.243$ & $0.0568$ & $0.0034$ & $0$ &  \\ 
\hline
\end{tabular}
\caption{\textbf{Proportion of number of violated local stabilisers in the experiments.}
}
    \label{tab:stabilizers}
\end{table}

\begin{figure}
    \centering
    \includegraphics[width=0.8\textwidth]{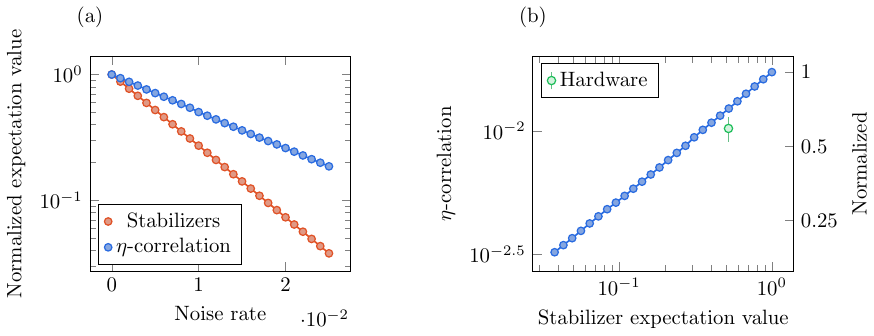}
    \caption{\textbf{Noisy simulation of the circuit for the $(1,1)$ state preparation on $6\times 6$, measuring the $\eta$-pairing correlation $\Delta_0^\dagger \Delta_1 + \mathrm{h.c.}$ and the local stabiliser sum $\frac{1}{18}\sum_j S_j$}. (a): expectation values normalized with their noiseless value, as a function of the depolarizing noise rate after every two-qubit gate. (b): $\eta$-pairing as a function of the stabiliser value on a log-log scale. Essentially perfect exponential decay of the signal (the ratio noisy/noiseless) is observed. The effect of hardware and depolarising noise of the same magnitude (measured by average stabiliser expectation value) on the observable is similar, despite the fact that hardware noise almost certainly is biased and contains non-Pauli error channels. The observable decay is thus much more dependent on the overall noise strength than the details of the noise model.
    }
    \label{fig:effect_of_noise}
\end{figure}

 In Fig \ref{fig:effect_of_noise}, we investigate the effect of noise on stabilisers and on local observables through numerical simulations. We simulate the circuit used for preparing the $(D_\mathrm{Heisenberg}, D_\mathrm{Hubbard})=(1,1)$ state with a Pauli string decomposition, see Section \ref{sec:psd}. We study the effect of depolarizing noise on the circuit, at a fixed Pauli string truncation threshold $\epsilon=0.1/2^{15}$. We observe that both the $\eta$-pairing correlation and the average stabiliser expectation value undergo an exponential decay with the noise rate. This very smooth behaviour of expectation value with depolarizing noise is to be put in contrast with truncation error in classical numerical simulations techniques (see for example section~\ref{sec:classicaldifficulty}). We see however that the stabilisers (which are weight $12$ Pauli strings) are damped significantly faster than the $\eta$-correlations (which are weight $4$). Such a strong effect of observable weight on noise sensitivity was observed and explained before~\cite{granet_dilution_2025}. We also compare these classical simulations with the values from the quantum computer for both $\eta$-pairing and  the stabilisers. The hardware value is compatible with the classical depolarising noise simulations within two standard errors on the mean.

\subsection{Zero-Wrong-Stabiliser extrapolation}
We now present and justify the noise mitigation we develop. It consists in using the information about the number of wrong stabilisers to perform a "zero-wrong-stabiliser" extrapolation, instead of the usual "zero-noise" extrapolation. Let us first note that there is no one-to-one correspondence between number of errors and number of wrong stabilisers. Indeed, different errors can flip a different number of stabilisers (for example, a $X$ error on the system qubits flips $2$ stabilisers, and a $Z$ error on an ancilla flips $4$ stabilisers), some errors are undetectable, and two errors can flip the same stabiliser twice. However, on average, at small error rates per qubit, the average number of errors $e$ can be approximated by a function of the form $e\approx a+bw$ with $w$ the average number of wrong stabilisers, and $a,b$ some fixed coefficients, because at small error rates per qubit it is unlikely that two errors flip the same stabiliser. Further assuming that the expectation value has a linear behaviour with the number of errors, we can thus assume that the expectation value $m$ has a linear behaviour with the number of wrong stabilisers
\begin{equation}
    m=\alpha+\beta w\,,
\end{equation}
with $\alpha,\beta$ some coefficients. The coefficient $\alpha$ is exactly the value of the expectation value post-selected onto $0$ wrong stabilisers. 

Let us bin the shots into different buckets indexed by $n$. We compute the expectation value of an observable $m_n$ with standard deviation $\sigma_n$ within bucket $n$, as well as an average number of wrong stabilisers $w_n$. To estimate the coefficients $\alpha,\beta$, one can then do a weighted fit on the $m_n$. This entails minimizing the cost function
\begin{equation}
    C=\sum_{n\geq 0} \frac{(\alpha+\beta w_n-m_n)^2}{\sigma_n^2}\,.
\end{equation}
The solution is
\begin{equation}
    \alpha=\frac{\sum_{n\geq 0}m_n \frac{\Sigma_2-w_n\Sigma_1}{\sigma_n^2}}{\Sigma_2\Sigma_0-\Sigma_1^2}\,,
\end{equation}
with
\begin{equation}
    \Sigma_k=\sum_{n\geq 0}\frac{w_n^k}{\sigma_n^2}\,.
\end{equation}
Neglecting the variance on the estimate of the variances $\sigma_n^2$, the standard deviation obtained on the estimate of $\alpha$ is
\begin{equation}
\begin{aligned}
    \sigma_\alpha^2&=\frac{1}{(\Sigma_2\Sigma_0-\Sigma_1^2)^2}\sum_{n\geq 0}\frac{(\Sigma_2-w_n\Sigma_0)^2}{\sigma_n^2}\\
    &=\frac{\Sigma_0 \Sigma_2(\Sigma_2-2\Sigma_1+\Sigma_0)}{(\Sigma_2\Sigma_0-\Sigma_1^2)^2}\,.
\end{aligned}
\end{equation}
Let us now further assume that $\sigma_n^2$ is proportional to $1/N_n$ with $N_n$ the number of shots in the bucket $n$, namely $\sigma_n^2=\lambda/N_n$ with some $\lambda$. Then
\begin{equation}
    \Sigma_0=\lambda \sum_{n\geq 0}N_n\,,\qquad \Sigma_1=\lambda \sum_{n\geq 0}w_n N_n
\end{equation}
are both independent of the choice of the binning, because $\Sigma_0/\lambda$ is the total number of shots, and $\Sigma_1/\lambda$ the total number of wrong stabilisers across all shots. However, $\Sigma_2$ depends on the choice of buckets, and so does the variance $\sigma_\alpha^2$. In practice, if there are too few shots per bucket, there will be significant imprecision on the estimates of $\sigma_n^2$, which will increase the variance $\sigma_\alpha^2$ compared to the expression above which neglects this effect.

To mitigate this variance on the estimated variance, we will consider the case of only two buckets $0$ and $1$. The two buckets are defined as follows: for a certain cutoff $0<\kappa<1$, bucket $0$ contains the proportion of shots $\kappa$ with the lowest number of wrong stabilisers, and bucket $1$ contains the remaining proportion of shots $1-\kappa$ with the largest number of wrong stabilisers. We thus have by construction $w_0\leq w_1$. In case of two buckets, we have the mitigated value $\alpha=\frac{w_1 m_0-w_0 m_1}{w_1-w_0}$, with standard deviation
\begin{equation}\label{stdmit}
    \sigma_\alpha^2=\sigma_0^2 \left(\frac{w_1}{w_1-w_0}\right)^2+\sigma_1^2 \left(\frac{w_0}{w_1-w_0}\right)^2\,.
\end{equation}
We recall that this standard deviation does \emph{not} take into account possible variation in the bucketing. The mitigation leads to reweighting one subset of shots by a given number $w_1/(w_1-w_0)$, and the rest of the shots by $-w_0/(w_1-w_0)$, and $\sigma_\alpha$ takes into account statistical variations of the expectation value within these subsets, not statistical variations in the definition of the subsets. We will check below that this latter effect is indeed negligible, and that $\sigma_\alpha$ gives a very good approximation of the actual standard deviation.

Let us denote $N_{0,1}$ the number of shots in bucket $0,1$ and $W_{0,1}$ the total number of wrong stabilisers across different shots within bucket $0,1$. Let us consider the action of removing a shot with $k$ wrong stabilisers from bucket $1$ and putting it in bucket $0$. The derivative of $\sigma_\alpha^2$ with respect to the number of such shots is
\begin{equation}
    \partial \sigma_\alpha^2=\frac{(W_0+W_1)^2(W_1 N_0+W_0 N_1-2k N_0 N_1)}{(W_0 N_1-W_1 N_0)^3}\,.
\end{equation}
Since $W_0/N_0< W_1/N_1$, the denominator is negative. It follows that the sign of $\partial \sigma_\alpha^2$ is the sign of
\begin{equation}
    s=2k-w_1-w_0\,.
\end{equation}
We can thus decrease the variance if $2k<w_0+w_1$. The generic optimization of the proportion $\kappa$ of shots is a complicated problem and will depend on the distribution of number of wrong stabilisers among the shots. To go forward, we must make some assumptions about this distribution. Since errors flip stabilisers in an independent way, the statistics of number of shots with a given number of wrong stabilisers is given by a sum of binomial distributions with different probabilities. For a large system, this sum will be approximated by a Gaussian distribution due to the law of large numbers. This distribution is thus symmetric around its mean. In that case, the value of $k$ that gives $s=0$ is given by the mean of this Gaussian distribution, and the cutoff is $\kappa=1/2$, because then we have $w_{0,1}=\mu\pm \sigma$ with $\mu,\sigma$ the mean and standard deviation of the distribution. The two buckets should be chosen so as to have the same number of shots each, one containing the shots with the lowest number of wrong stabilisers, and the other one the shots with the largest number of wrong stabilisers.

\subsection{Summary of the noise mitigation strategy}

The mitigation strategy that we defined and used in this manuscript works as follows.
\begin{enumerate}
    \item Compute the number of wrong stabilisers in each shot.
    \item Find the cutoff $w$ defined as the smallest integer such that there are more shots with $\leq w$ wrong stabilisers than shots with $>w$ wrong stabilisers. Define bucket $0$ as the set of shots with $\leq w$ wrong stabilisers and bucket $1$ as the set of shots with $>w$ wrong stabilisers.
    \item Compute the expectation value of an observable within these two buckets, obtaining $m_{0,1}$ with standard deviation $\sigma_{0,1}$. Compute the average number of wrong stabilisers $w_{0,1}$ within these two buckets.
    \item The mitigated value is
    \begin{equation}
        m_{\rm mit}=\frac{w_1 m_0-w_0 m_1}{w_1-w_0}\,,
    \end{equation}
    and the standard deviation is 
\begin{equation}\label{stdmit2}
    \sigma_{\rm mit}=\sqrt{\sigma_0^2 \left(\frac{w_1}{w_1-w_0}\right)^2+\sigma_1^2 \left(\frac{w_0}{w_1-w_0}\right)^2}\,.
\end{equation}
\end{enumerate}

\subsection{Numerical check of the mitigation variance \eqref{stdmit2}}

We now come back to the theoretical standard deviation $\sigma_\alpha$ given in \eqref{stdmit2}. We would like to check that the possible statistical variation in the definition of the buckets (not taken into account in this formula) is negligible. To that end, we would need in principle to repeat several times the same experiment on hardware with a given number of shots $N_S$, apply the noise mitigation on each of these experiments, and compare the theoretical variance \eqref{stdmit2} with the actual variance observed across the results of these experiments. In practice, it is of course unreasonable to repeat a large number of times the same experiment on hardware, given the scarcity of quantum computer runtime. Numerical simulations in tractable system sizes like $4\times 2$ ($20$ qubits) are not appropriate neither, because there are only three possible number of wrong stabilisers ($0$, $2$ or $4$) and the noise mitigation will be degenerate in this case. To overcome this problem, we note that the question that we want to solve is a question purely about statistics, and is unrelated to the fact that the data is obtained with a quantum computer. We thus proceed as follows. From a given experiment in Table \ref{tab:stabilizers}, we generate random values of number of wrong stabilisers $w=0,2,4,...,18$, drawn from their measured probability distribution, and for each such random value, we generate a random expectation value with mean $\alpha+\beta w$. Specifically we set arbitrary values $\alpha=-0.4$ and $\beta=0.022$, that in any case do not interfere with the bucketing. For a set of $N_S=200$ such values, we then apply our error mitigation and obtain a mitigated value $\bar{\alpha}$ and a theoretical standard deviation $\sigma_\alpha$ computed with \eqref{stdmit2}. This process can be repeated a large number of times. We then compare the standard deviation of $\bar{\alpha}$ obtained across different such processes, to the average value of $\sigma_\alpha$ obtained within each process. We show in Table \ref{tab:testmit} the results of these numerical tests when averaging over $10^4$ processes. We observe that the theoretical standard deviation is indeed very close to the actual standard deviation, and even slightly larger in all cases.

\begin{table}[h]
\label{table_stabilisers}
\begin{tabular}{ |c|c|c| } 
 \hline
 Stabilisers drawn from ... &\begin{tabular}{@{}c@{}}Average theoretical\\ standard deviation\end{tabular}   &\begin{tabular}{@{}c@{}}Actual\\ standard deviation\end{tabular}  \\ 
 \hline
 imbalance $1$ step & $0.02093$ & $0.02056$\\ 
   \hline
 imbalance $2$ steps & $0.02904$ & $0.02838$\\ 
   \hline
 imbalance $3$ steps & $0.03720$ & $0.03660$\\ 
   \hline
 imbalance $4$ steps & $0.04856$ & $0.04850$\\ 
   \hline
\end{tabular}
\caption{\textbf{Test of the precision of the theoretical standard deviation \eqref{stdmit2}.}}
\label{tab:testmit}
\end{table}

\newpage
\section{Trotter circuits for time evolution}

\label{sec_supplement_trotter_circuits}
In this section we discuss the primitives that carry out the evolution under the fermionic hopping with and without minimal coupling to an electromagnetic field ($-t \sum_{\langle ij \rangle \sigma} e^{iA} c^\dagger_{i \sigma} c_{j \sigma} + \mathrm{h.c.}$), on-site fermionic interaction ($U \sum_i n_{i \uparrow} n_{i \downarrow})$ and $A$-$B$-layer exchange interaction ($J \sum_i \mathbf{S}_{iA} \cdot \mathbf{S}_{iB} - n_{iA}n_{iB}/4$). 

\subsection{Hopping and interaction terms}

\begin{figure*}[b!]
    \centering
\includegraphics[width=0.99\linewidth]{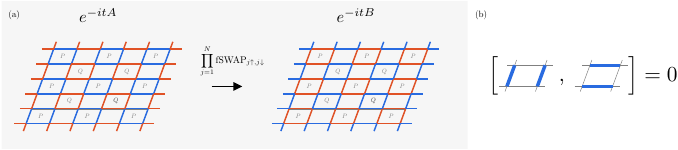}
\caption{\textbf{Trotter splitting of the hopping term. Blue colour denotes hopping of "up"-spins and red colour denotes hopping of "down"-spins.} (a) The lattice is decomposed into $P$-type and $Q$-type plaquettes. The simulation of the hopping Hamiltonian $A+B$ is split into two steps, which simulate hopping of "up"-spins on $P$ and "down"-spins $Q$ and vice versa in sequence. (b) $A$ and $B$ do not need to be further Trotterised because hopping within a plaquette can be simulated exactly, due to cancellations in the commutator.}
\label{supplement_fig_hopping_strategy}
\end{figure*}

\begin{figure*}[t]
    \centering
\includegraphics[width=\linewidth]{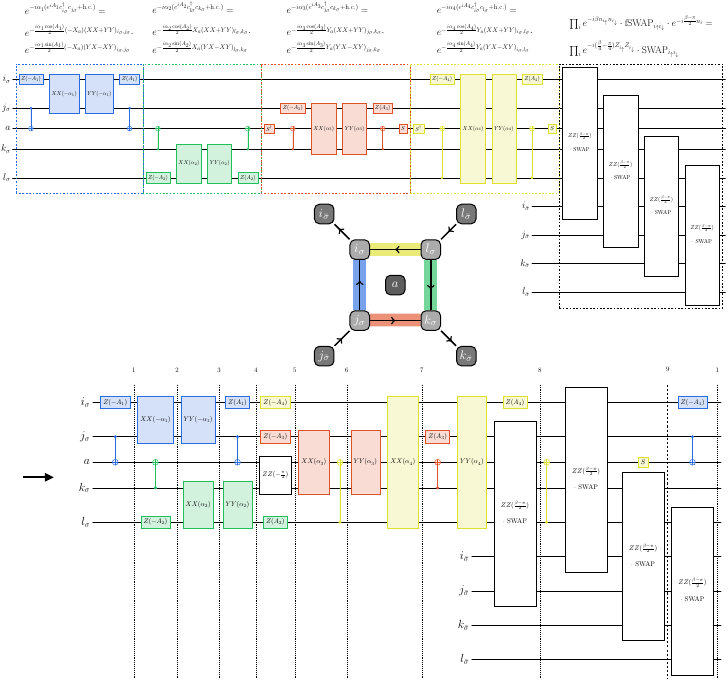}
\caption{\textbf{Circuit to implement the hopping operator for one plaquette $U_{\rm hop}^{\rm plaq}=\exp(-it \sum_{\langle ij \rangle \in \mathrm{plaquette}} c^\dagger_{i \sigma} c_{j \sigma} + \mathrm{h.c.})$.} Top: Direct implementation of the mapped operators via Pauli gadgets. Bottom: Reduction of two-qubit gate count and depth using circuit identities. Final gates (black) merge fermionic SWAP gates with interaction gates. One Hubbard-Trotter step requires two applications of the circuit. The total two-qubit gate count taking into account both hopping (coloured) as well as fSWAP/interaction gates (black) is thus $2 \times (15 N/2 + N) = 17 N$ which, for the $N=6 \times 6$ lattice evaluates to 612 two-qubit gates. The two-qubit gate depth is 18, independent of system size.
\label{supplement_fig_plaquette_hopping}}
\end{figure*}

In this section we describe the implementation of the time evolution of the hopping Hamiltonian
\begin{equation}\label{hhop}
   H_{\rm hop}= \sum_{\langle ij \rangle \sigma} c^\dagger_{i \sigma} c_{j \sigma} + c^\dagger_{j \sigma} c_{i \sigma}.
\end{equation}
The idea is shown in Fig.~\ref{supplement_fig_hopping_strategy} and Fig.~\ref{supplement_fig_plaquette_hopping}: The basic problem when Trotterising a sum of hopping terms on overlapping bonds is that an order must be chosen, i.e. one bond must be chosen before the other (alternatively particle number conservation can be broken~\cite{evered_probing_2025} which is unsuitable in this work since we rely on particle number conservation to measure pairing correlations). This breaking of time reversal-symmetry leads to error terms in the effective Hamiltonian that are currents from orbitals on the first bond to orbitals on the second bond
e.g., 
\begin{equation}
\begin{aligned}
    [c^\dagger_{i \uparrow} c_{j \uparrow} + c^\dagger_{j \uparrow} c_{i \uparrow}, 
     c^\dagger_{j \uparrow} c_{k \uparrow} + c^\dagger_{k \uparrow} c_{j \uparrow}] &=
    \left[ \vcenter{\hbox{\includegraphics[width=1.5cm]{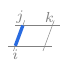}}}, \vcenter{\hbox{\includegraphics[width=1.5cm]{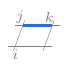}}}\right]\\
    &=
    \vcenter{\hbox{\includegraphics[width=1.5cm]{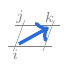}}} \\ 
    &= c^\dagger_{i \uparrow} c_{k \uparrow} - c^\dagger_{k \uparrow} c_{i \uparrow}.
\end{aligned}
\end{equation}
However, by simulating both vertical before the horizontal terms within one plaquette,
\begin{equation}
\begin{aligned}
    [c^\dagger_{i \uparrow} c_{j \uparrow} + c^\dagger_{j \uparrow} c_{i \uparrow} + c^\dagger_{k \uparrow} c_{l \uparrow} + c^\dagger_{l \uparrow} c_{k \uparrow}, 
     c^\dagger_{j \uparrow} c_{k \uparrow} + c^\dagger_{k \uparrow} c_{j \uparrow} +  c^\dagger_{i \uparrow} c_{l \uparrow} + c^\dagger_{l \uparrow} c_{i \uparrow}] &= \left[ \vcenter{\hbox{\includegraphics[width=1.5cm]{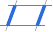}}}, \vcenter{\hbox{\includegraphics[width=1.5cm]{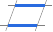}}}\right]\\
    &=  \vcenter{\hbox{\includegraphics[width=1.5cm]{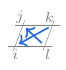}}} + \vcenter{\hbox{\includegraphics[width=1.5cm]{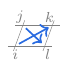}}} \\
    &= 0,
\end{aligned}
\end{equation}
we can cancel all commutators and implement hopping around $2 \times 2$ plaquettes exactly. Let us now describe the exact sequence of gates and implementation in our chosen fermionic encoding in more detail.

We first remind that the $\langle ij \rangle$ appearing in~\eqref{hhop} are \emph{not} the edges present in the representation of the fermionic encoding in Fig \ref{supplement_fig_qubit_labeling}a, but instead denote the edges on the \emph{original} $6\times 6$ square lattice, and link only fermions with identical spins. In Fig \ref{supplement_fig_qubit_labeling}a, they correspond for up spin (blue qubits) to edges such as $0-1$ and $0-6$, but also to edges such as $6-12$ and $7-8$, even though they are not linked by an edge in this fermionic encoding. For down spin (red qubits), they include for example $43-44$ and $43-49$, but also $44-45$ and $43-37$. They \emph{do not} include edges between different spins that appear in the fermionic encoding, such as $0-36$.

The terms appearing in \eqref{hhop} do not all commute with each other, so that further Trotterization is required to implement the exponential of $H_{\rm hop}$. In order to explain our choice of Trotterization, let us denote a "$\sigma$-spin plaquette" (with $\sigma=\uparrow,\downarrow$) a group of $4$ qubits with same spin located in a square pattern around an ancilla, either directly around it or not directly around it. Namely, there are exactly two plaquettes per ancilla, one with spin up and one with spin down qubits. For example, in Fig.~\ref{supplement_fig_qubit_labeling} (a), the plaquette with up spin around ancilla $72$ is comprised of qubits $0,1,6,7$, and the plaquette with down spin around ancilla $72$ is comprised of qubits $36,37,42,43$. In section \ref{sec_supplement_fermionic_encoding}, we further distinguished two types of plaquettes: we call $P$-plaquettes those associated to ancillas on even rows (equivalently, ancillas with up-spin qubits directly around them), and $Q$-plaquettes those associated to ancillas on odd rows (equivalently, ancillas with down-spin qubits directly around them). For example in Fig \ref{supplement_fig_qubit_labeling} (a), the plaquette $0,1,6,7$ is a up-spin $P$-plaquette, the plaquette $36,37,42,43$ is an down-spin $P$-plaquette, the plaquette $43,44,49,50$ is an down-spin $Q$-plaquette and the plaquette $7,8,13,14$ is a up-spin $Q$-plaquette. Note that each bond of the (periodic) square lattice is associated to one unique $P$-type or $Q$-type plaquette.

Let us consider one arbitrary fixed-spin plaquette, such as for example qubits $0$, $1$, $6$, $7$ in Fig \ref{supplement_fig_qubit_labeling} (a). As discussed above, the hopping terms on a fixed-spin plaquette can be implemented exactly without Trotter error through the following decomposition
\begin{equation}\label{uplaquette}
    e^{-it (c^\dagger_0 c_1+c^\dagger_1 c_7+c^\dagger_7 c_6+ c^\dagger_6 c_0 +h.c.)}=e^{-it(c^\dagger_6 c_7+c^\dagger_7 c_6)}e^{-it(c^\dagger_0 c_1+c^\dagger_1 c_0)}e^{-it(c^\dagger_7 c_1+c^\dagger_1 c_7)}e^{-it(c^\dagger_6 c_0+c^\dagger_0 c_6)}\,.
\end{equation}
Further, we note that the hopping terms of plaquettes of both the same type ($P$ or $Q$) and same spin always commute, because they apply on disjoint sets of qubits. Similarly, plaquettes that have both a different type and a different spin always commute. We Trotterize the entire hopping Hamiltonian by applying first the hoppings on all the up-spin $P$-plaquettes, then on all the down-spin $Q$-plaquettes, then on all the up-spin $Q$-plaquettes, and then on all the down-spin $P$-plaquettes. Namely
\begin{equation}
    U_{\rm hop}=\underbrace{\prod_{\substack{{\rm down-spin}\\{\rm P-plaquette}}} U_{\rm hop}^{\rm plaq} \prod_{\substack{{\rm up-spin}\\{\rm Q-plaquette}}}  U_{\rm hop}^{\rm plaq}}_{=:\exp(-i t B)} \underbrace{\prod_{\substack{{\rm down-spin}\\{\rm Q-plaquette}}} U_{\rm hop}^{\rm plaq} \prod_{\substack{{\rm up-spin}\\{\rm P-plaquette}}} U_{\rm hop}^{\rm plaq}}_{=:\exp(-i t A)}\,,
\end{equation}
where we defined 
\begin{equation}
    U_{\rm hop}^{\rm plaq}=\exp \left(-it \sum_{\substack{\langle ij \rangle \sigma\\\in {\rm plaquette}}} c^\dagger_{i \sigma} c_{j \sigma} + \mathrm{h.c.}\right)\,,
\end{equation}
itself decomposed into elementary hopping terms as in \eqref{uplaquette} and $H_\mathrm{hop} = A + B$.

Next, we explain how to implement each $e^{-it(c^\dagger_i c_j+c_j^\dagger c_i)}$ with the fermionic encoding. We first consider the case where the edge $\langle i,j\rangle$ \emph{does} appear in the fermionic encoding in Fig \ref{supplement_fig_qubit_labeling} (a). This is exactly the case of all the edges comprised in up-spin $P$-plaquettes and down-spin $Q$-plaquettes, such as for example $\langle 0,1 \rangle$ (adjacent to ancilla $72$) or $\langle 43,44\rangle$ (adjacent to ancilla $75$). In that case, we can write
\begin{equation}\label{exprsmallhopping}
    e^{-it(c^\dagger_i c_j+c_j^\dagger c_i)}=e^{-\frac{it}{2}(X_iX_j+Y_iY_j)P_a}\,,
\end{equation}
with $P_a$ the Pauli matrix that applies on the ancilla $a$ that is adjacent to the edge $\langle i,j\rangle$, that is $P_a=Y_a$ if the edge is a horizontal edge, that is $P_a=-X_a$ if the edge is a vertical edge on the left (resp. right) of a $P$ plaquette (resp. $Q$ plaquette), and that is $P_a=X_a$ if the edge is a vertical edge on the right (resp. left) of a $P$ plaquette (resp. $Q$ plaquette).

We now consider the case where the edge $\langle i,j\rangle$ appearing in \eqref{hhop} does \emph{not} appear in the fermionic encoding of Fig \ref{supplement_fig_qubit_labeling} (a). This is exactly the case of all the edges in the up-spin $Q$-plaquettes and down-spin $P$-plaquettes, such as for example $\langle 1,2 \rangle$ (adjacent to ancilla $87$) or $\langle 36,37\rangle$ (adjacent to ancilla $72$). In that case, we cannot use directly the expression \eqref{exprsmallhopping}. We must first implement a fermionic swap to bring the qubits $i,j$ on an edge appearing in the fermionic encoding of Fig.~\ref{supplement_fig_qubit_labeling} (a), then use \eqref{exprsmallhopping}, and then apply back the same fermionic swap on the same qubits. This can be done by applying a fermionic swap on all the oblique edges of Fig.~\ref{supplement_fig_qubit_labeling} (a), such as $0-36$ and $1-37$, namely on the fermions with different spins located on a same site. Performing a fermionic swap on a pair of such qubits denoted $j$ and $j'$ is done exactly by swapping the two qubits, and then applying a CZ gate on them. Since the ion-trap hardware is all-to-all connected, the swapping of the qubits can be done by just swapping their indices in memory. But for code simplicity, instead of swapping the indices of the qubits in memory, we swap the indices of the gates appearing in \eqref{exprsmallhopping} that we apply on these qubits. In practice, this entails to the following expression
\begin{equation}\label{exprsmallhopping2}
    e^{-it(c^\dagger_i c_j+c_j^\dagger c_i)}={\rm CZ}_{i,i'}{\rm CZ}_{j,j'}e^{-\frac{it}{2}(X_{i}X_{j}+Y_{i}Y_{j})P_a}{\rm CZ}_{i,i'}{\rm CZ}_{j,j'}\,,
\end{equation}
where $i'$ and $j'$ denote the qubits on the same physical sites as $i$ and $j$, but with opposite spins. For example, if $i=36$, then $i'=0$; if $i=7$ then $i'=43$. We note that in this expression, the hopping term in the middle is applied on the qubits $i,j$ (that are not linked by an edge in the fermionic encoding), not on the qubits $i',j'$ that are linked by an edge in the fermionic encoding. Gathering everything, we obtain the following expression for the circuit that implements one Trotter step under the hopping Hamiltonian:
\begin{equation}\label{uhopsimple}
    U_{\rm hop}= {\rm CZ}_{\rm all}\, U_{\rm hop}^{\rm sublattice-Q}\, {\rm CZ}_{\rm all}\, U_{\rm hop}^{\rm sublattice-P}\,,
\end{equation}
where ${\rm CZ}_{\rm all}$ denotes the operator
\begin{equation}
    {\rm CZ}_{\rm all}=\prod_{j} {\rm CZ}_{j\downarrow, j\uparrow}\,,
\end{equation}
with product over all the physical sites, and where $U_{\rm hop}^{\rm sublattice-P,Q}$ denotes

\begin{equation}
\begin{aligned}
    & U_{\rm hop}^{\rm sublattice-P}=\prod_{\substack{{\rm down-spin}\\{\rm Q-plaquette}}} U_{\rm hop}^{\rm plaq} \prod_{\substack{{\rm up-spin}\\{\rm P-plaquette}}} U_{\rm hop}^{\rm plaq}\\
    & U_{\rm hop}^{\rm sublattice-Q}=\prod_{\substack{{\rm down-spin}\\{\rm P-plaquette}}} U_{\rm hop}^{\rm plaq} \prod_{\substack{{\rm up-spin}\\{\rm Q-plaquette}}} U_{\rm hop}^{\rm plaq}\,.
\end{aligned}
\end{equation}

\begin{figure*}[t]
    \centering
\includegraphics[width=0.45\linewidth]{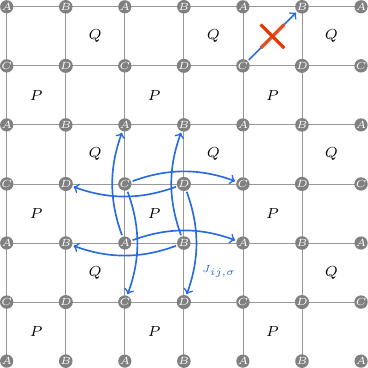}
\caption{\textbf{Error terms appearing in the first-order Floquet Hamiltonian for the hopping Trotterisation.} Only one spin species is shown, since hopping on different species commutes. The commutator between the two operators corresponding to $P$ and $Q$ comprises next-nearest-neighbour currents $J_{ij,\sigma} = c^\dagger_{i \sigma} c_{j \sigma} - \mathrm{h.c.}$, which break translation-invariance to a 2-site-shift symmetry, but all cross-plaquette currents cancel. This leads to relatively small Trotter errors even at $\tau=0.5$ (in units of inverse hopping). Sets of sites related by the 2-site-shift symmetry are labeled $A$, $B$, $C$, $D$ which are not to be confused with the operators $A$ and $B$ that make up the hopping Hamiltonian.
\label{supplement_fig_PQ_commutator}}
\end{figure*}

\begin{figure*}[t]
    \centering
\includegraphics[width=\linewidth]{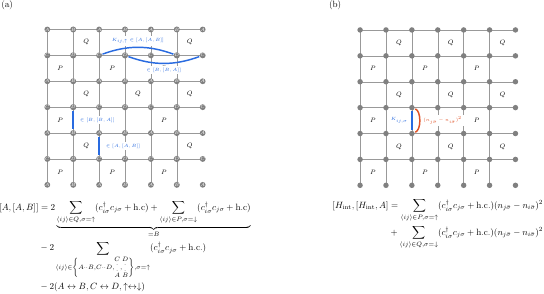}
\caption{\textbf{Dominant error terms appearing in the second-order Floquet Hamiltonian.} (a) Double commutators of the operators $A$ and $B$ that constitute the hopping Hamiltonian. The part of the double commutator that recovers the original hopping (first line) does not lead to error but only renormalises the hopping by $\mathcal{O}(\tau^2)$ (although in this work we have not taken into account this renormalisation). (b) For large on-site interaction $H_\mathrm{int} = U \sum_j n_{j \uparrow} n_{j \downarrow}$, the dominant term in the second-order Trotter step $U_\mathrm{int} U_\mathrm{hop} U_\mathrm{int}$ is given by the term $[H_\mathrm{int}, [H_\mathrm{int}, A]] = \mathcal{O}(U^2)$ (plus the same expression with $A$ substituted by $B$). This term is an imbalance-assisted-hopping, i.e. it leads to hopping $K_{ij, \sigma} = c^\dagger_{i \sigma} c_{j \sigma} + \mathrm{h.c.}$ on species $\sigma$ if there is an imbalance in the opposite species $n_{j \bar{\sigma}} - n_{i \bar{\sigma}} \neq 0$ on the given bond.
\label{supplement_2nd_order_Floquet}}
\end{figure*}

The overall unitary $U_\mathrm{hop}$ does not implement the kinetic term in the Hamiltonian exactly, since $e^{-itA} e^{-itB} \neq e^{-it(A+B)} = e^{-it H_\mathrm{hop}}$ due to $[A,B] \neq 0$. Nevertheless, in Fig.\ref{fig1}a, we see that this Trotter error is surprisingly small for the seemingly very large step size $\tau = 0.5$ (in units of inverse hopping). Two mechanisms are likely at play for the suprisingly good performance of the Trotterisation. First, as described above, the hopping within one plaquette is implemented exactly, due to fact that horizontal and vertical hopping terms commute. Second, the Floquet Hamiltonian arising from commutators of $A$ and $B$ is smaller than might be naively expected. This is due to the fact that many terms in the commutator of $A$ and $B$ might be expected to be currents on intermediate plaquettes that couple $P$ and $Q$-type plaquettes. However, as shown in Fig.~\ref{supplement_fig_PQ_commutator}, these terms cancel, which leave only next-nearest neighbour currents in the first-order commutator expansion
\begin{equation}
    [A,B] = \sum_{\overrightarrow{\langle \langle i j \rangle \rangle} , \sigma} (c^\dagger_{i \sigma} c_{j \sigma} - \mathrm{h.c.}).
\end{equation}
The main physical effect of this Trotterisation scheme is thus to break the full translation invariance to a symmetry under translation by two sites. Note that, for a periodic system with side length $L=4$, even the next-nearest-neighbour terms in Fig.~\ref{supplement_fig_PQ_commutator} cancel and thus, for a $4 \times 4$ system, our scheme implements the simulation of the hopping term exactly, without Trotter error. Furthermore, for arbitrary system sizes, the second-order Floquet Hamiltonian of the hopping Trotterisation (as well as the term that is dominant for strong interactions $U \gg t$) is shown in Fig.~\ref{supplement_2nd_order_Floquet}. It is noteworthy that part of the double-commutator recovers the original hopping Hamiltonian $H_\mathrm{hop} = A+B$. That commutator thus does not lead to an undesired term in the Floquet Hamiltonian, but merely renormalises the hopping $\sim(1+\tau^2)$ which could in principle be accounted for when targeting a specific evolution time (although we have not made use of this fact for the circuits in this work).

\subsection{Exchange terms}
The exchange term $J \sum_i \mathbf{S}_{i} \cdot \mathbf{S}_{j} - n_{i}n_{j}/4$ consists of a flip-flop term $S^x_{i} S^x_{j} + S^y_{i} S^y_{j}$ and a density-density interaction $(n_{i \uparrow }n_{j \downarrow} + n_{i \downarrow }n_{j \uparrow})/2$. While the density-density term can be implemented in a straightforward way using a depth-1 circuit consisting of two (native) ZZPhase gates, the flip-flop term is much more complicated. Indeed, after the fermion-to-qubit mapping~\eqref{eq_exchange_mapping}, this term becomes 
\begin{equation}
\label{eq_flip_flop_term}
    \begin{aligned}
        S^x_{i} S^x_{j} + S^y_{i} S^y_{j}
        & \rightarrow \frac{ (\sigma^+_{i \uparrow} \sigma^+_{j \downarrow})(\sigma^-_{j \uparrow} \sigma^-_{i \downarrow}) + \mathrm{h.c.}}{8}.
\end{aligned}
\end{equation}
Expanding $\sigma^{\pm}$ into Pauli matrices leads to 8 four-body Pauli terms (all terms with an even number of $Y$). Using the usual Pauli gadget construction to implement an $n$-body Pauli rotation using $2n-3$ (arbitrary-angle) two-qubit gates would thus lead to $8 \times 5 = 40$ two-qubit gates per flip-flop interaction, for a total two-qubit count of 42 for one exchange interaction. This would be impractical. Instead, consider the gadget $G$ defined as 
\begin{equation}
\begin{aligned}
\vcenter{\hbox{\includegraphics[]{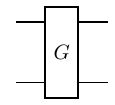}}}
\; := \;
\vcenter{\hbox{\includegraphics[]{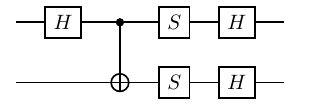}}}.
\end{aligned}
\end{equation}
The significance of this gadget is that, when acting on a pair of qubits corresponding to one fermionic site $(i \uparrow$, $i \downarrow)$, it implements the \textit{fermionic Hadamard gate}, in the sense that
\begin{equation}
    G^\dagger_{i \uparrow, i \downarrow} X_{i \uparrow} S^z_i X_{i \uparrow} G_{i \uparrow, i \downarrow} = S^x_i.
\end{equation}
Once the $S^x_i S^x_j$-part of the flip-flop term has been mapped to $S^z_i S^z_j$, the simulation is straightforward. Using
\begin{equation}
    e^{-i \tau S^z_i S^z_j} \rightarrow e^{-i \tau \left( Z_{i \uparrow} Z_{j \uparrow} + Z_{i \downarrow} Z_{j \downarrow} -Z_{i \uparrow} Z_{j \downarrow} - Z_{j \uparrow} Z_{i \downarrow} \right)/16},
\end{equation}
we find that
\begin{equation}
    e^{-i \tau S^x_i S^x_j} \rightarrow G^\dagger_{i \uparrow, i\downarrow} \otimes G^\dagger_{j \uparrow, j\downarrow} e^{-i \tau \left( Z_{i \uparrow} Z_{j \downarrow} + Z_{j \uparrow} Z_{i \downarrow} + Z_{i \uparrow} Z_{j \uparrow} + Z_{i \downarrow} Z_{j \downarrow} \right)/16} G_{i \uparrow, i\downarrow} \otimes G_{j \uparrow, j\downarrow} ,
\end{equation}
can be implemented with an overall two-qubit count of $2+4+2=8$ and two-qubit depth of $1+2+1=4$. When changing to the $S^y$-basis, one can make use of cancellations to write on overall expression
\begin{equation}
    e^{-i \tau (S^x_i S^x_j + S^y_i S^y_j)} = \vcenter{\hbox{\includegraphics[width=0.75\textwidth]{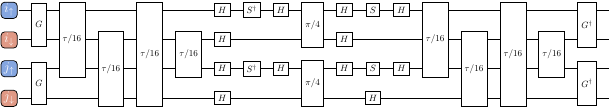}}}
\end{equation}
where boxes labeled by parameters only are $\exp(-i (\tau/16) ZZ)$ and $\exp(-i (\pi/4) \, ZZ)$. The full fermionic flip-flop interaction can thus be simulated using 14 two-qubit gates and a two-qubit-gate-depth of 7. Adding the two ZZPhase gates implementing the density-density interaction, we arrive at a total count of 16 two-qubit gates in depth 8 for a full simulation of $\mathbf{S}_{i} \cdot \mathbf{S}_{j} - n_{i}n_{j}/4$. While this implementation is reasonably efficient, we see why, for the half-filled single layer case, it is advantageous to start from the Heisenberg limit and use the injection technique if $U \gg t$ is desired: The exchange interaction in the all-singly occupied subspace requires 3 two-qubit gates instead of 16.

\newpage
\section{Toric Code preparation}
\label{sec_supplement_toric_code_preparation}
The fermion-to-qubit mapping introduces ancilla qubits, and thus the qubit Hilbert space is larger than the fermionic Fock space. Physically, this is due to the fact that the qubit system encodes all possible combinations of pairs of $\pi$-fluxes through even plaquettes of the lattice, associated with $S_j$ operators in Figure~\ref{supplement_fig_mapping_half_filled}. Since we are interested in the sector without any (topologically trivial) fluxes, we want to set $S_j=+1$ in the initial state ($S_j$ are preserved during the evolution). To achieve this, we decompose the operator into two parts $S_j = (Z^{\otimes 8})_\mathrm{system} (XYXY)_\mathrm{ancilla}$ and set both system and ancilla terms to $+1$ individually. On the system qubits (red and blue) $(Z^{\otimes 8})_\mathrm{system}=+1$ is achieved by initialising a computational basis eigenstate. The situation on the ancilla qubits is slightly more complicated since adjacent $S_j$ do not commute 1-locally. In fact, the $(XYXY)_\mathrm{ancilla}$ operators on the ancilla qubits are exactly the same as the canonical toric code operators ($Z^{\otimes 4}$ and $X^{\otimes 4}$) on the square lattice formed by the (green) ancilla qubits, up to single-qubit basis rotations.

\subsection{$6\times 6$ single layer}

In terms of the qubit labeling in Figure~\ref{supplement_fig_qubit_labeling}, for the $6 \times 6$ single layer, we initialise the toric code using the unitary $U_\mathrm{TC}^\mathrm{single \, layer}$, which is given in Listing ~\ref{listing_monolayer_toric_code}.

\definecolor{backcolour}{rgb}{0.90,0.90,0.90}
\lstdefinestyle{mystyle}{
    backgroundcolor=\color{backcolour},   
    basicstyle=\ttfamily\footnotesize,
    breakatwhitespace=false,         
    breaklines=true,                 
    captionpos=b,       
    keepspaces=true,
    numbersep=5pt,                  
    showspaces=false,
    showstringspaces=false,
    showtabs=false,
    tabsize=2
}
\begin{center}
\begin{minipage}{0.75\textwidth}
\begin{lstlisting}[style=mystyle, caption={\textbf{Unitary $U^{\mathrm{single \, layer}}_\mathrm{TC}$ to prepare the toric code on the ancilla qubits on the $6 \times 6$ single layer Hubbard model following the qubit labeling in Figure~\ref{supplement_fig_qubit_labeling}(a).}}, label={listing_monolayer_toric_code}]
H(72), H(73), H(74), H(75), H(76), H(77), H(78), H(79), H(80),
H(81), H(82), H(83), H(84), H(85), H(86), H(87), H(88), H(89),
CZ(78,75), CZ(84,83), CZ(86,82), CZ(85,81), CZ(76,79), CZ(88,87),
CZ(72,89), CZ(74,80),

CX(72,75), CZ(77,78), CZ(73,79), CX(85,86), CZ(84,81), CZ(89,87), 

CX(75,76), CZ(88,73), CZ(80,77), CZ(86,83), CX(84,78), CX(85,79),

CX(76,74), CX(86,80), CX(88,85), CX(89,84),
H(78), H(79), H(80), H(84), H(85), H(86), H(88), H(89), S(72), H(72), S(72),
S(73), H(73), S(73), S(74), H(74), S(74), Sdg(75), H(75), Sdg(76), H(76),
Sdg(77), H(77), S(78), H(78), S(78), S(79), H(79), S(79), S(80), H(80),
S(80), Sdg(81), H(81), Sdg(82), H(82), Sdg(83), H(83), S(84), H(84), S(84), 
S(85), H(85), S(85), S(86), H(86), S(86), Sdg(87), H(87), Sdg(88), H(88),
Sdg(89), H(89)
\end{lstlisting}
\end{minipage}
\end{center}
Note that this is a geometrically non-local circuit which allows us to prepare the toric code on the $3 \times 3$ periodic ancilla lattice using a two-qubit depth of 4. This is the reason that we are not using a scheme based on measurement and feed-forward: While such schemes offer circuit depths that are constant in system size, for this particular system size, the minimal depth of 4 can be achieved unitarily with a smaller number of overall gates ~\cite{iqbal_topological_2024, fossfeig2023experimentaldemonstrationadvantageadaptive}.

The circuit prepares a state with all $S_j=+1$, but the logical operators are initialised in a superposition of +1 and -1-states (which corresponds to the antiperiodic, APBC, and periodic, PBC, sectors, respectively), so that $\langle B_H \rangle=0$ and $\langle B_V\rangle =0$ (see Figure~\ref{supplement_fig_mapping_bilaer}). Since the observables $\mathcal{O}$ that we measure do not mix logical sectors, i.e., we have $\langle B_H, B_V = (h,v) | \mathcal{O} |B_H, B_V \neq (h,v)\rangle =0$ for $h,v \in \{+1, -1 \}$, the superposition over the four different boundary conditions (PBC-PBC, APBC-PBC, PBC-APBC, APBC-APBC) behaves the same as a classical mixture of the four states.

To simultaneously read out the value of all stabilisers, as well as the topologically non-trivial operators $B_H$ and $B_V$, we invert the toric code preparation at the end of the experiment, using another two-qubit-depth-4 circuit $U_\mathrm{in \, TC}^\mathrm{single \, layer}$, given in Listing~\ref{listing_monolayer_inverse_toric_code}.
\begin{center}
\begin{minipage}{0.75\textwidth}
\begin{lstlisting}[style=mystyle, caption={\textbf{Unitary $U^{\mathrm{single \, layer}}_\mathrm{inv \, TC}$ to invert the toric code on the ancilla qubits on the $6 \times 6$ single layer Hubbard model following the qubit labeling in Figure~\ref{supplement_fig_qubit_labeling}(a).}}, label={listing_monolayer_inverse_toric_code}]
Y(75), Y(76), Y(77), X(73), X(79), X(85), H(72), S(72), H(73), S(73), H(74), 
S(74), H(75), S(75), H(76), S(76), H(77), S(77), H(78), S(78), H(79), S(79), 
H(80), S(80), H(81), S(81), H(82), S(82), H(83), S(83), H(84), S(84), H(85),
S(85), H(86), S(86), H(87), S(87), H(88), S(88), H(89), S(89), H(72), H(73),
H(74), H(78), H(79), H(80), H(84), H(85), H(86), Sdg(72), Sdg(75), Sdg(80),
H(72), H(73), H(74), H(75), H(76), H(78), H(80), H(84), H(85), H(87),
Sdg(72), Sdg(73), Sdg(75), Sdg(82), Sdg(83), Sdg(86), Sdg(87), Z(72), Z(73),
Z(74), Z(81), Z(86), Y(76), Y(87), Y(88), X(78),
CX(84, 78), Sdg(89), CY(89, 77), CX(75, 72), Sdg(73), CY(73, 87), CX(82, 79),
CZ(88, 76), Sdg(83), CY(83, 80), CZ(81, 85), CX(83, 77),

CX(86, 76), Sdg(78), CY(78, 72),  CZ(88, 73), CZ(79, 85), CZ(84, 81), 
CZ(74, 89), CZ(75, 87), CZ(74, 86), CZ(82, 80), Sdg(83), CY(83, 89), 
CX(79, 72), CZ(76, 77), 

CZ(73, 81), CZ(78, 87),

CZ(89, 78), CZ(81, 87), CZ(83, 84), CZ(72, 77), 
CZ(76, 82), CZ(79, 73), CZ(86, 80), CZ(88, 74),
H(72), H(73), H(74), H(75), H(76), H(77), H(78), H(79), H(80), H(81), H(82),
H(83), H(84), H(85), H(86), H(87), H(88), H(89)

\end{lstlisting}
\end{minipage}
\end{center}
In the noiseless case, all stabilisers will be measured as $S_j=+1$, while for the mixed boundary conditions we implement in $U^\mathrm{single \, layer}_{\mathrm{TC}}$, we would find $B_H = \pm 1 = B_V$ with $50 \%$ probability, which indicates which of the 4 boundary condition sectors the shot should be assigned to.

In the following table, we list all the stabilisers of the fermionic encoding and their form after application of the inverse toric code preparation. In the "decoded" column, we indicate how the Pauli matrices shown in purple in the "stabiliser" column are transformed by the inverse toric code preparation. The remaining Pauli matrices shown in black in the "stabiliser" column are unmodified by the inverse toric code state preparation.

\begin{center}
\begin{tabular}{ |c|c| } 
 \hline
 stabiliser & decoded \\ 
   \hline
 $Z_{1}Z_{2}Z_{7}Z_{8}Z_{37}Z_{38}Z_{43}Z_{44}{\color{purple}X_{72}X_{73}Y_{75}Y_{87}}$&$Z_{75}$\\
 \hline
$Z_{3}Z_{4}Z_{9}Z_{10}Z_{39}Z_{40}Z_{45}Z_{46}{\color{purple}X_{73}X_{74}Y_{76}Y_{88}}$&$Z_{88}$\\
\hline
$Z_{0}Z_{5}Z_{6}Z_{11}Z_{36}Z_{41}Z_{42}Z_{47}{\color{purple}X_{72}X_{74}Y_{77}Y_{89}}$&$Z_{72}Z_{89}$\\
\hline
$Z_{6}Z_{7}Z_{12}Z_{13}Z_{42}Z_{43}Z_{48}Z_{49}{\color{purple}Y_{72}X_{75}X_{77}Y_{78}}$&$Z_{72}Z_{75}Z_{78}$\\
\hline
$Z_{8}Z_{9}Z_{14}Z_{15}Z_{44}Z_{45}Z_{50}Z_{51}{\color{purple}Y_{73}X_{75}X_{76}Y_{79}}$&$Z_{72}Z_{73}Z_{75}Z_{76}Z_{87}$\\
\hline
$Z_{10}Z_{11}Z_{16}Z_{17}Z_{46}Z_{47}Z_{52}Z_{53}{\color{purple}Y_{74}X_{76}X_{77}Y_{80}}$&$Z_{74}Z_{76}Z_{80}$\\
\hline
$Z_{13}Z_{14}Z_{19}Z_{20}Z_{49}Z_{50}Z_{55}Z_{56}{\color{purple}Y_{75}X_{78}X_{79}Y_{81}}$&$Z_{75}Z_{79}Z_{81}$\\
\hline
$Z_{15}Z_{16}Z_{21}Z_{22}Z_{51}Z_{52}Z_{57}Z_{58}{\color{purple}Y_{76}X_{79}X_{80}Y_{82}}$&$Z_{80}Z_{82}$\\
\hline
$Z_{12}Z_{17}Z_{18}Z_{23}Z_{48}Z_{53}Z_{54}Z_{59}{\color{purple}Y_{77}X_{78}X_{80}Y_{83}}$&$Z_{83}Z_{89}$\\
\hline
$Z_{18}Z_{19}Z_{24}Z_{25}Z_{54}Z_{55}Z_{60}Z_{61}{\color{purple}Y_{78}X_{81}X_{83}Y_{84}}$&$Z_{84}$\\
\hline
$Z_{20}Z_{21}Z_{26}Z_{27}Z_{56}Z_{57}Z_{62}Z_{63}{\color{purple}Y_{79}X_{81}X_{82}Y_{85}}$&$Z_{85}$\\
\hline
$Z_{22}Z_{23}Z_{28}Z_{29}Z_{58}Z_{59}Z_{64}Z_{65}{\color{purple}Y_{80}X_{82}X_{83}Y_{86}}$&$Z_{80}$\\
\hline
$Z_{25}Z_{26}Z_{31}Z_{32}Z_{61}Z_{62}Z_{67}Z_{68}{\color{purple}Y_{81}X_{84}X_{85}Y_{87}}$&$Z_{81}$\\
\hline
$Z_{27}Z_{28}Z_{33}Z_{34}Z_{63}Z_{64}Z_{69}Z_{70}{\color{purple}Y_{82}X_{85}X_{86}Y_{88}}$&$Z_{72}Z_{76}Z_{79}Z_{82}Z_{86}Z_{88}$\\
\hline
$Z_{24}Z_{29}Z_{30}Z_{35}Z_{60}Z_{65}Z_{66}Z_{71}{\color{purple}Y_{83}X_{84}X_{86}Y_{89}}$&$Z_{76}Z_{80}Z_{83}Z_{86}$\\
\hline
$Z_{0}Z_{1}Z_{30}Z_{31}Z_{36}Z_{37}Z_{66}Z_{67}{\color{purple}Y_{72}Y_{84}X_{87}X_{89}}$&$Z_{78}Z_{84}Z_{87}$\\
\hline
$Z_{2}Z_{3}Z_{32}Z_{33}Z_{38}Z_{39}Z_{68}Z_{69}{\color{purple}Y_{73}Y_{85}X_{87}X_{88}}$&$Z_{73}Z_{85}$\\
\hline
$Z_{4}Z_{5}Z_{34}Z_{35}Z_{40}Z_{41}Z_{70}Z_{71}{\color{purple}Y_{74}Y_{86}X_{88}X_{89}}$&$Z_{74}$\\
\hline
$Z_{1}Z_{7}Z_{13}Z_{19}Z_{25}Z_{31}Z_{37}Z_{43}Z_{49}Z_{55}Z_{61}Z_{67}{\color{purple}X_{72}X_{75}X_{78}X_{81}X_{84}X_{87}}$&$Z_{75}Z_{87}$\\
\hline
$Z_{6}Z_{7}Z_{8}Z_{9}Z_{10}Z_{11}Z_{42}Z_{43}Z_{44}Z_{45}Z_{46}Z_{47}{\color{purple}Y_{72}Y_{73}Y_{74}Y_{75}Y_{76}Y_{77}}$&$Z_{73}Z_{74}Z_{75}Z_{77}Z_{87}$\\
\hline
 
\end{tabular}
\end{center}

\subsection{$4\times 4$ bilayer}
The $4\times 4$ bilayer system can be seen as two juxtaposed copies of a $4\times 4$ single layer. The toric code state preparation and unpreparation is thus a tensor product of the toric code preparation on a $4\times 4$ system. The toric code state preparation on one of these layers is given in Listing \ref{listing_bilayer_toric_code}. The toric code state preparation on the other layer is identical, with all the qubit indices translated by $+40$.

\begin{center}
\begin{minipage}{0.75\textwidth}
\begin{lstlisting}[style=mystyle, caption={\textbf{One half of the unitary $U^{\mathrm{bilayer}}_\mathrm{TC}$ to prepare the toric code on the ancilla qubits on a $4 \times 4$ bilayer Hubbard model following the qubit labeling in Figure~\ref{supplement_fig_qubit_labeling}(a).} The other half is identical up to adding 40 to all qubit indices.}, label={listing_bilayer_toric_code}]
H(32), H(33), H(34), H(35), H(36), H(37), H(38), H(39), 
CZ(34,37), CZ(39,38), CZ(36,32),

CX(36,37), CZ(34,39), CZ(33,37),

CX(35,34), 
H(36), H(37), H(39), S(32), H(32), S(32), S(33), H(33), S(33), Sdg(34), H(34),
Sdg(35), H(35), S(36), H(36), S(36), S(37), H(37), S(37), Sdg(38), H(38),
Sdg(39), H(39)

\end{lstlisting}
\end{minipage}
\end{center}

The corresponding inverse toric code preparation is done with the code in Listing \ref{listing_bilayer_inverse_toric_code}.

\begin{center}
\begin{minipage}{0.75\textwidth}
\begin{lstlisting}[style=mystyle, caption={\textbf{Half of the unitary $U^{\mathrm{bilayer}}_\mathrm{inv \, TC}$ to invert the toric code on the ancilla qubits on a $4 \times 4$ bilayer Hubbard model following the qubit labeling in Figure~\ref{supplement_fig_qubit_labeling}(a).} The other half is identical up to adding 40 to all qubit indices.}, label={listing_bilayer_inverse_toric_code}]
Y(36), Y(37), X(38), X(34), Sdg(32), H(32), Sdg(32), Sdg(33), H(33),
Sdg(33), H(34), S(34), H(35), S(35), Sdg(36), H(36), Sdg(36), Sdg(37), H(37),
Sdg(37), H(38), S(38), H(39),  S(39), Sdg(35), H(35), H(36), H(37), H(39),
Sdg(35), Sdg(36), Sdg(37), Sdg(39), Z(34), Z(35), Y(38), Y(39),
X(36), X(37), Sdg(39),
CY(39, 35), CX(34, 38), Sdg(37), CY(37, 36), CX(32, 33),

CX(34, 32), CZ(38, 39), CZ(35, 36), CZ(33, 37),

CZ(38, 33), CZ(32, 36), 
H(32), H(33), H(34), H(35), H(36), H(37), H(38), H(39)

\end{lstlisting}
\end{minipage}
\end{center}

The following table lists all the stabilisers of one single $4\times 4$ layer. The decoded column indicates the value taken by the Pauli matrices in purple after the inverse toric code preparation, the Pauli matrices in black being unchanged. The stabilisers on the second layer of the bilayer are obtained by adding $40$ to all the qubit indices. 

\begin{center}
\begin{tabular}{ |c|c| } 
 \hline
 stabiliser & decoded \\ 
   \hline
$Z_{1}Z_{2}Z_{5}Z_{6}Z_{17}Z_{18}Z_{21}Z_{22}{\color{purple}X_{32}X_{33}Y_{34}Y_{38}}$&$Z_{34}$\\
 \hline
$Z_{0}Z_{3}Z_{4}Z_{7}Z_{16}Z_{19}Z_{20}Z_{23}{\color{purple}X_{32}X_{33}Y_{35}Y_{39}}$&$Z_{32}Z_{35}$\\
 \hline
$Z_{4}Z_{5}Z_{8}Z_{9}Z_{20}Z_{21}Z_{24}Z_{25}{\color{purple}Y_{32}X_{34}X_{35}Y_{36}}$&$Z_{35}Z_{36}$\\
 \hline
$Z_{6}Z_{7}Z_{10}Z_{11}Z_{22}Z_{23}Z_{26}Z_{27}{\color{purple}Y_{33}X_{34}X_{35}Y_{37}}$&$Z_{35}Z_{36}Z_{37}$\\
\hline
$Z_{9}Z_{10}Z_{13}Z_{14}Z_{25}Z_{26}Z_{29}Z_{30}{\color{purple}Y_{34}X_{36}X_{37}Y_{38}}$&$Z_{32}Z_{34}$\\
\hline
$Z_{8}Z_{11}Z_{12}Z_{15}Z_{24}Z_{27}Z_{28}Z_{31}{\color{purple}Y_{35}X_{36}X_{37}Y_{39}}$&$Z_{35}$\\
\hline
$Z_{0}Z_{1}Z_{12}Z_{13}Z_{16}Z_{17}Z_{28}Z_{29}{\color{purple}Y_{32}Y_{36}X_{38}X_{39}}$&$Z_{35}Z_{36}Z_{39}$\\
\hline
$Z_{2}Z_{3}Z_{14}Z_{15}Z_{18}Z_{19}Z_{30}Z_{31}{\color{purple}Y_{33}Y_{37}X_{38}X_{39}}$&$Z_{35}Z_{36}Z_{37}Z_{39}$\\
\hline
$Z_{1}Z_{5}Z_{9}Z_{13}Z_{17}Z_{21}Z_{25}Z_{29}{\color{purple}X_{32}X_{34}X_{36}X_{38}}$&$Z_{32}Z_{33}$\\
\hline
$Z_{4}Z_{5}Z_{6}Z_{7}Z_{20}Z_{21}Z_{22}Z_{23}{\color{purple}Y_{32}Y_{33}Y_{34}Y_{35}}$&$Z_{32}Z_{34}Z_{35}Z_{38}$\\
\hline

\end{tabular}
\end{center}

\newpage
\section{Effective Hamiltonians from Perturbation Theory}
\label{sec_supplement_perturbation_theory}
In all three experimental setups, we choose a strategy that relies on first preparing a state obtained from second-order degenerate perturbation theory. This state lives in a Hilbert space of qubits (rather than fermions) and comprises much fewer degrees of freedom which facilitates state preparation. Why choose the (unique) perturbative ground state rather than any of the many degenerate states in the perturbative limit? The perturbative ground state has the property that it behaves like the unique ground state of a local Hamiltonian, in the sense that it can be used as a starting point for adiabatic evolution (since all Hamiltonian matrix elements between the degenerate states are strictly zero in the perturbative limit). Thus, the ground state arising from perturbation theory is a valuable resource state for adiabatic exploration of the surrounding phases.

In all three perturbative limits discussed in the main text, we can identify a degenerate subspace $D$ of the Fock space with energy $E_0$ and a perturbation $V$ to some reference Hamiltonian $H_0$. Denoting an orthonormal basis of $D$ by the states $\ket{i}$ and a basis of its complement by $\ket{m}$, the formal expression for the effective Hamiltonian from second-order degenerate perturbation theory is
\begin{equation}
\label{eq_perturbation_theory}
    H^\mathrm{effective}_{kl} = \langle k | V |l \rangle + \sum_{m \notin D} \frac{\langle k | V |m \rangle \langle m | V | l \rangle }{E_0 - E_m}
\end{equation}
where $E_m = \langle m | H_0 | m \rangle $, and it will turn out that $\langle k | V |l \rangle=0$ in all cases we consider. In the remainder of this section, we will derive the effective Hamiltonians in the half-filled single layer, the doped plaquette model, and the strongly coupled bilayer model.
\subsection{Half-Filled Single Layer}
While the perturbation theory of the Heisenberg model is fairly standard, we go through the argument to prepare for the plaquette and bilayer cases. In the perturbative limit $t/U = 0$, all states in the space
\begin{equation}
    D = \{ \mathrm{all \, states \, singly \, occupied}\}
\end{equation}
are degenerate with energy $E_0=0$. The reference Hamiltonian and perturbation is
\begin{equation}
\begin{aligned}
    H_0 &= U\sum_{i} n_{i \uparrow} n_{i \downarrow} \\
    V &= -t \sum_{\langle ij \rangle, \sigma} c^\dagger_{i\sigma} c_{j\sigma}
\end{aligned}
\end{equation}
Since $\langle k | V |m \rangle \langle m | V | l \rangle=0$ for $\ket{k}$ and $\ket{l}$ that differ by more than two adjacent sites (i.e., $V$ acts locally on bonds), the effective Hamiltonian is a sum of nearest-neighbour interaction terms. This case is particularly simple, since for all states $m$ that fulfil $\langle m | V | l \rangle \neq 0$ we have $E_0-E_m=-U$, as well as $\sum_{m \notin D} \langle k | V |m \rangle \langle m | V | l \rangle = \langle k | V^2 | l \rangle$, and so we can simply compute the operator $V^2$ on the singly occupied subspace. We have
\begin{equation}
    V^2 = t^2 \left( \underbrace{\sum_{\sigma} (c^\dagger_{i \sigma} c_{j \sigma} c^\dagger_{j \sigma} c_{i \sigma} + 
    c^\dagger_{j \sigma} c_{i \sigma} c^\dagger_{i \sigma} c_{j \sigma})}_{-4 S^z_i S^z_j + n_i n_j + n_i + n_j} 
    + \underbrace{c^\dagger_{i \uparrow} c_{j \uparrow} c^\dagger_{j \downarrow} c_{i \downarrow} + c^\dagger_{i \downarrow} c_{j \downarrow} c^\dagger_{j \uparrow} c_{i \uparrow} }_{-4 (S^x_i S^x_j + S^y_i S^y_j)} \right)
\end{equation}
Dropping constants $n_i = n_j = n_i n_j = 1$ and bringing in the denominator $1/(E_0-E_m)=-1/U$ we arrive at $H^\mathrm{effective} = \frac{4 t^2}{U} \sum_{\langle ij \rangle} S^x_i S^x_j + S^y_i S^y_j + S^z_i S^z_j$. Expressing the above in terms of Pauli rather than spin-1/2 matrices, we obtain the usual form of the Heisenberg Hamiltonian
\begin{equation}
    H^\mathrm{effective} = \frac{t^2}{U} \sum_{\langle ij \rangle} X_i X_j + Y_i Y_j + Z_i Z_j.
\end{equation}

\subsection{Doped Single Layer}
\begin{figure*}[t]
    \centering
\includegraphics[width=0.9\linewidth]{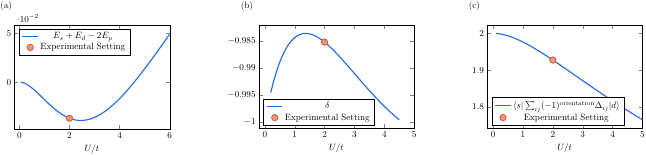}
    \caption{\textbf{Perturbation Theory in the doped checkerboard model.} The reference Hamiltonian $H_0$ describes decoupled $2 \times 2$ plaquettes and depends on the parameters $U/t$. (a) For $U/t \lesssim 4.6$, the lowest state of two decouled plaquettes in the sector with 3 up- and 3-down fermions is a state $\ket{sd}$ with one up- and one-down fermion on one plaquette ($s$) and two up- and two down-fermions on the other plaquette ($d$). The first excited state consists of $p$-plaquettes, which locally have $S^z \neq 0$. (b) Second-order degenerate perturbation theory leads to an XXZ model with anisotropy parameter $\delta$, which depends on $U/t$. Since the dependence is weak, correlators in the effective $3 \times 3$ model vary only extremely weakly with $U/t$ and we choose to approximate $\delta = -1$ for simplicity. (c) The perturbative ground state has $d$-wave pairing correlations, which depend on the matrix element of the singlet annihilation operator $|\langle s|\Delta|d \rangle|^2$ between the $s$- and $d$-plaquette states, which varies with $U/t$. In the experiment we choose $U/t=2$.
\label{supplement_fig_perturbation_theory}}
\end{figure*}

In the weakly coupled plaquette $6 \times 6$ Hubbard model at 1/6 doping, we have 
\begin{equation}
\begin{aligned}
    H_0 &= -t \sum_{\substack{\mathrm{strong \, bonds\,} \\ \langle ij \rangle, \sigma}} c^\dagger_{i\sigma} c_{j\sigma} + U\sum_{i} n_{i \uparrow} n_{i \downarrow} \\
    V &= -t' \sum_{\substack{\mathrm{weak \, bonds\,} \\ \langle ij \rangle, \sigma}} c^\dagger_{i\sigma} c_{j\sigma}.
\end{aligned}
\end{equation}
In this case, $H_0$ has a non-trivial dependence on the parameter $U/t$, and the degenerate subspace $D$ depends on the parameter regime. The scenario we consider is characterised by 15 up- and 15-down fermions in 36 sites and therefore it is a priori not clear how to distribute these particles across the 9 plaquettes. Intuitively, one should fill the plaquettes somewhat homogeneously: completely empty of filled plaquettes will incur energy penalties for both the kinetic and interaction terms. In practice, three plaquette states are energetically favourable for a range of values $U/t$: The $\ket{s}$-state, which is the ground state of the $2 \times 2$ Hubbard plaquette in the sector with 1 up- and 1 down-fermion (i.e. quarter fulling), a $\ket{p}$-state (2 up- and 1 down-fermions or vice versa) and the $\ket{d}$-state (2 up- and 2-down fermions, i.e. half-filling). Two strategies are compatible with 1/6-doping: Either fill the lattice with $6 \ket{p}$-states and $3 \ket{d}$-states, or use $6 \ket{d}$-states and $3 \ket{s}$-states. The overall energies are
\begin{equation}
\begin{aligned}
    \mathrm{Strategy \, 1}:& 6 E_p + 3 E_d \\ 
    \mathrm{Strategy \, 2}:& 6 E_d + 3 E_s
\end{aligned}
\end{equation}
Note that one can change from one strategy to the other by applying $\ket{pp} \rightarrow \ket{sd}$ 3 times. Whether that transformation lowers or increases the energy depends on $U/t$: As shown in~\cite{Trebst_dwave, rey_controlled_2009} and reproduced in Fig.~\ref{supplement_fig_perturbation_theory}, for $U/t \lesssim 4.6$, Strategy 2 will give the ground state. Since we choose $U/t=2$ in the experiment, the correct perturbative ground space is given by 
\begin{equation}
    D=\{ \mathrm{all \, states \, with \,} 6 \ket{d}-\mathrm{states \,and \,} 3 \ket{s}-\mathrm{states} \}.
\end{equation}
There are 
\begin{equation}
    \mathrm{dim}D = \binom{9}{3} = 84
\end{equation}
such states. In this case, the summation~\eqref{eq_perturbation_theory} is more complicated, but can easily be done numerically. In particular, we use the QuSpin package~\cite{weinberg_quspin_2017, weinberg_quspin_2019} to obtain states $\ket{ss}, \ket{sd}, \ket{ds}, \ket{dd}$ on nearest-neighbour plaquettes and then simply loop over all matrix elements generated by $V$, weighted by $E_0 - E_m$. For $U/t=2$, we obtain a ferromagnetic XXZ model
\begin{equation}
    H_\mathrm{XXZ} = -\sum_{\langle ij \rangle} X_i X_j + Y_i Y_j + \delta Z_i Z_j
\end{equation}
with $\delta =-0.985148 \approx 1$. Correlators in the XXZ model at this value of $\delta$ are virtually indistinguishable from $\delta=-1$, and so we approximate $\delta=-1$ for simplicity. The 84-dimensional subspace of the Hubbard model maps to the subspace of the $3 \times 3$ XXZ-model with $\sum Z_i = -3$. Note that, on a bipartite lattice like the square lattice, the sign of $\delta$ can be changed by applying Pauli-$Z$ to one sublattice and flipping the overall sign of the Hamiltonian. The convention introduced above is the one that was used to develop the circuits in a consistent way.

The perturbative ground state hosts $d$-wave pairing correlations. Consider for example two bonds $\langle ij \rangle$ and $\langle kl \rangle$ on (strong) plaquettes $(0,0)$ and $(x,y)$ with orientation $\alpha$ and $\beta$, where $\alpha, \beta \in \{\mathrm{left},\mathrm{right},\mathrm{top},\mathrm{bottom} \}$ denote the orientation and position of the bonds within their plaquettes. By considering the reduced density matrix of the perturbative ground state $\ket{\psi}$ on the two plaquettes, one sees that

\begin{equation}
    \langle \psi | \Delta^\dagger_{ij} \Delta_{kl} + \mathrm{h.c.} | \psi \rangle = \left \langle \frac{XX+YY}{2}\right \rangle_\mathrm{XXZ} \left( \langle s |\Delta_{\alpha} |d \rangle \langle d |\Delta^\dagger_{\beta} |s \rangle + \mathrm{h.c.} \right) 
\end{equation}
where the first term is the correlation function of the $3 \times 3$ XXZ model between \textit{qubits} $(0,0)$ and $(x,y)$ and the second is the product of matrix elements of the operators that annihilate and create singlets on the $\alpha$- and $\beta$-bonds of a plaquette, between the $s$- and $d$-states. The reason for the $d$-wave symmetry of the overall state is the fact that
\begin{equation}
    \langle s | \Delta_\mathrm{left} | d \rangle = \langle s | \Delta_\mathrm{right} | d \rangle = -\langle s | \Delta_\mathrm{top} | d \rangle = -\langle s | \Delta_\mathrm{bottom} | d \rangle.
\end{equation}

\subsection{Bilayer}
In the Bilayer Hubbard model the role of reference Hamiltonian and perturbation are played by
\begin{equation}
\begin{aligned}
    H_0 &= J \sum_i \mathbf{S}_{iA} \cdot \mathbf{S}_{iB} - \frac{n_{iA} n_{iB}}{4} \\
    V &= -t \sum_{\langle ij \rangle \in A,B, \sigma} c^\dagger_{i\sigma} c_{j\sigma}
\end{aligned}
\end{equation}
where each $\langle ij \rangle \in A,B$ refers to the bonds of the $A$- and $B$-layers, respectively. The on-site interaction $U$ in each Hubbard layer can also be directly taken into account, but does not qualitatively change the discussion since the matrix elements of terms in $V^2$ that include on-site interactions between elements of $D$ will turn out to be zero. Locally, $H_0$ favours singlet states $\Delta^\dagger_{Ai Bi}|\mathrm{vac} \rangle = \frac{1}{\sqrt{2}} ( c^\dagger_{Ai \uparrow} c^\dagger_{Bi \downarrow}  -  c^\dagger_{Ai \downarrow} c^\dagger_{Bi \uparrow}) |\mathrm{vac} \rangle$ which are eigenstates of the local term $\mathbf{S}_{iA} \cdot \mathbf{S}_{iB} - (n_{iA} n_{iB})/4$ with eigenvalue -1.
In the quarter-filled sector on the $4 \times 4$ lattice, the degenerate ground space is given by
\begin{equation}
    D = \{ \mathrm{all \, states \,with \,8 \,singlets \,and \,8 \,hole \, pairs \,across \,the \,16 \, rungs}\}
\end{equation}
and has energy $E_0 = -8$ with respect to $H_0$ and is $\mathrm{dim}D=\binom{16}{8} = 12870$-dimensional. We can again perform the summation~\eqref{eq_perturbation_theory} and obtain a ferromagnetic XXZ-model on a single $4 \times 4$ square lattice of qubits
\begin{equation}
    H_{\mathrm{XXZ}}=-\sum_{\langle ij  \rangle} X_i X_j + Y_i Y_j + \delta Z_i Z_j,
\end{equation}
this time with $\delta = -2/3$ exactly. The discussion of the bilayer perturbation theory is similar to the one in~\cite{schlomer_local_2024}, although in that reference an exchange coupling between two layers of $t-J$-models was considered, whereas here we allow for the perturbation $V$ to create virtual states with doubly-occupied sites and only turn on the on-site interaction during the adiabatic evolution later on.

\newpage
\section{State Preparation Circuits}
\label{sec_supplement_state_preparation_circuits}
In all three sets of experiments (half-filled single layer, doped single layer and bilayer), we initialise some non-trivial initial state with locally fixed fermionic parity. In this section we go over the different circuits used to do so.
\subsection{Half-filled Single Layer}
The half-filled single-layer Hubbard model at $U/t=8$ can be approached from the limit $U/t \rightarrow \infty$ at which it becomes the isotropic Heisenberg model
\begin{equation}
    H=\sum_{\langle ij \rangle} X_i X_j + Y_i Y_j + Z_i Z_j.
\end{equation}
To prepare a low-energy state of the $6 \times 6$ Heisenberg model, we select a subset of 36 qubits, namely all the "up" (blue) qubits, labeled $0$ through $35$ in Figure~\ref{supplement_fig_qubit_labeling}, and apply a classically optimised adiabatic circuit to it. Namely, we initialise the "up" qubits in the $X$-basis N\'eel state
\begin{equation}
    U_{X-\mathrm{Neel}} |0 \rangle= \ket{-+ \dots -+}
\end{equation}
by applying $H$ to all 36 sites and $Z$ to all qubits in the sublattice that contains our qubit labeled 0. The state $\ket{-+ \dots -+}$ is the ground state of the parent Hamiltonian
\begin{equation}
    H_x = \sum_i (-1)^\mathrm{sublattice} X_i.
\end{equation}
As such, we can write down one step of the Trotterised adiabatic evolution towards the Heisenberg model
\begin{equation}
    U^\mathrm{Heisenberg}(\boldsymbol{\theta}) = e^{-i \theta_{xx} H_{xx}} e^{-i\theta_{x_2} H_{x}} e^{-i \theta_{yy} H_{yy}} e^{-i \theta_{x_1} H_{x}} e^{-i \theta_{zz} H_{zz}}
\end{equation}
where $H_{xx} = \sum_{\langle ij \rangle } X_i X_j$, $H_{yy} = \sum_{\langle ij \rangle } Y_i Y_j$ and $H_{zz} = \sum_{\langle ij \rangle } Z_i Z_j$. This unitary is repeated $D_\mathrm{Heisenberg}$ times
\begin{equation}
    U^\mathrm{Heisenberg}(\boldsymbol{\theta}^{D_\mathrm{Heisenberg}}) \dots U^\mathrm{Heisenberg}(\boldsymbol{\theta}^1).
\end{equation}
The coefficients $\boldsymbol{\theta}$ can in principle be obtained from e.g., a linear adiabatic path, but in this work we use coefficients obtained by from a classically tractable $4 \times 4$ system:

\renewcommand{\arraystretch}{1.5}
\begin{center}
\begin{tabular}{ |c|c|c| } 
 \hline
  & $D_\mathrm{Heisenberg}=1$ & $D_\mathrm{Heisenberg}=2$ \\ 
   \hline
 $\theta^1_{zz}$ & 0.12060681 &  -0.10841615 \\ 
 $\theta^1_{x_1}$ & 0.08211975 & -0.05390048 \\ 
 $\theta^1_{yy}$ & 0.05188793 & 0.25322733 \\
 $\theta^1_{x_2}$ & 0.33385306 & 0.42241679 \\
 $\theta^1_{xx}$ & 0 & 0.05271746 \\ 
 $\theta^2_{zz}$ &  & 0.129924 \\ 
 $\theta^2_{x_1}$ &  &  -0.33777063 \\ 
 $\theta^2_{yy}$ &  &  0.07656301 \\ 
 $\theta^2_{x_2}$ &  &  0.35528938 \\ 
 $\theta^2_{xx}$ &  &  0.00924239 \\ 
 \hline
\end{tabular}
\end{center}
The Heisenberg energy density of the state obtained on $4\times 4$ is $-2.6502$ for $D_\mathrm{Heisenberg}=1$ and $-2.7197$ for $D_\mathrm{Heisenberg}=2$, the exact ground state energy density being $-2.8071$.

At this point, we have encoded a low-energy state of the $6 \times 6$ Heisenberg model on the qubits $0 \dots 36$, while all other qubits are still in the $\ket{0}$ state. To steer the (optimised) adiabatic evolution towards $U/t < \infty$, we would now like to evolve with the hopping operator $-t \sum_{\langle ij \rangle \sigma} c^\dagger_{i \sigma} c_{j \sigma}$. Since the Hilbert space on the 36 representative qubits does not include any space for holons or doublons, we need to \textit{inject} the spin state into the Octagon Fermionic Encoding. That is, we need to implement the isometry
\begin{equation}
\begin{aligned}
\vcenter{\hbox{\includegraphics[]{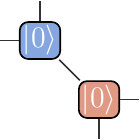}}}
\;\rightarrow\;
\vcenter{\hbox{\includegraphics[]{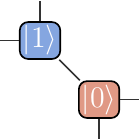}}} \quad \quad \quad \quad
\vcenter{\hbox{\includegraphics[]{SupplementFigures/10.pdf}}}
\;\rightarrow\;
\vcenter{\hbox{\includegraphics[]{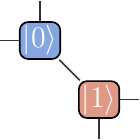}}}.
\end{aligned}
\end{equation}
This injection is achieved via the (two-qubit-)depth-1 sequence $X_{\uparrow} CX_{\uparrow \rightarrow \downarrow}$ on all diagonal edges. With respect to the qubit labeling in Figure~\ref{supplement_fig_qubit_labeling}, the injection $U^\mathrm{half-filled}_\mathrm{inject}$ is given in Listing~\ref{listing_injection_half-filled}. The isometry is the same on each diagonal bond independent of the arrow orientation of on the diagonal edge. In practice, since the Heisenberg state preparation and injection act on the blue and red qubits only, the toric code state preparation on the green qubits is carried out in parallel.

\begin{center}
\begin{minipage}{0.75\textwidth}
\begin{lstlisting}[style=mystyle, caption={\textbf{Unitary $U^\mathrm{half-filled}_\mathrm{inject}$ to inject the low-energy state of the $6 \times 6$ single Heisenberg model into the subspace of the fermionic Fock space with only singly occupied sites, following the qubit labeling in Figure~\ref{supplement_fig_qubit_labeling}(a).}}, label={listing_injection_half-filled}]
CX(0, 36), CX(1, 37), CX(2, 38), CX(3, 39), CX(4, 40), CX(5, 41), CX(6, 42),
CX(7, 43), CX(8, 44), CX(9, 45), CX(10, 46), CX(11, 47), CX(12, 48),
CX(13, 49), CX(14, 50), CX(15, 51), CX(16, 52), CX(17, 53), CX(18, 54),
CX(19, 55), CX(20, 56), CX(21, 57), CX(22, 58), CX(23, 59), CX(24, 60),
CX(25, 61), CX(26, 62), CX(27, 63), CX(28, 64), CX(29, 65), CX(30, 66),
CX(31, 67), CX(32, 68), CX(33, 69), CX(34, 70), CX(35, 71),
X(0), X(1), X(2), X(3), X(4), X(5), X(6), X(7), X(8), X(9), X(10), X(11), 
X(12), X(13), X(14), X(15), X(16), X(17), X(18), X(19), X(20), X(21), X(22),
X(23), X(24), X(25), X(26), X(27), X(28), X(29), X(30), X(31), X(32), X(33),
X(34), X(35)
\end{lstlisting}
\end{minipage}
\end{center}

Right after the injection, the energy density of the state obtained with respect to the Hubbard Hamiltonian is the same as the N\'eel state, namely $-U/4$. Indeed, after the injection, every site is either occupied by a down-spin fermion or a up-spin fermion. The interaction energy density is thus $-U/4$. As for the kinetic energy, applying a hopping term $c^\dagger_{j\sigma}c_{i\sigma}$ on the state will either annihilate the state if there are no fermion with spin $\sigma$ on site $i$, or create a hole on site $i$ if there was a fermion with spin $\sigma$. The inner product with the same state thus vanishes, resulting in exactly zero kinetic energy.

To lower further the energy, we now apply Hubbard-like steps on top of the state. We write it as
\begin{equation}
    U^{\rm Hubbard}(\pmb{\theta})=e^{i \theta_{S^x} \sum_{i} (-1)^\mathrm{sublattice} S^x_i } U_\mathrm{int}(\theta_\mathrm{int}) U_\mathrm{hop}(\theta_\mathrm{hop})\,.
\end{equation}
As for the Heisenberg step, we optimize the parameters $\pmb{\theta}$ on a $4\times 4$ system. We obtain the following parameters:
\begin{center}
\begin{tabular}{ |c|c|c| } 
 \hline
  & $D_\mathrm{Hubbard}=1$ & $D_\mathrm{Hubbard}=2$ \\ 
   \hline
 $\theta^1_{\rm hop}$ & -0.10667747 & -0.09238994 \\ 
 $\theta^1_{\rm int}$ & -0.97794924 &-1.50738096 \\ 
 $\theta^1_{S^x}$ & -0.04714107 & 0.05233736 \\
 $\theta^2_{\rm hop}$ &  &  -0.16975937 \\ 
 $\theta^2_{\rm int}$ &  & -0.7777884 \\ 
 $\theta^2_{S^x}$ &  & -0.05063808 \\
 \hline
\end{tabular}
\end{center}

\subsection{Doped Single Layer}
The target initial state of the doped single layer is the unique ground state of the checkerboard model in the weakly coupled limit $t'\rightarrow 0$, where $t'$ denotes the hopping strength between the strong $2\times 2$ plaquettes. In this case, we approximately prepare the ground state in three stages. First, we prepare the approximate ground state of an effective ferromagnetic XXZ model
\begin{equation}
    H = -\sum_{\langle ij \rangle} X_i X_j + Y_i Y_j + \delta Z_i Z_j
\end{equation}
with $\delta=-1$ on a periodic lattice of $3\times 3$ qubits. The qubit states $\ket{0}$ and $\ket{1}$ are then mapped to representative plaquette states in the fermionic Fock space
\begin{equation}
    \begin{aligned}
        \ket{0} &\rightarrow \ket{\tilde{0}} \quad \mathrm{(quarter-filled)} \\
       \ket{1} &\rightarrow \ket{\tilde{1}} \quad \mathrm{(half-filled)} \\
    \end{aligned}
\end{equation}
Finally, we perform a basis transformation circuit to map these representative states to the ground states of the Hubbard plaquette at $U/t=2$
\begin{equation}
\label{eq_plaquette_preparation}
    \begin{aligned}
        \ket{\tilde{0}} &\rightarrow \ket{s}  \\
       \ket{\tilde{1}} &\rightarrow \ket{d}.
    \end{aligned}
\end{equation}
Each step is a sub-circuit that can be obtained separately. The full circuit is stitched together and has a fan-out structure. The details of each step are explained below.

To establish convention, each of the 9 plaquettes of the $6 \times 6$ checkerboard model can be assigned a coordinate $(x,y)$. We choose to assign qubits to individual plaquettes as shown in Listing~\ref{domonolayer_plaquette_convention}.

\begin{center}
\begin{minipage}{0.75\textwidth}
\begin{lstlisting}[style=mystyle, caption={\textbf{Assignment of qubits to strong plaquettes in the checkerboard model.}}, label={domonolayer_plaquette_convention}]
  (0, 0): [ 1, 37,  2, 38,  7, 43,  8, 44,] + [72, 87, 73, 75],
  (0, 1): [13, 49, 14, 50, 19, 55, 20, 56,] + [78, 75, 79, 81],
  (0, 2): [25, 61, 26, 62, 31, 67, 32, 68,] + [84, 81, 85, 87],
  (1, 0): [ 3, 39,  4, 40,  9, 45, 10, 46,] + [73, 88, 74, 76],
  (1, 1): [15, 51, 16, 52, 21, 57, 22, 58,] + [79, 76, 80, 82],
  (1, 2): [27, 63, 28, 64, 33, 69, 34, 70,] + [85, 82, 86, 88],
  (2, 0): [ 5, 41,  0, 36, 11, 47,  6, 42,] + [74, 89, 72, 77],
  (2, 1): [17, 53, 12, 48, 23, 59, 18, 54,] + [80, 77, 78, 83],
  (2, 2): [29, 65, 24, 60, 35, 71, 30, 66,] + [86, 83, 84, 89]
\end{lstlisting}
\end{minipage}
\end{center}
The two lists given for each plaquette consist of (i) the qubit indiced within the corresponding plaquettes as given in Fig.~\ref{supplement_fig_qubit_labeling} and (ii) the ancilla qubits surrounding the plaquette on the left, right, down, and up (in that order). Notice that the plaquettes defined here are \textit{not} the (odd) $P$- and $Q$-plaquettes used for the Trotterisation of the hopping operator. Instead, all strong plaquettes are \textit{even} (i.e they correspond to the large octagons in Fig.~\ref{supplement_fig_mapping_half_filled}). This choice enables roughly a factor 2 shallower circuits for plaquette preparation~\eqref{eq_plaquette_preparation}, since more ancilla are available per plaquette.

\subsubsection*{Sub-Circuit 1a: XXZ over 9 qubits}
The circuit preparing the approximate XXZ ground state is obtained through variational fidelity maximization. We first obtain the ground state of the XXZ model with the given parameter $\delta=-1$ and spin sector $\sum_i Z_i = -3$ on the $3\times 3$ lattice with periodic boundary conditions.
The $3 \times 3$ lattice is made up of each lower-right qubit representing the spin-up fermion within each plaquette. This corresponds to qubit indices, $[2, 4, 0, 14, 16, 12, 26, 28, 24]$. All other qubits are initialised in the $\ket{0}$ (vacuum) state.

\begin{figure*}
    \centering
\includegraphics[scale=1.6]{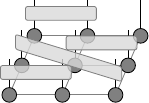}
\hspace{2em} 
\includegraphics[scale=1.6]{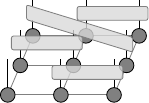}
\\[2em]
\caption{\textbf{Variational circuit layout for state preparation of the $3\times 3$ XXZ model.}
    The circuit consists of alternative application of parallel gates in the layout of odd layers (left) and even layers (right).
    All gates have independent variational parameters as given in the list below. 
\label{supplement_single_layer_XXZ_state_prep_circuit}}
\end{figure*}

The variational circuit is a 1D brickwall circuit with qubit ordering row by row. The circuit has 10 layers in total, as shown in Fig.~\ref{supplement_single_layer_XXZ_state_prep_circuit}.
\begin{center}
\begin{minipage}{0.75\textwidth}
\begin{lstlisting}[style=mystyle, caption={}, label={brickwall_circuit}]
layer-1: [U(2, 4), U( 0, 14), U(16, 12), U(26, 28)]
layer-2: [U(4, 0), U(14, 16), U(12, 26), U(28, 24)]
...
\end{lstlisting}
\end{minipage}
\end{center}
where each $U$ is a general $U(1)$ symmetry-preserving two-qubit gate
\begin{equation}
\label{eq:U1_gate}
U_{i,j} = e^{-i c_1 Z_i} e^{-i c_2 Z_j} e^{-i c_3 X_i X_j} e^{-i c_4 Y_i Y_j} e^{-i c_5 Z_i Z_j} e^{-i c_6 Z_i} e^{-i c_7 Z_j}.    
\end{equation}
The objective function to maximize is the overlap between the target state and the state generated by the circuit. The optimization over the circuit parameters is carried out through gradient descent over the infidelity.  
Specifically, we utilize the quasi-second order gradient descent method, L-BFGS algorithm, from SciPy~\cite{2020SciPy-NMeth}.
With the 10-layer circuit, we obtain the fidelity $\mathcal{F}=0.99478$ with the best circuit realization from 30 random initializations.
Notice each general $U(1)$ symmetry-preserving gate requires 3 elementary entangling 2-qubit gates, e.g., ZZPhase on our hardware. Due to a cancellation at the first layer, the total depth in terms of elemantary entangling 2-qubit gate of the XXZ state preparation circuit is 29. The parameters for all $10 \times 4 = 40$ general $U(1)$-preserving gates used in the quantum experiments are given in Listing~\ref{domonolayer_XXZ_params}.

\begin{center}
\begin{minipage}{0.75\textwidth}
\begin{lstlisting}[style=mystyle, caption={\textbf{Parameters for XXZ ground state preparation for the checkerboard model}. Parameters correspond to 10 layers with 4 general $U(1)$ symmetry-preserving two-qubit gates each $U_{i,j} = e^{-i c_1 Z_i} e^{-i c_2 Z_j} e^{-i c_3 X_i X_j} e^{-i c_4 Y_i Y_j} e^{-i c_5 Z_i Z_j} e^{-i c_6 Z_i} e^{-i c_7 Z_j} $. }, label={domonolayer_XXZ_params}]

================  Id  |   Z1  |  Z2  |  XX  |  YY  |  ZZ  |  Z1  |  Z2  |==
coefficients:  [-0.    -0.87   1.249 -0.403 -0.403 -0.    -0.725  0.346]
coefficients:  [-0.    -0.339  0.718 -0.    -0.    -0.    -1.232  0.853]
coefficients:  [-0.    -0.18   0.56  -0.425 -0.425 -0.    -0.894  0.515]
coefficients:  [-0.    -0.339  0.718  0.     0.    -0.    -1.232  0.853]
coefficients:  [-0.     0.138  0.26  -0.105 -0.105 -0.037 -1.616  1.255]
coefficients:  [-0.     0.532 -0.252 -0.366 -0.366  0.199  0.408 -0.886]
coefficients:  [-0.    -0.314  0.793  0.57   0.57  -0.199 -0.947  0.668]
coefficients:  [-0.    -0.045  0.424  0.675  0.675 -0.    -0.715  0.336]
coefficients:  [-0.     0.509  0.028 -1.137 -1.137 -0.315 -0.401  0.18 ]
coefficients:  [-0.    -0.431  0.476 -0.424 -0.424 -1.596 -0.803  0.089]
coefficients:  [-0.    -0.061  0.44  -0.093 -0.093  0.001 -1.284  0.904]
coefficients:  [-0.     1.041  0.214 -0.394 -0.394 -0.01   1.017 -0.52 ]
coefficients:  [-0.     0.203  0.172 -0.702 -0.702  0.159 -1.408  1.025]
coefficients:  [-0.     0.062  0.299  0.686  0.686 -0.787 -1.27   0.873]
coefficients:  [ 1.571  1.035 -0.459 -0.456 -0.456 -0.14  -0.463  0.281]
coefficients:  [-1.571  0.591 -0.344  0.401  0.401 -0.204 -0.855  0.344]
coefficients:  [ 1.571  0.375 -0.096 -0.072 -0.072 -0.516 -0.748  0.268]
coefficients:  [-0.    -0.877  0.841 -0.336 -0.336  0.208 -1.096  0.302]
coefficients:  [ 1.571 -0.013  0.569 -0.756 -0.756 -0.557 -0.393  0.191]
coefficients:  [-0.     0.134  0.348  0.543  0.543 -0.038 -0.582  0.306]
coefficients:  [-0.     0.397 -0.246 -0.294 -0.294  0.21  -0.495 -0.113]
coefficients:  [-0.     0.492 -0.597  0.651  0.651 -1.434 -0.811 -0.053]
coefficients:  [-0.    -0.656  0.522 -0.678 -0.678 -0.802 -0.88  -0.013]
coefficients:  [-0.     0.678 -0.313 -0.687 -0.687  0.264  0.059 -0.452]
coefficients:  [ 1.571  0.034  0.177 -0.591 -0.591 -0.649 -0.805  0.258]
coefficients:  [-0.    -0.684  0.371 -0.695 -0.695 -1.402 -1.095  0.024]
coefficients:  [-0.    -0.64   0.316 -0.703 -0.703 -0.753 -1.337  0.255]
coefficients:  [-0.     0.297  0.182 -0.616 -0.616 -1.423 -1.163  0.884]
coefficients:  [ 1.571  0.919 -0.51   0.706  0.706  0.068 -0.534  0.184]
coefficients:  [-0.    -0.265  0.492 -0.642 -0.642  0.72  -0.374 -0.157]
coefficients:  [-0.    -0.519  0.603  0.284  0.284  0.28  -1.335  0.66 ]
coefficients:  [-0.     0.575 -0.154 -1.191 -1.191  0.073 -0.944  0.607]
coefficients:  [ 1.571  0.341  0.657 -0.09  -0.09  -0.654  0.278 -0.039]
coefficients:  [-0.     0.148  0.073 -1.135 -1.135  0.67  -0.607  0.069]
coefficients:  [-0.    -0.677  0.677  0.115  0.115 -0.184 -1.303  0.545]
coefficients:  [-0.    -0.152  0.579  0.225  0.225 -0.011 -1.451  1.119]
coefficients:  [-0.    -0.514  0.157 -0.619 -0.619 -0.706 -1.38   0.264]
coefficients:  [-0.    -0.551  0.237  0.284  0.284 -0.828 -1.447  0.374]
coefficients:  [-0.     0.739  0.736  0.536  0.536  0.566 -0.439  1.156]
coefficients:  [-0.    -0.544  0.106 -0.275 -0.275 -0.678 -1.967  0.77 ]
================  Id  |   Z1  |  Z2  |  XX  |  YY  |  ZZ  |  Z1  |  Z2  |==
\end{lstlisting}
\end{minipage}
\end{center}

\subsubsection*{Sub-Circuit 1b: Toric code preparation}
During the XXZ state preparation, the toric code preparation takes place in parallel. We use exactly the same circuit as in the half-filled case, see Section~\ref{sec_supplement_toric_code_preparation}. We add a barrier to align the toric code state preparation to end at the same time as the XXZ state preparation. This is meant to reduce memory errors from qubit idling. Assuming homogeneous magnetic field imperfections $h$, a state $\ket{\psi}$ idling for a short time $s$ would incur infidelity
\begin{equation}
    \left| \left \langle \psi | e^{-ish \sum_i Z_i} | \psi \right \rangle \right|^2 = 1-(sh)^2 \left( \left \langle \left(\sum Z_i \right)^2 \right \rangle - \left \langle \sum Z_i \right \rangle^2 \right) + \mathcal{O}((sh)^3)
\end{equation}
which is zero for the initial $\ket{0}$-state, but $\mathcal{O}(N)$ for the toric code.

\subsubsection*{Sub-Circuit 2: Injection}

Recall the injection of the 36-qubit spin state into 72 the 72 fermionic orbitals represented by the Octagon Fermionic Encoding in previous section. This local update rule acts on one qubit from the spin state and one qubit from the fermionic vacuum:
\begin{equation}
\begin{aligned}
    |0\rangle \otimes |0\rangle \rightarrow |1 0\rangle\\
    |1\rangle \otimes |0\rangle \rightarrow |0 1\rangle    
\end{aligned}
\end{equation}
One can verify the fermionic parity is respected after the local update.

In the checkerboard model case, we have a spin state over 9 qubits and would like to expand to 72 fermionic orbitals. We consider the following local update rule, defined on one qubit from the spin state and 7 qubits from the fermionic vacuum. In our convention, the qubit from the spin state is placed at index-2 position:
\begin{equation}
\begin{aligned}
    |0\rangle^{\otimes2}\otimes  |0\rangle \otimes |0\rangle^{\otimes5} \rightarrow |01001000\rangle = |\tilde{0}\rangle \\
    |0\rangle^{\otimes2}\otimes  |1\rangle \otimes |0\rangle^{\otimes5} \rightarrow |01101001\rangle = |\tilde{1}\rangle
\end{aligned}
\end{equation}
which is graphically represented as
\begin{equation}
\begin{aligned}
\vcenter{\hbox{\includegraphics[width=3.5cm]{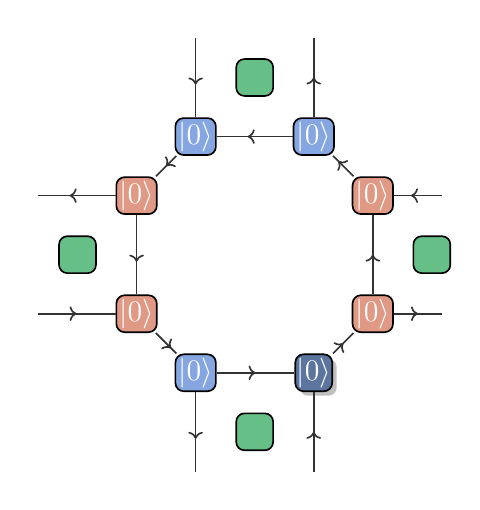}}}
\;&\rightarrow\;
\vcenter{\hbox{\includegraphics[width=3.5cm]{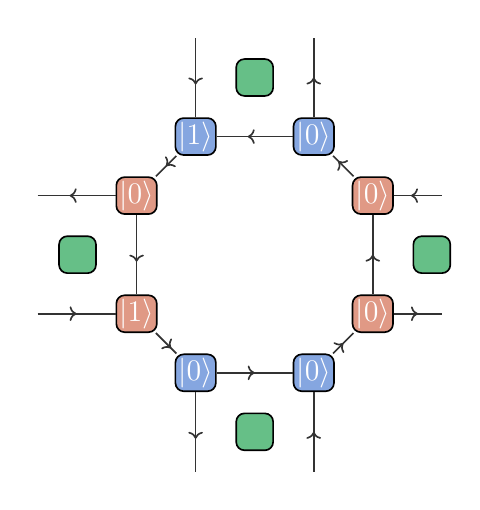}}} 
&=
\vcenter{\hbox{\includegraphics[width=1.5cm]{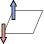}}} \\
\vcenter{\hbox{\includegraphics[width=3.5cm]{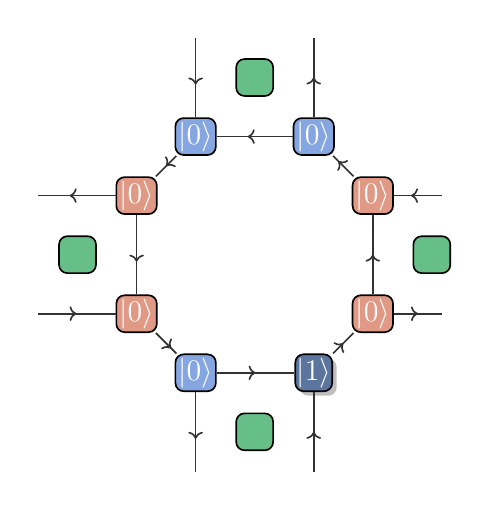}}}
\;&\rightarrow\;
\vcenter{\hbox{\includegraphics[width=3.5cm]{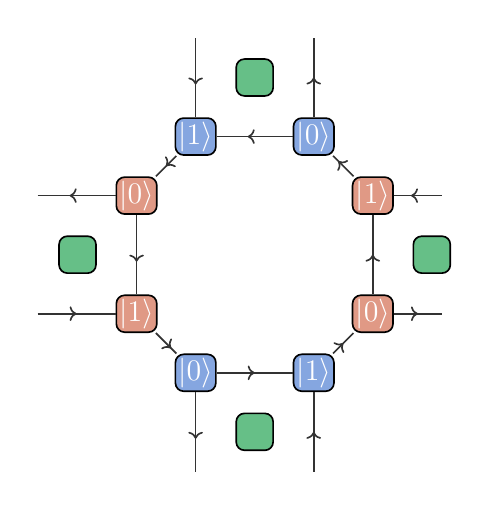}}}
&=
\vcenter{\hbox{\includegraphics[width=1.5cm]{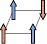}}},
\end{aligned}
\end{equation}
where the shaded qubit is the representative qubit which carries the information of the XXZ state. A sketch of the update rule is provided in Listing~\ref{domonolayer_injection}.
\begin{center}
\begin{minipage}{0.75\textwidth}
\begin{lstlisting}[style=mystyle, caption={\textbf{Sub-Circuit 2 of the checkerboard model state preparation.} This snippet coherenly maps qubit states to representative plaquette states for the quarter- and half-filled sectors.}, label={domonolayer_injection}]

    # The input plaquette_qubits are qubits given according to the order:
    # [1ower-left-spin-up, 1ower-left-spin-down,
    #  lower-right-spin-up, lower-right-spin-down,
    #  upper-left-spin-up, upper-left-spin-down,
    #  upper-right-spin-up, upper-right-spin-down,
    #  ancilla_left, ancilla_down, ancilla_right, ancilla_up]
    #
    # The spin state is already at plaquette_qubits[2], i.e.,
    # lower-right-spin-up.
    #
    # |0000 0000> -> |0100 1000>
    # |0010 0000> -> |0110 1001>
    
    # Prepare the initial product state
    circ.X(plaquette_qubits[1])
    circ.X(plaquette_qubits[4])
    
    # Perform the injection
    circ.CX(plaquette_qubits[2], plaquette_qubits[7])

    # We have to fix the parity of the right stabilisers
    circ.CX(plaquette_qubits[2], plaquette_qubits[10])
    
\end{lstlisting}
\end{minipage}
\end{center}
Notice that both states $|01001000\rangle$ and $|01101001\rangle$ have even fermionic parity. Therefore, the fermionic parity is respected within the plaquette. That is the stabiliser condition is satisfied within the plaquette, e.g., $Z_{1} Z_{2} Z_{7} Z_{8} Z_{37} Z_{38} Z_{43} Z_{44} Y_{72} Y_{73} X_{75} X_{87}=1$.
However, the stabiliser condition can be violated on the upper and lower right  face if a given plaquette is in the $|\tilde{1}\rangle$ state. That is for example the stabiliser of the upper right face, e.g., $Z_{8} Z_{9} Z_{14} Z_{15} Z_{44} Z_{45} Z_{50} Z_{51} Y_{73} X_{75} X_{76} Y_{79} = -1$ can be violated by the presence of a $|\tilde{1} \rangle$ state on the lower left plaquette, resulting on flipping $Z_{44}$.
To restore the stabiliser conditions on the two right faces, we perform an additional CX gate from the spin qubit to the ancilla qubit on the right, which fixes the stabiliser condition when the final state is $|\tilde{1}\rangle$.
The total depth of the injection circuit is $2$ as there are two CX involved per plaquette.

\subsubsection*{Sub-Circuit 3: Basis transformation}

The last step of the state preparation is to coherently transform the two initial fermionic states $|\tilde{0}\rangle$ and $|\tilde{1}\rangle$ into the target fermionic states $|s\rangle$ and $|d\rangle$. Recall the $|s\rangle$ and $|d\rangle$ states are respectively the ground states from the 2-particle and 4-particle sectors of a $2\times 2$ Fermi-Hubbard plaquette with $U/t=2$.

\begin{equation}
\begin{aligned}
    |\tilde{0}\rangle &= |01001000\rangle \longrightarrow |s\rangle  \\
    |\tilde{1}\rangle &= |01101001\rangle \longrightarrow |d\rangle,
\end{aligned}
\end{equation}
that is,
\begin{equation}
\begin{aligned}
\vcenter{\hbox{\includegraphics[width=1.5cm]{SupplementFigures/plaquette_injection_07.pdf}}}
\;&\rightarrow\;
\vcenter{\hbox{\includegraphics[width=1.5cm]{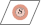}}}
\\
\vcenter{\hbox{\includegraphics[width=1.5cm]{SupplementFigures/plaquette_injection_06.pdf}}}
\;&\rightarrow\;
\vcenter{\hbox{\includegraphics[width=1.5cm]{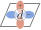}}}
\end{aligned}
\end{equation}
Since all fermionic states live in the Hilbert space of same size and both the initial states and the target states are orthonormal, i.e., $\langle\tilde{0} | \tilde{1} \rangle = 0 = \langle s | d \rangle$, the transformation can be realized by unitaries generated by some fermionic (and fermion-number-preserving) Hamiltonian. We can formulate this as a variational problem to find a \emph{fermionic} quantum circuit $U_f$ that maps simultaneously and most accurately the two initial \emph{fermionic} states to the two target \emph{fermionic} states. The optimization objective is to maximize the sum of the fidelity or equivalently minimizing the sum of the infidelity
\begin{equation}
C = 2 - |\langle s | U_f | \tilde{0}\rangle |^2 - |\langle d | U_f | \tilde{1}\rangle |^2
\end{equation}

The layout for the variational fermionic circuit consisting of general number-preserving rotations on two fermionic modes gates is shown in Listing~\ref{domonolayer_basis_trans_circuit}. The indices are the positions in the list order given in listing~\ref{domonolayer_plaquette_convention}.
\begin{center}
\begin{minipage}{0.75\textwidth}
\begin{lstlisting}[style=mystyle, caption={\textbf{Layout of the fermionic variational circuit to map initial states to Hubbard plaquette ground states.}}, label={domonolayer_basis_trans_circuit}]
list_of_indices = [
                   (0, 1), (2, 3), (4, 5), (6, 7),
                   (1, 3), (5, 7), (0, 4), (2, 6),
                   (0, 1), (2, 3), (4, 5), (6, 7),
                   (0, 2), (4, 6), (1, 5), (3, 7),
                   ] * 2

\end{lstlisting}
\end{minipage}
\end{center}
The fermionic circuit has in total 8 layers of gates.
Each layer has 4 gates that can be applied in parallel.
The first layer of the circuit, i.e., $[(0, 1), (2, 3), (4, 5), (6, 7)]$, is composed of unitaries acting between the on-site fermions with up spin and down spin, i.e. one may think of them as terms like $n_{\uparrow} n_{\downarrow}$, as well as $S^x$ and $S^z$.
The second layer of the circuit, i.e., $[(1, 3), (5, 7), (0, 4), (2, 6)]$, is composed of unitaries acting between the fermions with down spin horizontally and the fermions with up spin vertically, i.e. one may think of these these terms as implemeting fermionic hopping and currents on the vertical (horizontal) bonds for the up- (down-)spins.
The third layer repeats the same layout as the first layer.
The fourth layer of the circuit, i.e., $[(0, 2), (4, 6), (1, 5), (3, 7)]$, is composed of unitaries acting between the fermions with up spin horizontally and the fermions with down spin vertically.
The fifth to the eighth layers repeat the same layout as the first to the fourth layers.
Notice that this layout resembles the circuit of the hopping terms in the time evolution circuit. In fact, the fermionic circuit can be viewed as a time evolution circuit with hopping terms interleaving with chemical potential and onsite interaction where the parameters in the Hamiltonian are the variational parameters.
Each on-site general $U(1)$ symmetry-preserving two-qubit gate in general requires depth-3 in terms of the elementary 2-qubit entangling gates.
While the general two-qubit gate between sites, when interpreted as fermionic operation and translated to the Octagon Encoding circuit, requires depth-6 in terms of the elementary 2-qubit entangling gates.

To carry out the optimization of finding parameters of the fermionic circuits, here we utilize the Jordan-Wigner transformation for simplicity and efficiency. The Jordan-Wigner ordering we choose follows the qubit indices ordering. We first obtained both the $|s\rangle$ and $|d\rangle$ states with the Jordan-Wigner transformation. Then, we transform the variational fermionic circuit correspondingly according to the Jordan-Wigner ordering into a variational circuit over qubits by inserting fSWAPs. That is for any fermionic gate acting over not adjacent fermionic modes in the Jordan-Wigner ordering, we insert the fSWAPs. For example, the $U^f_{(1,3)}=\textrm{fSWAP}_{(1,2)} U_{(2,3)} \textrm{fSWAP}_{(1,2)}$. For a fermionic gate acting over adjacent fermionic modes, we apply no additional transformation.
Now the problem reduces back to the standard variational quantum circuit optimization. We optimize the circuit with the quasi-second order gradient descent method, L-BFGS algorithm, from SciPy~\cite{2020SciPy-NMeth} over 30 different random initialization and select the best result. The best average fidelity is $(2-C)/2=0.99446$.

After the optimization, we can readout parameters for the general $U(1)$ symmetry-preserving gate in the circuit which gives exactly the corresponding parameters in the fermionic circuit. We can then use these parameters in the fermionic circuit to construct the circuit in the Octagon Encoding.
The list of the parameters is given in Listing~\ref{domonolayer_basis_trans_circuit_params}.
\begin{center}
\begin{minipage}{0.75\textwidth}
\begin{lstlisting}[style=mystyle, caption={\textbf{Parameters of the circuit that maps fermionic computational eigenstates to the ground state of the Hubbard plaquettes. }The decomposition coefficients of the serialized general $U(1)$ symmetry-preserving two-qubit gates $U_{i,j} = e^{-i c_1 Z_i} e^{-i c_2 Z_j} e^{-i c_3 X_i X_j} e^{-i c_4 Y_i Y_j} e^{-i c_5 Z_i Z_j} e^{-i c_6 Z_i} e^{-i c_7 Z_j} $. The coefficients of the first layer of gate cancelled out and have no contribution to the final circuit.}, label={domonolayer_basis_trans_circuit_params}]

================  Id  |   Z1  |  Z2  |  XX  |  YY  |  ZZ  |  Z1  |  Z2  |==
coefficients:  [-0.     0.446 -0.067  0.    -0.    -0.    -0.446  0.067]
coefficients:  [ 0.     0.446 -0.067 -0.    -0.    -0.    -0.446  0.067]
coefficients:  [ 0.     0.446 -0.067 -0.    -0.    -0.    -0.446  0.067]
coefficients:  [-0.     0.446 -0.067 -0.    -0.     0.    -0.446  0.067]
coefficients:  [ 0.009  0.517 -0.144 -0.391 -0.391  0.004 -0.533  0.148]
coefficients:  [ 0.003  0.653 -0.281  0.393  0.393  0.015 -0.624  0.238]
coefficients:  [ 0.017  0.484 -0.11   0.278  0.278 -0.004 -0.535  0.15 ]
coefficients:  [-0.018  0.751 -0.376 -0.493 -0.493  0.01  -0.669  0.285]
coefficients:  [-0.     0.416 -0.071 -0.    -0.    -0.048 -0.477  0.063]
coefficients:  [-0.     0.497 -0.063 -0.    -0.     0.026 -0.396  0.071]
coefficients:  [-0.     0.476 -0.047 -0.    -0.    -0.336 -0.416  0.087]
coefficients:  [ 0.     0.396 -0.087 -0.    -0.    -0.455 -0.497  0.047]
coefficients:  [-0.005  0.504 -0.096  0.412  0.412 -0.275 -0.692  0.341]
coefficients:  [ 0.     0.343  0.007 -0.381 -0.381 -0.657 -0.36  -0.049]
coefficients:  [-0.018  0.686 -0.288 -0.063 -0.063  0.109 -0.674  0.314]
coefficients:  [ 0.025  0.262  0.107  0.715  0.715 -0.085 -0.854  0.465]
coefficients:  [-0.     0.417 -0.064  0.    -0.     0.057 -0.476  0.07 ]
coefficients:  [-0.     0.496 -0.128 -0.    -0.     0.57  -0.397  0.007]
coefficients:  [-0.     0.415 -0.054 -0.    -0.     0.877 -0.478  0.08 ]
coefficients:  [ 0.     0.458 -0.022 -0.    -0.     0.024 -0.435  0.112]
coefficients:  [ 0.012  0.568 -0.254 -0.382 -0.382 -0.076 -0.519  0.075]
coefficients:  [ 0.007  0.7   -0.271 -0.388 -0.388  0.112 -0.778  0.449]
coefficients:  [ 0.006  0.562 -0.25   0.338  0.338 -0.409 -0.636  0.189]
coefficients:  [ 0.001  0.565 -0.134  0.328  0.328 -0.357 -0.462  0.135]
coefficients:  [-0.     0.42  -0.065 -0.    -0.    -0.114 -0.473  0.069]
coefficients:  [ 0.     0.485 -0.127  0.    -0.     0.148 -0.407  0.007]
coefficients:  [ 0.     0.412 -0.1   -0.    -0.    -0.056 -0.481  0.034]
coefficients:  [ 0.     0.468  0.024 -0.    -0.     0.059 -0.424  0.158]
coefficients:  [ 0.016  0.347  0.049 -0.029 -0.029  0.079 -0.405  0.043]
coefficients:  [-0.021  0.066  0.31  -0.029 -0.029 -0.45  -0.1   -0.282]
coefficients:  [ 0.002  0.967 -0.618 -0.388 -0.388  0.031 -0.743  0.333]
coefficients:  [ 0.006  0.211  0.186 -0.411 -0.411 -0.018 -0.613  0.251]
================  Id  |   Z1  |  Z2  |  XX  |  YY  |  ZZ  |  Z1  |  Z2  |==

\end{lstlisting}
\end{minipage}
\end{center}

Similar to the discussion in section~\ref{sec_supplement_fermionic_encoding}, we can now associate each coefficient $c_i$ in the gate decomposition to a coefficient $\theta_i$ in front of the corresponding fermionic operator.
\begin{align*}
    (X_iX_j + Y_iX_j)/2 &\longrightarrow c^\dagger_{i \sigma} c_{j \sigma} + \mathrm{h.c.}\\
    (1-Z_{i \sigma})/2 &\longrightarrow n_{i \sigma}
\end{align*}
The mapping is different for the hopping term from the Octagon Encoding here as it comes from the Jordan-Wigner transformation.
We observe that from the variational circuit, the current terms have zero coefficients.

We observe that for all gate decomposition, $c_3 = c_4$. We can therefore identify the parameters.
\begin{align}
   &e^{-i c_1 Z_i} e^{-i c_2 Z_j} e^{-i \times 2c_3 (X_i X_j + Y_i Y_j)/2} e^{-i c_5 Z_i Z_j} e^{-i c_6 Z_i} e^{-i c_7 Z_j} \nonumber \\
   &\qquad \longrightarrow e^{-i c_1 (2n_i-1)} e^{-i c_2 (2n_j-1)} e^{-i \times 2c_3 (c^\dagger_{i \sigma} c_{j \sigma} + \mathrm{h.c.})} e^{-i c_5 (2n_i-1)(2n_j-1)} e^{-i c_6 (2n_i-1)} e^{-i c_7 (2n_j-1)} 
\end{align}
and then reconstruct the circuit in Octagon Encoding. Note that the only non-trivial difference between the Jordan-Wigner encoding used during the variational optimisation and the Octagon Encoding occurs during the hopping terms, since number operators $n_i$ in both encodings map to the same operation.

Sub-Circuits 2 and 3 can be applied in parallel to all 9 plaquettes of the $6 \times 6$ lattice. The full basis transformation completes the (approximate) preparation of the unique ground state of the checkerboard model in the weakly coupled limit $t'/t \rightarrow 0$.

\subsection{Bilayer}
In order to prepare the perturbative ground state of the exchange-coupled bilayer Fermi-Hubbard model with $U=0$ and $t/J \rightarrow 0$, we proceed in two steps: first we prepare the ground state of the 4x4 XXZ model with $\delta = -\frac{2}{3}$  and then inject that state into the fermionic Fock space.

\subsubsection{Ground state preparation of the $4 \times 4$ XXZ model.}

\begin{figure*}[h]
    \centering
\includegraphics[scale=1.6]{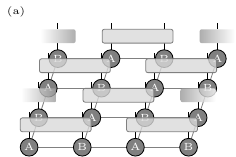}
\hspace{2em} 
\includegraphics[scale=1.6]{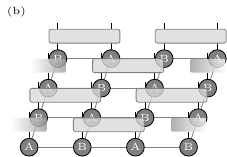}
\\[2em]
\includegraphics[scale=1.6]{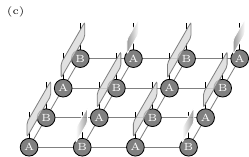}
\hspace{2em} 
\includegraphics[scale=1.6]{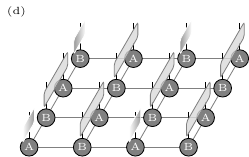}
\caption{\textbf{Variational circuit layout for state preparation of the $4\times 4$ XXZ model.}
    (a) Layer with horizontal gates between qubits on sub-lattices A and B.
    (b) Layer with horizontal gates between qubits on sub-lattices B and A.
    (c) Layer with vertical gates between qubits on sub-lattices A and B.
    (d) Layer with vertical gates between qubits on sub-lattices B and A.
    Due to the periodic boundary condition, there are gates acting across the boundary.
    All gates within the same layer are identical. 
\label{supplement_bilayer_XXZ_state_prep_circuit}}
\end{figure*}

The variational circuit we used for the ground state preparation here is a translationally invariant brickwall circuit ansatz with a unit-cell size $2\times 2$. Gates within a given layer are applied in parallel over all lattice sites.
In the first layer, we apply the same gate on all horizontal bonds between qubits on sub-lattices A and B, as shown in Fig.~\ref{supplement_bilayer_XXZ_state_prep_circuit}(a).
In the second layer, gates are applied is applied between all qubits on sub-lattices B and A, as shown in Fig.~\ref{supplement_bilayer_XXZ_state_prep_circuit}(b). All horizontal bonds have been acted on at this stage.
In the third layer, we apply the same vertical gates between all the qubits on sub-lattices A and B, as shown in Fig.~\ref{supplement_bilayer_XXZ_state_prep_circuit}(c).
Finally, in the fourth layer, we apply the same vertical gates between all qubits on sub-lattices B and A, as shown in Fig.~\ref{supplement_bilayer_XXZ_state_prep_circuit}(d).
The application of these four layers is called a "round". We repeat the same construction for two more rounds, resulting in a circuit of 12 layers (3 rounds). While all gates considered are general $U(1)$-symmetry-preserving two-site gate, the total depth in terms of elementary two-qubit entangling gates is 36.
The circuit ansatz respects two-site translational invariance both horizontally and vertically by two sites. While the ansatz does not preserve the C\(_4\) symmetry, we observe after the optimization the final state is approximately C\(_4\) symmetric.

We use exact state vector simulation to obtain the exact energy gradient of all parameters. The optimization is carried out with the quasi-second order gradient descent method, L-BFGS algorithm, from SciPy~\cite{2020SciPy-NMeth} over 30 random initializations. 

The classically optimized approximate ground-state of the $4 \times 4$ XXZ model is prepared on the up-spins of the $A$-layer, that is, the first 16 qubits according to the ordering presented in Fig. \ref{supplement_fig_qubit_labeling}(b). All other qubits are initialised in the $|0\rangle$ (vacuum) state. For completeness, in Listing~\ref{brickwall_circuit}, we give the exact construction of the circuit for the first four layers (first round).
\begin{center}
\begin{minipage}{0.8\textwidth}
\begin{lstlisting}[style=mystyle, caption={\textbf{Layout of one round of the XXZ preparation circuit}. The next eight layers are the repetition of the same structure. Note that in each layer, all gates are the same to ensure the translational invariance.}, label={brickwall_circuit}]
layer-1: [U1(0, 1), U1(8, 9), U1(5, 6), U1(13, 14),
          U1(2, 3), U1(10, 11), U1(7, 4), U1(15, 12)]
layer-2: [U2(4, 5), U2(12, 13), U2(1, 2), U2(9, 10),
          U2(6, 7), U2(14, 15), U2(3, 0), U2(11, 8)]
layer-3: [U3(0, 4), U3(8, 12), U3(5, 9), U3(13, 1),
          U3(2, 6), U3(10, 14), U3(7, 11), U3(15, 3)]
layer-4: [U4(4, 8), U4(12, 0), U4(1, 5), U4(9, 13),
          U4(6, 10), U4(14, 2), U4(3, 7), U4(11, 15)]
...
\end{lstlisting}
\end{minipage}
\end{center}
Each $U$ is a general $U(1)$ symmetry-preserving two-qubit gate, see Eq.~\eqref{eq:U1_gate}. The final coefficients obtained from the classical optimization used for all the experiments reported in the main text are given in Listing~\ref{bilayer_XXZ_params}.
\begin{center}
\begin{minipage}{0.75\textwidth}
\begin{lstlisting}[style=mystyle, caption={\textbf{Parameters for the approximate ground state preparation of the $4 \times 4$ XXZ model.} The coefficients of the serialized general $U(1)$ symmetry-preserving two-qubit gates follow the convention $U_{i,j} = e^{-i c_1 Z_i} e^{-i c_2 Z_j} e^{-i c_3 X_i X_j} e^{-i c_4 Y_i Y_j} e^{-i c_5 Z_i Z_j} e^{-i c_6 Z_i} e^{-i c_7 Z_j} $. The 12 sets of coefficients correspond to the 12 layers (3 rounds) of the state preparation circuit. The same gate is applied across all bonds in one layer.}, label={bilayer_XXZ_params}]

================  Id  |   Z1  |  Z2  |  XX  |  YY  |  ZZ  |  Z1  |  Z2  |==
coefficients:  [-0.    -0.823  1.203 -0.191 -0.191 -0.    -0.712  0.333]
coefficients:  [-0.     0.674 -0.295 -0.872 -0.872 -0.005 -0.486  0.107]
coefficients:  [-0.    -0.148  0.527 -0.182 -0.182 -0.018 -1.446  1.067]
coefficients:  [-0.     0.388 -0.009 -0.06  -0.06  -0.012 -0.38   0.001]
coefficients:  [-0.     1.406 -1.027 -0.277 -0.277 -0.027 -0.238 -0.141]
coefficients:  [-0.     0.565 -0.186 -0.16  -0.16  -0.036 -0.638  0.259]
coefficients:  [-0.    -0.02   0.4    0.215  0.215 -0.065 -0.742  0.363]
coefficients:  [-0.    -0.476  0.855 -0.092 -0.092 -0.068 -1.201  0.822]
coefficients:  [ 0.     1.005 -0.626  0.155  0.155 -0.076 -0.216 -0.163]
coefficients:  [-0.    -0.578  0.957  0.137  0.137 -0.068 -1.01   0.631]
coefficients:  [-0.    -0.504  0.883  0.143  0.143 -0.046 -0.914  0.535]
coefficients:  [-0.     1.284 -0.905 -0.048 -0.048 -0.04  -0.254 -0.125]
================  Id  |   Z1  |  Z2  |  XX  |  YY  |  ZZ  |  Z1  |  Z2  |==

\end{lstlisting}
\end{minipage}
\end{center}

\subsubsection{Injection circuit.}
Following the perturbation theory in section~\ref{sec_supplement_perturbation_theory}, we need to implement a mapping from the qubit states $\ket{0}$ and $\ket{1}$ to the rung states with either a singlet or a hole pair:
\begin{align}
    |1\rangle &\rightarrow \frac{1}{\sqrt{2}} (c_{Ai \uparrow}^\dagger c_{Bi \downarrow}^\dagger - c_{Ai \downarrow}^\dagger c_{Bi \uparrow}^\dagger ) |\text{vacuum}\rangle\\
    |0\rangle &\rightarrow |\text{vacuum} \rangle.
\end{align}
The injection circuit realizing the above transformation is given by
\begin{equation}
CX(Ai\uparrow, Bi\downarrow) CX( Bi\downarrow, Ai\downarrow) CH(Bi\downarrow, Ai \downarrow) CX(Ai\downarrow, Bi\downarrow) CX(Ai\downarrow,Bi\uparrow) CX(Bi\uparrow,Ai\uparrow)
\end{equation}
on each rung $i$ in parallel. Due to the edge convention in the Octagon encoding, edges point from up- to down-fermions on one sublattice and vice-versa on the other sublattice. Thus, the injection transforms $\ket{1}$ into a singlet on one sublattice and into $(-1)\times$singlet on the other sublattice. Since we are preparing a coherent superposition of singlet coverings, these relative signs are important and we take care of them by applying a Pauli-$Z$-rotation on the sublattice of the $4 \times 4$ XXZ model that contains qubit $0$ before the injection sequence above.

\newpage
\section{Measurement Circuits}
\label{sec_supplement_measurement_circuits}
\subsection{Kinetic energy}
To measure the kinetic energy of a state, one needs to measure $X_{i \sigma} X_{j \sigma} P_a$ and $Y_{i \sigma} Y_{j \sigma} P_a$ for every edge $\langle i,j\rangle, \sigma$ with Pauli operator $P_a$ on the ancilla $a$ adjacent to the edge. Since these operators do not all commute with each other, we need to measure them in different shots. We define four different measurement setting. Each measurement setting consists of a choice of non-overlapping bonds $\{ \langle ij \rangle \sigma \}$. On each of these bonds we measure $X_{i \sigma} X_{j \sigma} P_a$ and $Y_{i \sigma} Y_{j \sigma} P_a$ simultaneously. In the first (second) measurement basis, we measure all the horizontal (vertical) edges of the $P$ plaquettes for the up-spins and $Q$ plaquettes for the down-spins. In the third (fourth) measurement basis, we measure all the horizontal (vertical) edges of the $Q$ plaquettes for the up-spins and $P$ plaquettes for the down-spins. The measurement gadget that we apply is $CX(j \sigma ,a)CX(i \sigma,j \sigma)H(i \sigma)$ for vertical edges and $S^\dagger(a) CX(j \sigma,a) S(a) CX(i \sigma,j \sigma)H(i \sigma)$ for horizontal edges (in circuit ordering). After this local basis transformation, the quantity $X_{i \sigma} X_{j \sigma} P_a$ is obtained by reading out $Z_{i \sigma}$, and the quantity $Y_{i \sigma} Y_{j \sigma} P_a$ by reading out $-Z_{i \sigma}Z_{j \sigma}$. For the third and fourth measurement bases, these gadgets are preceded by applying $CZ$ on all pairs $(i\downarrow,i\uparrow)$ for $i=1,...,L$.

\subsection{Eta correlations}
\label{sec_measurement_eta_correlations}
We recall from section \ref{sec_supplement_fermionic_encoding} that the expectation value of the $\eta$ correlation between sites $i,j$ is
\begin{equation}
   \langle \Delta_i^\dagger \Delta_j + \mathrm{h.c} \rangle= \pm\frac{(-1)^{i+j}}{4}\langle(X_{i\uparrow}X_{i\downarrow}-Y_{i\uparrow}Y_{i\downarrow})(X_{j\uparrow}X_{j\downarrow}-Y_{j\uparrow}Y_{j\downarrow})\rangle\,,
\label{eq_eta_measurement}
\end{equation}
where $(-1)^{i,j}$ takes value $\pm 1$ according to whether sites $i,j$ are odd or even in a checkerboard pattern. This operator can be measured by applying the gadget $G = H(i\downarrow)H(i\uparrow)S(i\downarrow)S(i\uparrow) CX(i\downarrow,i\uparrow)H(i\downarrow)$ that was introduced in section~\ref{sec_supplement_trotter_circuits}. Measuring $Z_{i\downarrow}$ after this gadget yields $Y_{i\uparrow}Y_{i\downarrow}$, and measuring $Z_{i\uparrow}$ yields $X_{i\uparrow}X_{i\downarrow}$(and similarly for site $j$).

\subsection{$d$-wave pairing correlation}

In this subsection we describe the procedure that we use to sample from the bond-bond singlet pairing correlator
\begin{equation}
    P_b(x,y) = \frac{4}{N} \sum_{\langle ij \rangle} \Delta_{ij} \Delta^\dagger_{ij+(x,y)} + \mathrm{h.c.},
\end{equation}
the results of which are shown in Fig.~\ref{fig2}d. 

\subsubsection*{Measuring singlet pairing correlations}
We have to measure singlet pairing correlations $\langle \Delta^\dagger_{ij} \Delta_{kl} + \mathrm{h.c.} \rangle$, where $(i,j)$ and $(k,l)$ are nearest-neighbor sites either on a vertical or horizontal links. As described in section~\ref{sec_supplement_fermionic_encoding}, it is equivalently to measure
\begin{equation*}
\langle (\Delta_{ij} + \Delta^\dagger_{ij}) (\Delta_{kl} + \Delta^\dagger_{kl}) \rangle,
\end{equation*}
Since the work in a sector with a fixed number of particles, the additional terms $\langle \Delta_{ij} \Delta_{kl} \rangle$ and $\langle \Delta^\dagger_{ij} \Delta^\dagger_{kl}) \rangle$ are zero. The measurement of 
\begin{equation}
    \Delta_{ij} + \Delta_{ij}^\dagger = \frac{1}{\sqrt{2}}\left( \hat{c}_{j\downarrow} \hat{c}_{i\uparrow} - \hat{c}_{j\uparrow} \hat{c}_{i\downarrow} \right) + \frac{1}{\sqrt{2}}\left( \hat{c}^\dagger_{i\uparrow}\hat{c}^\dagger_{j\downarrow} -  \hat{c}^\dagger_{i\downarrow}\hat{c}^\dagger_{j\uparrow}  \right) 
\end{equation}
can be simplified provided $(i,j)$ are nearest neighbor sites, (which is the case for the $d$-wave bond-bond pairing that we are interested in).
If $i$ and $j$ are vertical nearest neighbours, we first apply $\textrm{fSWAP}$ between $(i\uparrow, j\uparrow)$ and obtain 
\begin{equation}\label{single_pairing_correlation_after_fswap_vertical}
    \textrm{fSWAP}_{(i\uparrow, j\uparrow)} (\Delta_{ij} + \Delta_{ij}^\dagger) \textrm{fSWAP}_{(i\uparrow, j\uparrow)} = \frac{1}{\sqrt{2}}\left( \hat{c}_{j\downarrow} \hat{c}_{j\uparrow} - \hat{c}_{i\uparrow} \hat{c}_{i\downarrow} \right) + \frac{1}{\sqrt{2}}\left( \hat{c}^\dagger_{j\uparrow}\hat{c}^\dagger_{j\downarrow} -  \hat{c}^\dagger_{i\downarrow}\hat{c}^\dagger_{i\uparrow}  \right).
\end{equation}
The fermionic SWAP 
\begin{equation}
\begin{aligned}
    \mathrm{fSWAP}_{(i\uparrow, j\uparrow)} &= c^\dagger_{i\uparrow} c_{j\uparrow} + c^\dagger_{j \uparrow} c_{i \uparrow} - n_{i\uparrow} - n_{j\uparrow} + \mathds{1} \\
    &\propto \exp(\frac{i \pi}{2} [c^\dagger_{i\uparrow} c_{j\uparrow} + c^\dagger_{j \uparrow} c_{i \uparrow}]) \exp(-\frac{i \pi}{2} [n_{i\uparrow} + n_{j\uparrow}]) \\
    &\rightarrow \exp(\frac{i \pi}{4} [X_{i \uparrow}X_{j \uparrow}+Y_{i \uparrow}Y_{j \uparrow}]) S_{i \uparrow} S_{j \uparrow}
\end{aligned}
\end{equation}
can be easily realized in the Octagon encoding. In particular, if a hopping term is simulated right before the measurement, the $\textrm{fSWAP}_{(i\uparrow, j\uparrow)}$ operator can be merged with the existing hopping simply by changing the phase of the XXPhase and YYPhase gates and adding single-qubit $S$-gates.

The right hand side of Eq.~\eqref{single_pairing_correlation_after_fswap_vertical} now is a combination of $\pm \frac{1}{2\sqrt{2}} (X_{i\uparrow} X_{i\downarrow} -Y_{i\uparrow} Y_{i\downarrow}) \mp (X_{j\uparrow} X_{j\downarrow} -Y_{j\uparrow} Y_{j\downarrow}) $ where the overall $\pm$ sign depends on the fermionic ordering, i.e., the arrow direction between the sites $(i,j)$.
If $(i,j)$ follows the arrow direction, i.e., $i \rightarrow j$ and $(i,j)$ is a vertical bond, then we have a $+$ in front the $i$ terms and $-$ in front the $j$ terms.

If $(i, j)$ are on a horizontal bond, the same approach can be applied except that we apply the $\textrm{fSWAP}$ now between $(i\downarrow, j\downarrow)$ as they are the neighbouring qubits in the Octagon Encoding. We have
\begin{align}\label{single_pairing_correlation_after_fswap_horizontal}
    \textrm{fSWAP}_{(i\downarrow, j\downarrow)} (\Delta_{ij} + \Delta_{ij}^\dagger) \textrm{fSWAP}_{(i\downarrow, j\downarrow)} &= \frac{1}{\sqrt{2}}\left( \hat{c}_{i\downarrow} \hat{c}_{i\uparrow} - \hat{c}_{j\uparrow} \hat{c}_{j\downarrow} \right) + \frac{1}{\sqrt{2}}\left( \hat{c}^\dagger_{i\uparrow}\hat{c}^\dagger_{i\downarrow} -  \hat{c}^\dagger_{j\downarrow}\hat{c}^\dagger_{j\uparrow}  \right) \nonumber \\
    &= \frac{1}{\sqrt{2}}\left( \hat{c}_{j\downarrow} \hat{c}_{j\uparrow} - \hat{c}_{i\uparrow} \hat{c}_{i\downarrow} \right) + \frac{1}{\sqrt{2}}\left( \hat{c}^\dagger_{j\uparrow}\hat{c}^\dagger_{j\downarrow} -  \hat{c}^\dagger_{i\downarrow}\hat{c}^\dagger_{i\uparrow}  \right)
\end{align}
recovering the same fermionic operator as above.
If $(i,j)$ follows the arrow direction, i.e., $i \rightarrow j$ and $(i,j)$ is a horizontal bond, then we have a $-$ in front the $i$ terms and $+$ in front the $j$ terms. The sign is flipped compared to the vertical bond because we choose to prepare the strong plaquettes on the large octagons in Fig.~\ref{supplement_fig_qubit_labeling}(a) and as such, the vertical bonds are ordered $(i\downarrow )\rightarrow (i\uparrow) \rightarrow (j\uparrow )\rightarrow (j\downarrow)$ while the horizontal bonds have the opposite ordering $(i\uparrow )\rightarrow (i\downarrow) \rightarrow (j\downarrow )\rightarrow (j\uparrow)$.

The full expectation value becomes
\begin{align}
    \langle (\Delta_{ij} + \Delta^\dagger_{ij}) (\Delta_{kl} + \Delta^\dagger_{kl}) \rangle &= \left( \pm \frac{1}{2\sqrt{2}} (X_{i\uparrow} X_{i\downarrow} -Y_{i\uparrow} Y_{i\downarrow}) \mp (X_{j\uparrow} X_{j\downarrow} -Y_{j\uparrow} Y_{j\downarrow}) \right) \times \nonumber \\
    &\qquad \left( \pm \frac{1}{2\sqrt{2}} (X_{k\uparrow} X_{k\downarrow} -Y_{k\uparrow} Y_{k\downarrow}) \mp (X_{l\uparrow} X_{l\downarrow} -Y_{l\uparrow} Y_{l\downarrow}) \right) .
\end{align}
The expression is now be measured following the same approach as in measuring $\eta$-correlations. See section~\ref{sec_measurement_eta_correlations}.

\begin{figure*}[t]
    \centering
\includegraphics[width=0.85\linewidth]{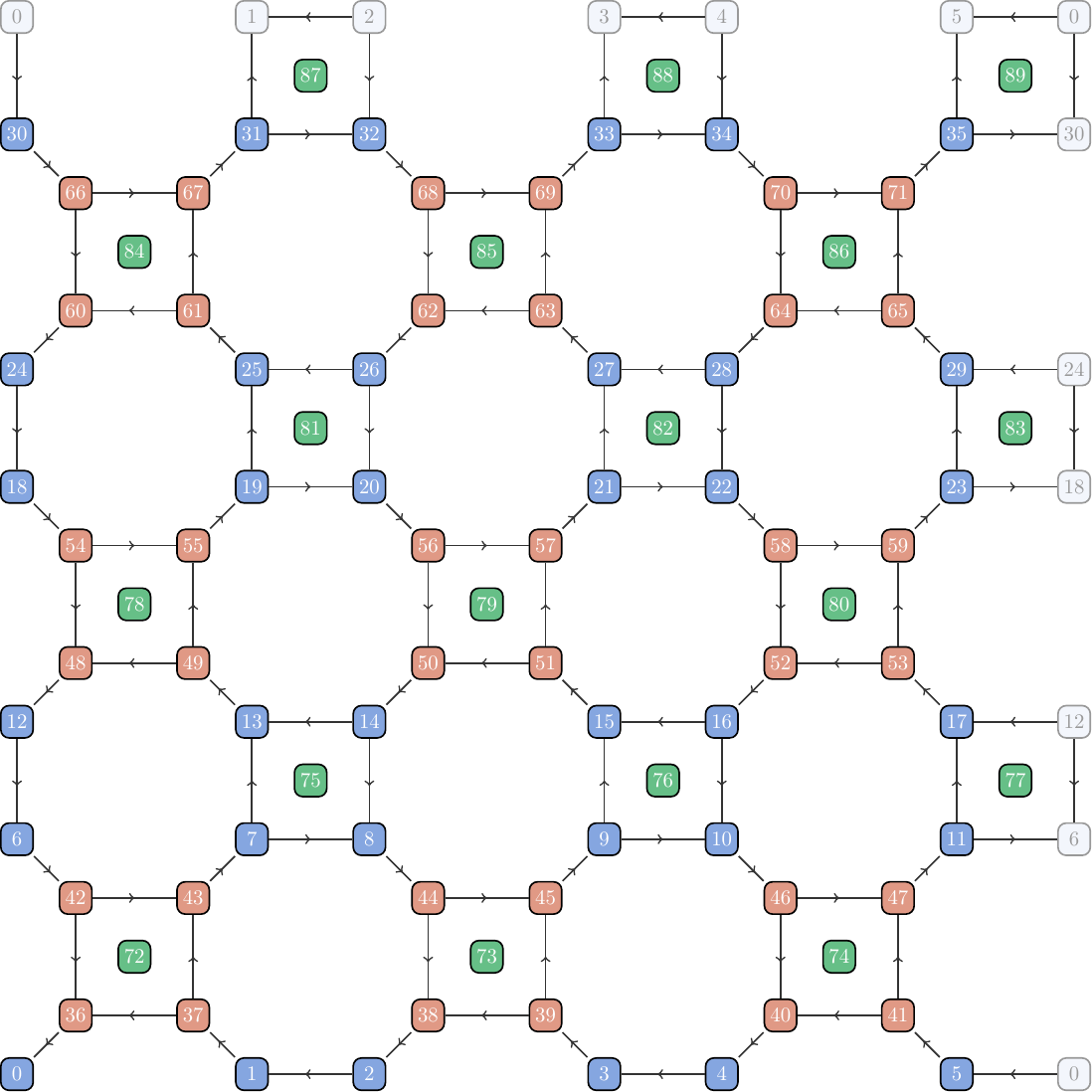}
    \caption{\textbf{Qubit Labeling after a round of CZ is applied on all diagonal edges}. The CZ gates swap the on-site fermionic ordering between qubits $(i)\leftrightarrow(i+36)$. This can be viewed as reordering the qubits in the Octagon Encoding accordingly.
    \label{supplement_fig_qubit_labeling_CZ}}
\end{figure*}

\subsubsection*{fSWAP patterns for $d$-wave pairing}
Since the vertical and horizontal singlet pairing operators do not commute, to measure the $d$-wave pairing correlation over the strong bonds in the plaquettes, we have to perform measurement in two different settings.

In each setting, we assign each plaquette to either measure vertical bonds (V) or horizontal bonds (H). For a measurement of the vertical bonds, we apply two fSWAP operators vertically and then measure the singlet pairing on these vertical bonds. For a measurement of the horizontal bonds, we apply two fSWAP horizontally and then measure the singlet pairing horizontally.

We use the two specific patterns shown below in Listing~\ref{single_layer_fSWAP}.
For each measurement setting, we can choose one of the left vertical bonds in the "V"-plaquette to be the reference bond $(i,j)$ and compute the pairing correlation with respect to all other possible bonds $(k,l)$. The data reported are an average over all possible reference bonds in the two settings.

\begin{center}
\begin{minipage}{0.75\textwidth}
\begin{lstlisting}[style=mystyle, caption={The V, H, patterns for the two measurement settings. The V and H plots on the top give an overview idea of how the V and H are distributed over the 9 plaquettes. The code below realizes the corresponding pattern when combining with the state preparation circuit.}, label={single_layer_fSWAP}]
# Realizing vertical (V) and horizontal (H) measuring pattern 1:
# ========== 
# V H V
# H V H
# V H V
# ========== 

fSWAP(13, 14), fSWAP(19, 20), fSWAP(9, 10), fSWAP(27, 28), fSWAP(3, 4)
fSWAP(33, 34), fSWAP(17, 12), fSWAP(23, 18), fSWAP(37, 43), fSWAP(36, 42)
fSWAP(61, 67), fSWAP(60, 66), fSWAP(38, 44), fSWAP(51, 57), fSWAP(62, 68)
fSWAP(41, 47), fSWAP(52, 58), fSWAP(65, 71)

# Realizing vertical (V) and horizontal (H) measuring pattern 2:
# ========== 
# H V H
# V V V
# H V H
# ========== 

fSWAP(7, 8), fSWAP(25, 26), fSWAP(1, 2), fSWAP(31, 32), fSWAP(11, 6),
fSWAP(29, 24), fSWAP(5, 0), fSWAP(35, 30), fSWAP(49, 55), fSWAP(48, 54),
fSWAP(39, 45), fSWAP(51, 57), fSWAP(50, 56), fSWAP(63, 69), fSWAP(40, 46),
fSWAP(53, 59), fSWAP(52, 58), fSWAP(64, 70)
\end{lstlisting}
\end{minipage}
\end{center}

In the last step of the doped single layer state preparation, we simulate hopping terms over the down-spin $Q$-plaquettes and up-spin $P$-plaquettes.
To this end, one round of CZ have are applied over all sites, resulting in the effectively "CZ-reordered Octagon Encoding", as shown in Fig~\ref{supplement_fig_qubit_labeling_CZ}.

To measure the pairing correlation at this stage, instead of applying CZ over all sites to recover the original ordering and then applying additional fSWAPs to measure pairing correlators, we apply fSWAPs directly on the CZ-reordered Octagon Encoding, resulting in the fSWAPs given in Listing~\ref{single_layer_fSWAP}.
The reason to do this is to save the cost of applying CZ and fSWAP, as now the fSWAP can be merged with the hopping operations. We reduce the circuit depth by 4 with using this technique.

One can follow the same derivation in the previous section with minor change to the qubit ordering to the CZ-reordered Octagon Encoding, resulting in an overall sign flip for both the vertical and horizontal setting. As this overall sign flip shows up twice, the final expression does not change. Finally, we comment that if one wants to include the measurement for inter-plaquette bonds, additional measurement settings have to be considered.

\subsection{Bilayer pairing measurement circuits}

We are interested in inter-layer singlet paring correlations $\langle \Delta_{AiBi} \Delta_{AjBj}^\dagger \rangle$, which, due to particle-number conservation, is equivalent to measuring $\langle (\Delta_{AiBi} + \Delta^\dagger_{AiBi}) (\Delta_{AjBj} + \Delta_{AjBj}^\dagger) \rangle$.
In the Bilayer Octagon encoding (cf. Fig.~\ref{supplement_fig_mapping_bilaer}), qubits $Ai \uparrow$ and $Bi \downarrow$ are already adjacent on an edge, and so the measurement reduces to the measurement of $\eta$-type correlations~\eqref{eq_eta_measurement}.

\newpage
\section{More details on the $\eta$-superconductivity setup\label{sec_etasupplement_details}}
The light pulse is modelled by a uniform and time-dependent electric field $\pmb{E}(s)=E(s) \pmb{u}$ on the system, with $\pmb{u}$ a fixed vector, and a vanishing magnetic field $\pmb{B}=0$. These can be expressed in terms of the potentials $\varphi$ and $\pmb{A}$ as
\begin{equation}
   \pmb{E}=-\pmb{\nabla}\varphi-\partial_s \pmb{A}\,,\qquad \pmb{B}=\pmb{\nabla}\times \pmb{A}\,.
\end{equation}
In our case we can thus choose $\varphi=0$ and a uniform vector potential
\begin{equation}
    \pmb{A}=A(s)\pmb{u}\,,\qquad A(s)\equiv-\int_0^s E(s'){\rm d}s'\,.
\end{equation}
We implement this vector potential through the Peierls substitution, which is done by replacing any hopping term from site $\pmb{x}$ to site $\pmb{x}+\pmb{v}$, namely $c^\dagger_{\pmb{x}+\pmb{v}}c_{\pmb{x}}$, by
\begin{equation}
    c^\dagger_{\pmb{x}+\pmb{v}}c_{\pmb{x}}\longrightarrow e^{i \pmb{A}(s)\cdot \pmb{v}}c^\dagger_{\pmb{x}+\pmb{v}}c_{\pmb{x}}\,.
\end{equation}
Let us now consider $U$ the operator that implements a time evolution between $s$ and $s+\tau$ for the time-dependent Hubbard model $H(s)$. Implementing a second-order Trotterization, we want to approximate it as
    \begin{equation}
\begin{aligned}
    &U\approx \exp \left(\frac{i U\tau}{2}\sum_i n_{i\uparrow}n_{i\downarrow}\right)\\
    &\times\exp \left(-i t\tau \sum_{\langle ij\rangle,\sigma}\alpha_{\langle ij\rangle}c_{i\sigma}^\dagger c_{j\sigma}\right) \exp \left(\frac{iU\tau}{2}\sum_i n_{i\uparrow}n_{i\downarrow}\right)\,,
\end{aligned}
\end{equation}
at leading order in $\tau$, with some appropriate coefficients $\alpha_{\langle ij\rangle}$. At leading order in $\tau$, we have
\begin{equation}
    \alpha_{\langle ij\rangle}=\frac{1}{\tau}\int_{s}^{s+\tau}e^{i\pmb{A}(s')\cdot \pmb{v}}{\rm d}s'\,,
\end{equation}
with $\pmb{v}$ the vector that points from $j$ to $i$, and with $s$ the time that corresponds to the beginning of the step. Namely we can decompose into hopping and current
\begin{equation}
    \begin{aligned}
        &\alpha_{\langle ij\rangle}c_{i\sigma}^\dagger c_{j\sigma}+\alpha_{\langle ji\rangle}c_{j\sigma}^\dagger c_{i\sigma}=\\
        &K_{\langle ij\rangle}(c_{i\sigma}^\dagger c_{j\sigma}+c_{j\sigma}^\dagger c_{i\sigma})+J_{\langle ij\rangle}i(c_{i\sigma}^\dagger c_{j\sigma}-c_{j\sigma}^\dagger c_{i\sigma})\,,
    \end{aligned}
\end{equation}
with
\begin{equation}
    K_{\langle ij\rangle}=\frac{1}{\tau}\int_{s}^{s+\tau}\cos(\pmb{A}(s')\cdot \pmb{v}){\rm d}s'\,,
\end{equation}
and
\begin{equation}
    J_{\langle ij\rangle}=\frac{1}{\tau}\int_{s}^{s+\tau}\sin(\pmb{A}(s')\cdot \pmb{v}){\rm d}s'\,.
\end{equation}
Let us now specify an expression for $\pmb{E}(s)$. We fix the vector $\pmb{u}$ to be $(1,0)$ in the horizontal direction, and set
\begin{equation}
    E(s)=A_0 \omega \sin(\omega s)\,,\qquad 0\leq s \leq \frac{\pi}{\omega}\,,
\end{equation}
and $E(s)=0$ for other values of $s$, with some parameters $\omega$ and $A_0$. We thus have
\begin{equation}
    A(s)=A_0(1-\cos(\omega s))\,,\qquad 0\leq s \leq \frac{\pi}{\omega}\,,
\end{equation}
and $A(s)=2A_0$ for $s> \pi/\omega$. In order to simplify measurement of observables at the end of the pulse, we impose that there is no Peierls phase in the system at the end of the pulse. This implies that $2A_0$ must be a multiple of $\pi$. We set thus
\begin{equation}
    A_0=\frac{\pi}{2}\,.
\end{equation}
We choose to implement the Trotterization of the time-dependent Hubbard model with $2$ steps of equal time. This gives the relation between $\tau$ and $\omega$
\begin{equation}
    \tau=\frac{\pi}{2\omega}\,.
\end{equation}
The coefficients $K,J$ become independent of $\tau$. When $\langle i,j\rangle$ is a vertical bond, we have
\begin{equation}
    K_{\langle ij\rangle}=1\,,\qquad J_{\langle ij\rangle}=0\,.
\end{equation}
When $\langle i,j\rangle$ is a horizontal bond, we have for the two Trotter steps
\begin{equation}
    K_{\langle ij\rangle}=\pm\frac{4}{\pi}\int_0^{\pi/4}\cos(\pi \sin^2 s){\rm d}s\approx 0.75\,,
\end{equation}
and
\begin{equation}
    J_{\langle ij\rangle}=\frac{4}{\pi}\int_0^{\pi/4}\sin(\pi \sin^2 s){\rm d}s\approx 0.47\,.
\end{equation}
Namely, the first Trotter step is done with coefficients $(K,J)\approx (0.75,0.47)$ and the second Trotter step with coefficients $(K,J)\approx (-0.75,0.47)$. This leaves in total a single parameter $\tau$ in the protocol. For a field oriented in the diagonal direction, we have $\pmb{u}=(1,1)$. In that case these expressions for $K_{\langle ij\rangle}$ and $J_{\langle ij\rangle}$ hold for all edges $\langle ij\rangle$.\\

Let us now write this hopping term in terms of the compact encoding. When the arrow of the compact encoding points from $i$ to $j$, we have
\begin{equation}
\begin{aligned}
    c_{i\sigma}^\dagger c_{j\sigma}+c_{j\sigma}^\dagger c_{i\sigma}&=\frac{X_iX_j+Y_iY_j}{2}P_a\\
    i(c_{i\sigma}^\dagger c_{j\sigma}-c_{j\sigma}^\dagger c_{i\sigma})&=\frac{X_iY_j-Y_iX_j}{2}P_a\,,
\end{aligned}
\end{equation}
where $P_a$ is the appropriate Pauli operator applied on the ancilla adjacent to the edge $\langle i,j\rangle$. Noting that
\begin{equation}
\begin{aligned}
     &e^{i\theta Z_i}\frac{X_iX_j+Y_iY_j}{2} e^{-i\theta Z_i}=\\
     &\cos(2\theta)\frac{X_iX_j+Y_iY_j}{2}+\sin(2\theta)\frac{X_iY_j-Y_iX_j}{2}\,,
\end{aligned}
\end{equation}
we can implement the hopping term as
\begin{equation}
    \begin{aligned}
        &\exp \left(-t\tau (\alpha_{\langle ij\rangle}c_{i\sigma}^\dagger c_{j\sigma}+\alpha_{\langle ji\rangle}c_{j\sigma}^\dagger c_{i\sigma})\right)=\\
        &e^{i\theta Z_i}\exp \left(-t\tau_{\rm eff} \frac{X_iX_j+Y_iY_j}{2}P_a\right)e^{-i\theta Z_i}\,,
    \end{aligned}
\end{equation}
with
\begin{equation}
    \theta=\frac{1}{2}\arctan \frac{J}{K}\,,\qquad \Delta \tau_{\rm eff}=\tau \sqrt{K^2+J^2}\,{\rm sign}(K)\,.
\end{equation}
Numerically, we have $\theta\approx 0.28$ and $\tau_{\rm eff}\approx 0.89\,{\rm sign}(K)  \tau $. We recall that the value of $\tau$ chosen in the protocol is $\tau=0.375$.

\newpage
\section{More hardware data on the imbalance benchmark}
\label{supp:moredataeta_imbalance}

\subsection{Benchmark score}
We presented in the main text a benchmark of the hardware for the simulation of fermionic systems. The benchmark relied on the case $U=0$ where there is no interaction between the fermions, for which the imbalance can be computed in polynomial time. Such benchmark protocol was defined in Ref.~\cite{granet2025appqsim}. There, was introduced the notion of distinguishability cost to evaluate the quality of the output of a quantum hardware. This distinguishability cost is defined as the minimal amount of resources that an ideal perfect quantum computer has to spent to certify that the output of the tested hardware is incorrect. This amount of resources can be measured in shots or in number of gates to implement for example. In case of just one output $m$ with standard deviation $\sigma$ obtained from the hardware, and ideal value $m_0$ with standard deviation per shot $\sigma_0$ on the perfect hardware, this distinguishability cost is defined as
\begin{equation}
    \delta=\lceil \frac{9\sigma_0^2 }{\max((m-m_0)^2,\sigma^2)}\rceil N_{\rm gates}  \,,
\end{equation}
where $N_{\rm gates}$ is the number of two-qubit gates in the circuit tested, and $\lceil \cdot \rceil$ the rounding to the nearest larger integer. This quantity $\delta$ corresponds exactly to the total number of two-qubit gates to implement (across different shots) on a perfect quantum computer to be confident by three standard deviations that the outcome of the tested hardware, whose precision is bounded by $\sigma$, is incorrect. The exact values $m_0$ and the variance per shot $\sigma_0$ can be computed exactly for this free-fermion benchmark. In case of multiple time points, the definition of the distinguishability cost is more involved and requires a $\chi^2$ test, see \cite{granet2025appqsim}. The interpretation of this score is that before running $\delta$ two-qubit gates, the outcome of the tested hardware is statistically compatible with that of a perfect hardware. The score automatically balances shot noise and hardware noise and puts them on same footing.

We display the values of distinguishability cost in Figure \ref{tab:distinguishcost}. The first column displays single-time-point cost, whereas the cumulated column for step $n$ displays the distinguishability cost of all the time points $1,...,n$ taken together. As a comparison, we perform noisy simulations of the circuits with Pauli string simulations, for the $3$-step case, with varying depolarizing noise amplitude after every two-qubit gate. The relatively fast convergence of the simulation despite the circuit depth is only made possible by the free fermionic nature of the system, from which the Pauli string simulation automatically benefits by having to keep track of a polynomial number of strings. In Fig \ref{fig:distinguishcost} we plot the distinguishability cost obtained for these simulations, imposing the same standard deviation as measured on hardware, for the raw data and the mitigated data. We see that both curves saturate at low noise rate, indicating that in this regime, the precision is shot-limited and not noise-limited. The curve with mitigated standard deviation takes lower maximal values, because the shot noise is larger. We see that the mitigated hardware data coincides with the plateau, indicating that more shots would improve the precision. It matches simulation values with noise rate $\leq 8\times 10^{-4}$. On the other hand, the score of the raw hardware data is limited by hardware noise and not by shot noise. The corresponding effective two-qubit gate noise rate is around $3.8 \times 10^{-3}$. These values cannot be compared to two-qubit gate fidelities of the ion-trap, because other sources of noise such as memory error are not taken into account in these simulations, while they contribute to degrading the signal.

\begin{figure}[h]
\begin{tabular}{ |c|c|c|c|c|} 
 \hline
  & distinguishability cost & (mitigated) &\begin{tabular}{@{}c@{}}cumulated\\ distinguishability cost\end{tabular}  & (mitigated)  \\ 
   \hline
$1$ step & \num{1238640} & \num{369408} &  \num{1238640} & \num{369408}\\ 
\hline
$2$ steps & \num{104412}& \num{297077} &   \num{136730}  &  \num{390302}  \\ 
\hline
$3$ steps & \num{16632} & \num{262416} &    \num{25872} &   \num{412104}  \\ 
\hline
$4$ steps & \num{17220}  & \num{137760} &   \num{27060} &    \num{248460} \\ 
\hline
\end{tabular}

\caption{\textbf{Values of distinguishability costs obtained for the imbalance benchmark experiments on hardware.}}
\label{tab:distinguishcost}
\end{figure}

\begin{figure}[b]
    \centering
    \includegraphics[scale=0.8]{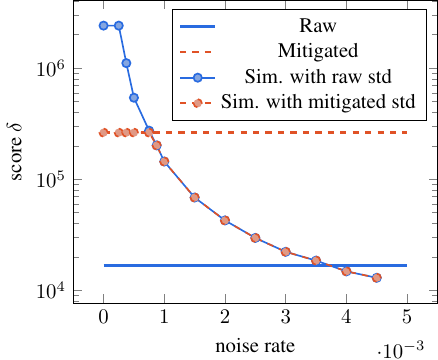}
    \caption{\textbf{Distinguishability score obtained for noisy simulation of the $3$-step imbalance setup, as a function of depolarizing noise rate after every two-qubit gate, calculated with the standard deviation of raw hardware (blue bullets) and of mitigated hardware (red bullets).} The horizontal lines indicate the score obtained from hardware data.}
    \label{fig:distinguishcost}
\end{figure}

\begin{figure}
    \centering
    \includegraphics[scale=0.82]{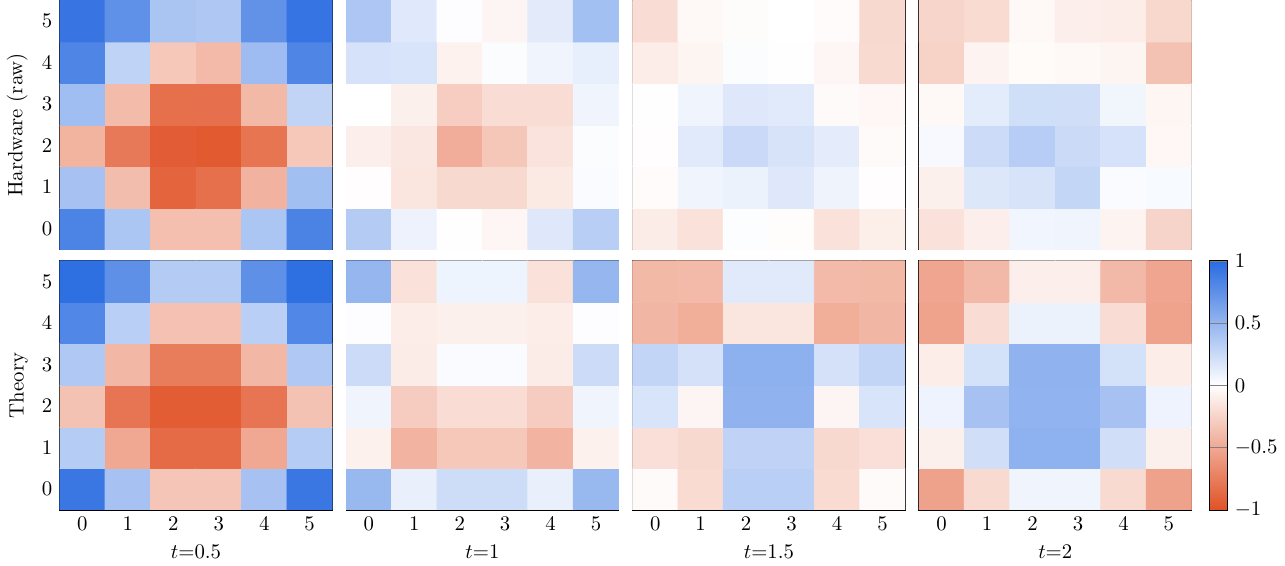}
    \caption{\textbf{Site-resolved imbalance $1-n_{j\uparrow}-n_{j\downarrow}$ as a function of lattice site for different times in the imbalance benchmark experiment, comparing raw hardware data and theory.}}
    \label{fig:imbalance_site_resolved}
\end{figure}

\subsection{Site-resolved magnetization}
In Fig \ref{fig:imbalance_site_resolved} we show the site-resolved imbalance $1-n_{j\uparrow}-n_{j\downarrow}$ (i.e., the difference between local particle density and average particle density) averaged over shots, comparing hardware results and exact theoretical values. We see that the general behaviour of the experiment - the fermions moving back and forth inside and outside of the region $\mathcal{A}$ - is correctly reproduced by the hardware. We observe a global attenuation of the signal on hardware, especially at larger times. However, we also observe some more pronounced differences locally, beyond a simple attenuation.

\subsection{Correlations}
\begin{figure}[b!]
    \centering
    \includegraphics[scale=0.82]{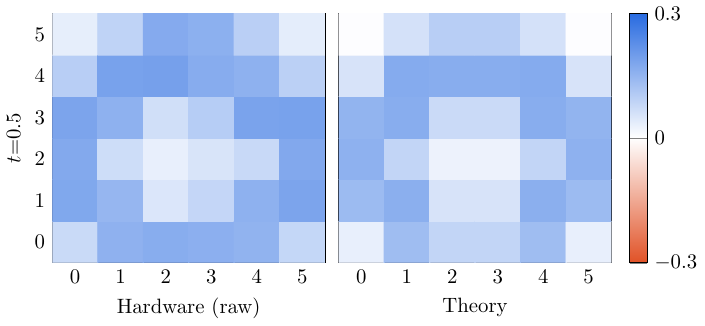}
    \caption{\textbf{Local kinetic average $K_i$ for the imbalance after one step, comparing raw hardware data and theory.}}
    \label{fig:local_kinetic}
\end{figure}

Free-fermion states satisfy Wick's theorem, which means that all correlation functions of fermions can be written as products of two-point functions. In particular, for a free-fermion state that conserves particle number, we have for two sites $i,j$
\begin{equation}\label{wick}
    \langle c_i^\dagger c_i c_j^\dagger c_j\rangle=\langle c_i^\dagger c_i\rangle \langle c_j^\dagger c_j\rangle-\langle c_i^\dagger c_j\rangle \langle c_j^\dagger c_i\rangle\,.
\end{equation}
Hence we can estimate the amplitude of the hopping term $c_i^\dagger c_j$ (but not the sign) from just measuring density expectation values
\begin{equation}
    |\langle c_i^\dagger c_j\rangle|=\frac{1}{2}\sqrt{|\langle (1-Z_i)(1-Z_j) \rangle-\langle 1-Z_i\rangle\langle 1-Z_j \rangle|}\,.
\end{equation}
These values can then be compared to their exact value. In our case, the system is equipped with mixed boundary conditions, which means a superposition of four different boundary conditions. Relation \eqref{wick} holds then separately in each of these four boundary conditions. However, after one step of the imbalance benchmark, the light-cone of the circuit has not reached the boundaries, so that the observables are identical in all four boundary conditions settings. In Fig \ref{fig:local_kinetic} we plot the following quantity
\begin{equation}
    K_{i}=\frac{1}{4}\sum_{\langle i,j\rangle}  |\langle c_i^\dagger c_j\rangle|\,,
\end{equation}
that is the average of the hopping amplitude of sites neighbouring site $i$, comparing hardware and theory, after one step. This quantity would be $0$ for a product state, and is a signature of entanglement in the system. We observe non-zero values and excellent agreement with the theory.

\newpage

\section{More hardware data on the $\eta$-superconductivity setup \label{supp:moredataeta}}
In this section, we give additional hardware data for the $\eta$-superconductivity setup that was not presented in the main text. 

\subsection{Doublon density}
We start by giving the values of doublon density in the different experimental setups. Because of the measurement gadget we do to measure the $\eta$-pairing correlations, we can only measure $c^\dagger_{i\uparrow}c^\dagger_{i\downarrow}c_{i\downarrow}c_{i\uparrow}+c_{i\downarrow}c_{i\uparrow}c^\dagger_{i\uparrow}c^\dagger_{i\downarrow}$, which gives only access to the site-resolved density of doublon + holes, not just to doublon density. However, assuming particle number conservation, we can deduce from it the total number of doublons in the system. At half-filling, hardware noise is unlikely to modify the expectation value of particle number because the value already corresponds to infinite temperature. We show the values obtained in Table \ref{tab:doublondensity}. We recall that infinite temperature has doublon density $0.25$. We observe that the doublon density is low before pulse, which is expected from the low-energy character of the state and the on-site Coulomb repulsion. After the pulse, the doublon density is significantly enhanced, reaching values larger than infinite temperature. After relaxation, the doublon density goes to values close to $0.25$.

\begin{table}[h]
\begin{tabular}{ |c|c|c|} 
 \hline
 experiment & average doublon density & mitigated \\ 
   \hline
before pulse & $0.0475 \pm 0.0015$ & $0.0378 \pm 0.0032$ \\ 
\hline
after pulse & $0.279 \pm 0.0024$ & $0.292\pm 0.0081$ \\ 
\hline
$+1$ extra step & $0.256 \pm 0.0031$ &  $0.252 \pm 0.012$  \\ 
\hline
$+2$ extra steps &  $0.259 \pm 0.0032$ &  $0.261 \pm 0.014$ \\ 
\hline
\end{tabular}
\caption{\textbf{Average doublon density in different experimental setups.}}
\label{tab:doublondensity}
\end{table}

\subsection{Mitigated staggered $\eta$-correlations}
\begin{figure}[b!]
    \centering
    \includegraphics[scale=0.9]{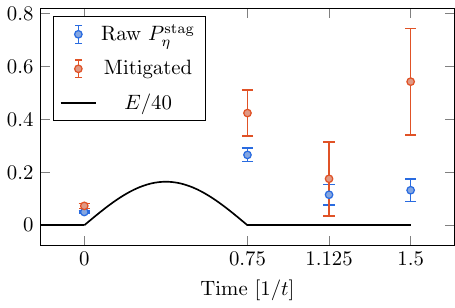}
    \caption{\textbf{Data for the staggered $\eta$-correlations from Fig.~\ref{fig1}g, shown together with the results of error mitigation.}}
    \label{supplement_fig_eta_stag_mitigated}
\end{figure}

We show in Fig.~\ref{supplement_fig_eta_stag_mitigated} the mitigated values of the staggered $\eta$-correlations shown in Fig.~\ref{fig1}g in the main text. We observe significant amplification of the error bars by the mitigation, especially for the larger time points $t=3\times 0.375$ and $t=4\times 0.375$. This is expected for particularly deep circuits where the stabiliser outcome distribution is significantly impacted by hardware noise, as we see in Fig \ref{fig:stabilizers}. We also observe a systematic increase of the expectation value after the mitigation.

\subsection{Two extra steps and horizontal field \label{sec:horizontal}}
Next, we show in Fig \ref{fig:horizontal_2steps}  the $\eta$-pairing correlations (i) after $2$ extra Trotter steps are performed after the light pulse, without magnetic field (in the main text, only $1$ extra Trotter step is shown) and (ii) after the light pulse, but when the electromagnetic field is in the horizontal direction instead of the diagonal direction. 

In case (i), we observe similar quasi-alternating pattern of blue and red squares as for only $1$ extra Trotter step. We observe a particularly strong signal for the farthest square, at distance $(3,3)$. 

In case (ii), we see that the $\eta$-pairing is enhanced only in the direction of the electromagnetic field. This shows that the $\eta$-pairing amplification is not just due to time evolution under the $H$, but requires this light pulse to occur. The amplification observed is significantly larger than for the diagonal field case, by a factor around $2$. Moreover, we see some intriguing horizontal "stripe" features, albeit with small amplitude. It would be interesting to investigate further these features.

\subsection{Distance-resolved staggered average}
\begin{figure}[t!]
    \centering
    \includegraphics[scale=0.79]{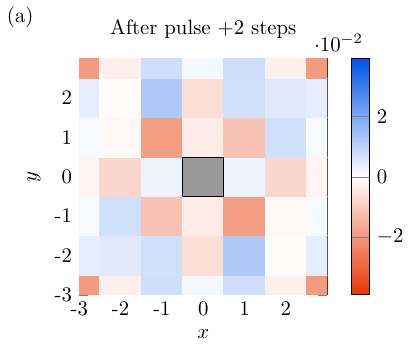}
    \includegraphics[scale=0.79]{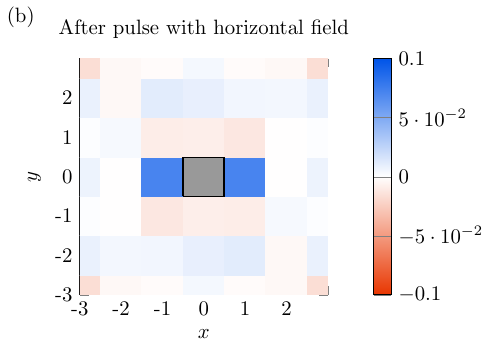}
    \includegraphics[scale=0.79]{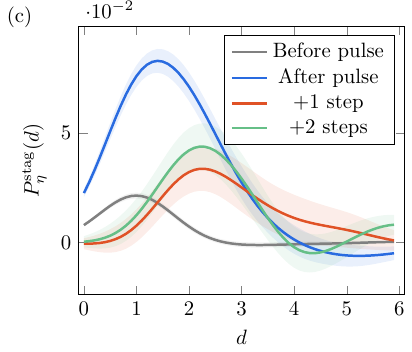}\\
    \caption{\textbf{Additional hardware data on the light-pulse experiment.} (a): $\eta$-pairing correlations, after $2$ Trotter steps following the light pulse, and (b): right after the light pulse when the field is in the horizontal direction. (c): Distance-resolved staggered average $P_{\eta}^{\rm stag}(d)$ in the diagonal field case, with the shades indicating one error bar.}
    \label{fig:horizontal_2steps}
\end{figure}

In the main text, we presented the value of the staggered average $P_{\eta}^{\rm stag}=\sum_{x,y\neq 0,0}(-1)^{\rm sublattice} P_{\eta}(x,y)$, before pulse, after pulse, and after relaxation, which was taking non-zero values. In the $\eta$-pairing plots, it is clear that this staggered $\eta$ pairing is short-ranged right after the pulse, but looks more spread over the entire lattice after relaxation. To probe this effect, we define the distance-resolved staggered average
\begin{equation}
    P_{\eta}^{\rm stag}(d)=\sum_{x,y\neq 0,0}(-1)^{\rm sublattice} P_{\eta}(x,y) \frac{e^{-(d-(|x|+|y|))^2/(2\sigma)}}{\sqrt{2\pi}\sigma}\,,
\end{equation}
with $\sigma=1$. This filters the sum to only terms located at distance $d$ in Manhattan distance. We present the results in Fig \ref{fig:horizontal_2steps}c. We observe that as expected, the staggered $\eta$-pairing is strongly short-ranged right after the pulse, whereas after relaxation it vanishes at $d=0$, peaks at $d=2.3$ and has non-zero tails at larger distance. This increase of long-distance staggered $\eta$-pairing correlations after relaxation was also observed in a one-dimensional setting \cite{kaneko_photoinduced_2019}.

\subsection{Spin clusters \label{sec:spincluster}}
We now present hardware results about shot-based observables that cannot be written in terms of the expectation value of local observables written in terms of a small number of Pauli strings. These observables should be challenging to compute with any Heisenberg-picture-based classical simulation method. 

\begin{figure}[b!]
    \centering
    \includegraphics[scale=0.7]{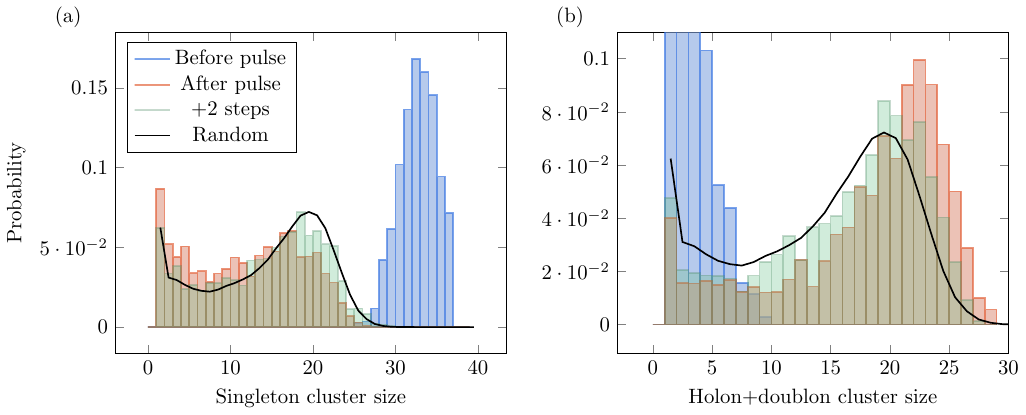}
\includegraphics[scale=0.8]{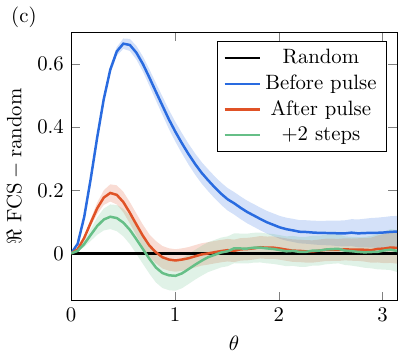}\\
    \caption{\textbf{Hardware data for shot-based statistics and observables.} (a) and (b): Histograms of cluster size of singletons for (a) and holon+doublon for (b), in different experiments. (c): real part of ${\rm FCS}(\theta)$ for the observable \eqref{oalternating} as a function of $\theta$, minus the FCS obtained for a random output.}
    \label{fig:spincluster}
\end{figure}

In the $\eta$-superconductivity setup, we measured $X_{j\uparrow}X_{j\downarrow}$ and $Y_{j\uparrow}Y_{j\downarrow}$ on all sites $j$. This gives access to $Z_{j\uparrow}Z_{j\downarrow}$, namely the parity of the number of particles per site. We can thus know whether each site is occupied by a singleton (namely, one single particle of either spin) or by a holon or doublon (without being able to distinguish holon from doublon). On each shot, we identify the clusters of singletons, namely sets of sites occupied by a singleton and that are connected sets (allowing for connection through the periodic boundaries). We then collect the statistics of the size of these clusters across all shots. For a fair comparison with a random state, we assign a random value of $Z_{j\uparrow}Z_{j\downarrow}$ to sites $j$ where one of the two qubits have leaked. To remove the randomness associated to these assignments, we average the resulting histograms over a large number of realizations. We present the results in Fig \ref{fig:spincluster}. We plot histograms of the probability that a singleton sampled at random belongs to a cluster of size $n$ (this enhances thus the probability of large clusters compared to the bare cluster size distribution). We observe that before the pulse, the typical size of the singleton cluster is very large $\gtrapprox 30$, and conversely that the typical size of holon+doublon clusters is very small. This is explained by the fact that at low energy, the double occupation number is small. At half-filling, most of the sites are thus occupied by singletons, and the typical cluster size is therefore large. In contrast, after the pulse, the histograms take a shape that is closer to what is obtained for a random state. However, the histograms still show significant difference with the random state. Small clusters of holon+doublon are clearly less probable, and large clusters are significantly enhanced compared to a random state. After $2$ extra Trotter steps of relaxation, the histograms appear to be approximately half-way between the random state and the state after the pulse.

\subsection{Full-counting statistics}
Full-counting statistics (FCS) of an observable $\mathcal{O}$ refers to the expectation value of the moment generating function
\begin{equation}
    {\rm FCS}(\theta)=\langle e^{i\theta \mathcal{O}}\rangle\,.
\end{equation}
This function gives further insights beyond the mean value of the observable $\mathcal{O}$, since for example its higher-order moments can be obtained by differentiation with respect to $\theta$. From the shots obtained from hardware, this quantity is straightforward to compute. However, similarly to information about clusters in section \ref{sec:spincluster}, it would be difficult to compute for Heisenberg-picture-based classical simulation techniques, because it involves a large number of long Pauli strings. We consider the observable
\begin{equation}\label{oalternating}
    \mathcal{O}=\sum_j (-1)^{\rm sublattice}(c_{j\uparrow} c_{j\downarrow}+c_{j\downarrow}^\dagger c_{j\uparrow}^\dagger)\,.
\end{equation}
Similarly to the spin cluster observables, we will compare the ${\rm FCS}$ measured on hardware to what is obtained from a random state where all bits are $0$ or $1$ with probability $1/2$. We will again for fairness of the comparison interpret the leaked qubits L on hardware data as  $0$ or $1$ with probability $1/2$. To remove the randomness introduced by this assignment, we then average the ${\rm FCS}$ obtained over $1000$ realizations. We show the hardware results in Fig \ref{fig:spincluster}c. We see that the three curves before pulse, after pulse and at the end of the relaxation all display statistically significant differences with the FCS obtained for random shots. The light pulse strongly modifies the FCS. In contrast, the relaxation has little effect apart from an attenuation of the signal. The real part of the FCS obtained at small $\theta$ is larger than the FCS for a random output, indicating that the variance of the observable $\mathcal{O}$ is smaller in the hardware than for random outputs.

\newpage
\section{Classical Difficulty of Simulating the Circuits \label{sec:classicaldifficulty}}

Since quantum computer time remains a valuable resource, it is prudent to attempt to classically simulate circuits both for logical debugging and to set experimental expectations. While we have succeeded to simulate some of the circuits discussed in the main text (and in fact used these circuits to benchmark the quantum hardware), other circuits proved exceedingly difficult, especially in the light-induced $\eta$-pairing setup (Fig.~\ref{fig1}g). In this section we document our classical simulation efforts and estimate what classical resources would be needed for reliable simulation of the $\eta$-pairing circuits.

\subsection{Matrix-Product States}

One powerful method to simulate quantum circuits is to find high-fidelity matrix-product state (MPS) approximations~\cite{ayral_density-matrix_2023}. In particular, in a recent quantum computer study of real-time dynamics in the transverse-field Ising model, the circuit-DRMG method was found to be among the most reliable classical circuit simulation methods~\cite{haghshenas_digital_2025}. Specifically, an $8 \times 7$ periodic square lattice of qubits was initialised in a product state $    \left( \cos(\theta/2) \ket{0} + \sin(\theta/2) \ket{1} \right)^{\otimes 56}$, where $\theta = \pi/18$. Subsequently
$s$ second-order Trotter steps $U = e^{-i h \tau/2 \sum_i X_i} e^{-i J \tau \sum_{\langle ij \rangle} Z_i Z_j} e^{-i h \tau/2 \sum_i X_i}$ with $J=1$, $h=2$, $\tau=0.25$ were applied, and the total magnetisation $\left( \sum_i Z_i \right)^2$ measured.

On these circuits, Fermioiniq's Ava emulator~\cite{noauthor_ava_nodate} was used to obtain reliable data for circuits up to $s=8$ Trotter steps (depth 32). The emulator automatically partitions the input circuit into $K$ subcircuits $U_1, U_2, \dots U_K$, and attempts to sequentially find (normalised) matrix-product states $\ket{\phi_0}, \ket{\phi_1}, \dots, \ket{\phi_K}$ by maximising the $k$-th partial fidelity $f_k = |\langle \phi_{k} | U_k |\phi_{k-1} \rangle|^2$, subject to the constraint that the bond dimension of each $\ket{\phi_k}$ never exceeds an input value $\chi$. For a given value of $\chi$ one may compute both the observable of interest $O$, as well as the simulation fidelity
\begin{equation}
    F = \prod_{k=1}^K f_k.
    \label{eq:fidelity}
\end{equation}
Linearly extrapolating $\langle O(F) \rangle$ using the data points with the largest simulation fidelity was found to yield the most accurate estimators (more accurate than e.g. extrapolating in $1/\chi$). An interesting feature of these extrapolations is their dependence on the maximum simulation fidelity $F$ used for extrapolation: For maximum simulation fidelities of order 1, reliable extrapolations could be carried out, whereas for $F \ll 50 \%$, one enters a chaotic regime in which extrapolations yield essentially random results, independent of system size. One consequence of these findings is that, while the method works extremely well for low-entanglement quenches, obtaining reliable results for the deepest circuits in that experiment ($s=20$, two-qubit depth 80) would likely require unpractically large bond dimensions $\chi > 2^{16}= 65536$.

\begin{figure*}[t]
    \centering
\includegraphics[width=\linewidth]{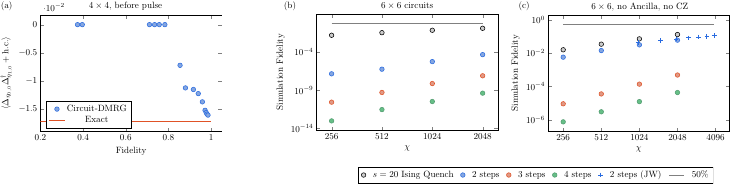}
\caption{\textbf{Circuit Simulation with Fermioniq's Ava circuit-DMRG emulator on the light-matter interaction circuits in Fig.~\ref{fig1}g.}  (a) Verification on the exactly solvable $4 \times 4$ case, after application of the 2-Trotter step light pulse, before any further relaxation. The exact nearest-neighbour $\Delta_\eta$ correlation is obtained by translating the circuit to a 32-qubit Jordan-Wigner circuit and computing the exact statevector result. Compatible with the findings in~\cite{haghshenas_digital_2025}, (extrapolated) results from circuit-DMRG are reliable when a simulation fidelity $F$ of order 1 can be achieved, and are essentially random for $F \ll 50\%$. (b) Simulation fidelities on the $6 \times 6$ circuits on 90 qubits, right after the light pulse (2 steps) and after up to two Hubbard-Trotter relaxation steps (3 and 4 steps), compared with the "intermediate temperature" ($\Delta \theta = 2\pi/9$) quench circuit with $s=20$ Ising-Trotter steps in~\cite{haghshenas_digital_2025}. (c) Circuit fidelities on 72-qubit circuits that are obtained by deleting all gates on the ancilla qubits, as well as any maximum angle C$Z$ gates, which are responsible for implementing the fermion-to-qubit encoding. The result is a lower bound on the cost of natively fermionic circuit-DMRG.
\label{supplement_fig_fermioniq}}
\end{figure*}

How do the circuits in the present work compare to those Ising circuits in terms of difficulty for the circuit-DMRG method? To address this, we apply Fermioniq Ava to the light-matter interaction circuits shown in Fig.~\ref{fig1}g. First, we focus on the situation right after the light pulse has been applied to the  $(D_\mathrm{Heisenberg}, D_\mathrm{Hubbard}) = (1,1)$ initial state (2 steps), and additionally restrict the system size to a $4 \times 4$ system. Nearest-neighbour $\eta$-correlations can be computed exactly in this case, and compared with circuit-DMRG outputs at various simulation fidelities (Fig.~\ref{supplement_fig_fermioniq}a). The data indicates that, similar to the Ising quench, circuit-DRMG requires fidelities  not much smaller than $50\%$ in order to obtain reliable (extrapolated) observables on the light-matter interaction circuit.

The key question is then: how difficult, in terms of bond dimension is it to achieve order-one fidelities using the circuit-DMRG method, for the larger $6 \times 6$ system? To answer this question, we show simulation fidelity as a function of bond dimension in Fig.~\ref{supplement_fig_fermioniq}b. Even the shallowest (2 steps) circuit achieves a fidelity roughly 3000 times lower that of the hardest Ising quench considered in~\cite{haghshenas_digital_2025} at bond dimension $\chi=2048$.

The extremely small fidelity achieved for these circuits is not very surprising: A significant fraction of two-qubit gates are Clifford gates, most of which are used to couple the system qubits to the ancilla qubits during the fermionic hopping gadget. While these are necessary to enforce the fermion-to-qubit mapping, classical methods can achieve the correct fermionic statistics in other ways.
For example, one can associate the physical degree of freedom of each MPS tensor to a fermionic mode and enforce the fermionic parity by introducing a $Z_2$ symmetry~\cite{cirac2021matrix}.
To lower bound the cost of such an approach, we have also considered circuits in which all ancilla qubits (and gates acting on them) are removed, as well as all other gates that are responsible for the fermion-to-qubit encoding. The resulting simulation fidelities, shown in Fig.~\ref{supplement_fig_fermioniq}c, are much larger than those where the ancilla degrees of freedom are kept track of explicitly. In particular, the shallowest (2 steps) circuit now becomes comparable in difficulty to the $s=20$ Ising quench discussed above. As a result, we expect that a fermionic circuit-DMRG method utilizing additional symmetries, for example $SU(2) \times U(1)$ or $U(1) \times U(1)$ over spin and number of particles, may be able to capture the nearest-neighbour correlations accurately~\cite{liu2025accurate}.
The situation on the deeper circuits (3 steps and 4 steps), where after the light pulse additional Hubbard-Trotter steps are applied to relax the state, remain out of reach.

In addition to the circuit-DMRG results obtained from the FermioniQ Ava emulator, we have also run some simulations of the circuit written with the Jordan-Wigner (JW) encoding, alleviating the difficulty of simulating the compact encoding, at the price of adding long range JW strings. The JW string order is chosen to match the ordering of MPS tensors, making this approach similar to the fermionic MPS approach without having to utilize an extra $Z_2$ symmetry. These MPS simulations were performed using a slightly simpler approach where quantum gates are contracted into the MPS one at a time. The algorithm supports native application of two-qubit gates between non-adjacent qubits, as well as multi-qubit gates of the form $\exp(i \theta P)$ for arbitrary Pauli strings. These are applied by first obtaining their exact representation as an MPO of bond dimension 2, which is then contracted into the MPS following the zip-up algorithm from~\cite{Stoudenmire_2010}, with one caveat: the zip-up algorithm is carried out using QR decompositions, rather than SVD. After the gate has been contracted into the MPS, the latter is put in canonical form with the center at the leftmost bond affected by the gate, followed by SVD truncation down to the target bond dimension; this is repeated on all affected bonds from left to right. As discussed in~\cite{Zhou2020}, the sum of the squared singular values we keep determines the fidelity of each single truncation, and the product of all of these is a standard estimate of the fidelity of the final state. The algorithm runs on GPUs, implemented using NVIDIA cuTensorNet and is publicly available at~\cite{PabloAndresCQ} under the name MPSxGate.
The results can be seen (blue crosses) in Fig.~\ref{supplement_fig_fermioniq}c. Interestingly, the fidelity match closely the circuit simulations where all ancilla have been removed.
This indicates the naive estimate by removing gates related to ancilla gives a reasonable estimate of the complexity of simulating with fermionic MPS.

\subsection{Sparse Pauli decomposition \label{sec:psd}}
\subsubsection{Setup}
We now study the simulability of the circuits with Sparse Pauli Decomposition (SPD). This is a promising simulation technique that has been attracting attention recently~\cite{beguvsic2025real,broers2025scalable,lin2025utility,gharibyan2025practical,rudolph2025pauli} and which we have found most practical for the development and testing of quantum circuits in this work. Indeed, it displays a fast and precise convergence for observables with few enough gates in their light-cone for an arbitrary number of qubits (and in general, for circuits with finite magic~\cite{dowling2025magic,dowling2025bridging}). This simulation technique works in the Heisenberg picture, and consists in expanding the observable to be measured into strings of Pauli matrices
\begin{equation}\label{on}
    \mathcal{O}=\sum_{P_n \in \{I,X,Y,Z\}^{\otimes N}}c_n P_n\,,
\end{equation}
where the sum runs over all the $4^N$ possible Pauli strings $P_n$ that are tensor products of Pauli matrices $I,X,Y,Z$ at each site, and with $c_n$ real coefficients. Then, one computes the application of each gate $g$ of the circuit on this observable, starting from the last gate of the circuit, and updating the list of coefficients $c_n$. When the gates are written as exponentials of Pauli strings $Q$, this is done through the equation
\begin{equation}\label{eipq}
    e^{i\theta Q}P e^{-i\theta Q}=\begin{cases}
        P\,,\quad \text{if $P$ and $Q$ commute}\\
        \cos(2\theta)P+i\sin(2\theta)QP\,,\quad \text{if $P$ and $Q$ anticommute}\,.
    \end{cases}
\end{equation}
The observable after $m$ gates is thus decomposed as
\begin{equation}\label{onm}
    \mathcal{O}(m)=\sum_{n}c_n(m) P_n\,,
\end{equation}
with coefficients $c_n(m)$ that are expressed in terms of $c_{n'}(m-1)$. At the end of the circuit with $M$ gates, one obtains a list of coefficients $c_n(M)$, and the expectation value of the observable in the circuit is obtained as
\begin{equation}
    \langle \mathcal{O} \rangle=\sum_{P_n \in \{I,Z\}^{\otimes N}}c_n(M)\,,
\end{equation}
where the sum runs only over Pauli strings $P_n$ that contain only $I$ or $Z$ Pauli matrices.

A generic operator to be decomposed as in~\eqref{on} would require $4^N$ terms, which becomes intractable for $N\gtrapprox 20$. Starting from a local observable that involves only a few Pauli matrices, we can keep perform the simulation by storing only the non-zero terms in the decomposition~\eqref{onm}.
However, the number of gates to keep track of grows exponentially (generically, is multiplied by $2$) after every gate application. The exact bookkeeping of all the non-zero terms ends up thus being unfeasible. As a result, different approximations can be made to get a certain precision on the result. The approximation that we will consider consist in systematically truncating, after every application of gates, all the terms $c_n$ that are below some threshold $\epsilon$ in absolute value. Namely, one retains in the sum~\eqref{onm} only the terms $c_n$ such that $|c_n|\geq \epsilon$.
The complexity of the simulation is related to the non-stabilizerness of the operator, which can be quantified by the operator stabilizer entropy~\cite{dowling2025magic}. 
This simulation technique is particularly appealing for finite-magic circuits and for code testing, as it is not limited by the number of qubits and the circuit geometry. This truncation scheme has been used in the largest-scale SPD numerical simulation to date~\cite{broers2025scalable}.

In practice, the quality of the approximation can be measured by computing the norm squared of the time-evolved observable
\begin{equation}
    ||\mathcal{O}(m)||^2=\sum_n c_n(m)^2\,.
\end{equation}
Since the evolution is unitary, in absence of approximations, the norm should be conserved $||\mathcal{O}(m)||^2=||\mathcal{O}(0)||^2$. Because of the truncation $\epsilon>0$, the computed norm $||\mathcal{O}(m)||^2$ will differ from the initial value $||\mathcal{O}(m)||^2<||\mathcal{O}(0)||^2$ for $m$ large enough.

In this resource estimation study, we will first focus on the memory requirement, namely on the number of Pauli strings required in the SPD to obtain a certain precision on the observable. We will then discuss an estimation of the actual runtime based on comparison with previous works. We note that Majorana fermion decomposition (see Ref. \cite{miller2025simulation}) instead of Pauli string decompositions are not expected to give an advantage in terms of memory, because within the fermionic encoding there should be a one-to-one correspondence between Pauli strings and Majorana strings.

We will consider the following circuits. We will designate by "$6\times 6$, $j$ steps" for $j=0,1,2,3,4$ the exact circuits that we ran on hardware (except for $j=1$ that we have not run), on a $6\times 6$ lattice, namely $90$ qubits. $j=0$ step designates the circuit with only the preparation of the $(1,1)$ state. $j=1$ step designates the circuit with just one Trotter step of the light pulse on top of the preparation of the $(1,1)$ state. $j=2$ steps means the circuit with the full light pulse on top of the preparation of the $(1,1)$ state. $j=3$ steps means one extra Trotter step on top of the light pulse with no electric field, and $j=4$ steps means two extra Trotter steps on top of the light pulse with no electric field. We will designate by "$4\times 4$, $j$ steps" the same circuits ran on a $4\times 4$ lattice composed of $40$ qubits. This number of qubits is already challenging to simulate classically with state-vector simulations. However, by using the Jordan-Wigner (JW) encoding instead of the compact encoding used in this paper, one requires only $32$ qubits, which can be routinely  simulated classically on dedicated machines. This smaller system size will thus be used to compare the exact expectation value to the one obtained with SPD. For ease of comparison with the JW simulations, we will impose periodic boundary conditions on the $4\times 4$ lattice only, instead of the mixed boundary conditions. On these circuits, we will measure the $\eta$-pairing correlation functions between site $(0,0)$ and site $(x,y)$, that after the measurement gadget implemented in the circuit take the form
\begin{equation}
    \Delta_\eta(x,y)=\frac{(-1)^{x+y}}{4}(Z_{0}-Z_L)(Z_{x+yL_x}-Z_{x+yL_x+L})
\end{equation}
for different values of $x,y=0,...,L_{x,y}-1$.

\subsubsection{Relation between truncation, norm and number of Pauli strings}
\begin{figure}
    \centering
    \includegraphics[width=0.32\linewidth]{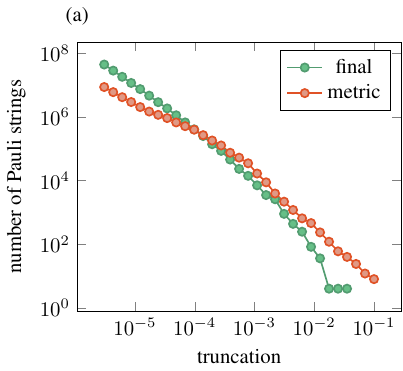}
    \includegraphics[width=0.32\linewidth]{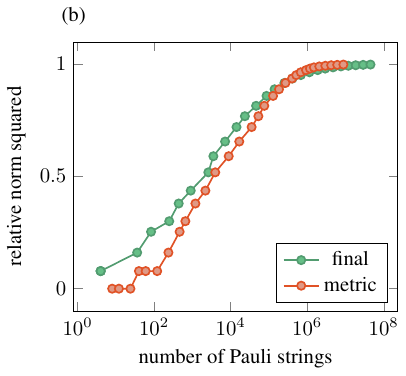}
    \includegraphics[width=0.32\linewidth]{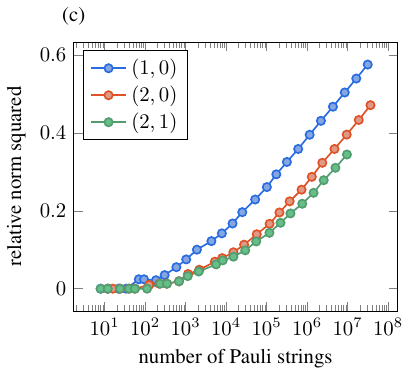}
    \caption{\textbf{Relation between truncation error, number of Pauli strings and norm.} (a): number of Pauli strings at the end of the circuit $\mathcal{S}(M)$ and maximal value before injection $\mathcal{S}_{\rm metric}$ as a function of truncation $\epsilon$, for the $4\times 4$ circuit with $1$ step when measuring $\Delta_\eta(1,1)$. (b): relative norm squared $||\mathcal{O}(m)||^2/||\mathcal{O}(0)||^2$ as a function of number of Pauli strings in the same setting. (c): relative norm squared as a function of number of Pauli strings $\mathcal{S}_{\rm metric}$ for three different correlations $\Delta_\eta(x,y)$.}
    \label{psdfig:1}
\end{figure}
Firstly, we investigate the relation between truncation $\epsilon$, norm $||\mathcal{O}(m)||^2$ and number of Pauli strings $\mathcal{S}$ to take into account in the SPD. As a function of the gate number $m$ in the circuit, the number of Pauli strings to keep $\mathcal{S}(m)$ generally grows with $m$. However, it can happen that this number decreases with $m$, especially towards the end of the circuit when the coefficients are of order $\gtrapprox \epsilon$, because after application of \eqref{eipq} it can be that the coefficients in front of both $P$ and $QP$ are below $\epsilon$ and so get truncated away. For this reason, the right indicator of the amount of memory needed to perform the computation is not the number of strings at the end $\mathcal{S}(M)$, but the maximum number of strings reached during the computation
\begin{equation}
    \mathcal{S}_{\rm max}=\underset{0\leq m \leq M}{\max}\mathcal{S}(m)\,.
\end{equation}
This is a rather standard metric for estimating the difficulty of the simulation, that has been employed before \cite{gharibyan2025practical}. However, in our case, the circuit has a particular structure. Before the "injection", the gates at the beginning of the circuit only act on the first $L=36$ qubits, for $90$ qubits in total. This means that in the Heisenberg picture, after the injection is passed, one can keep in the SPD only the strings that have either $I$ or $Z$ Pauli matrices on all the remaining $54$ qubits. This significantly reduces the number of strings to keep track of, but this also modifies the norm, because this projection is non-unitary. Since we would like to know the norm at the end of the circuit, we do not perform this projection. Therefore, for a fairer resource estimate, we will consider the maximal number of strings only \emph{before} (in the Heisenberg picture) the injection, namely
\begin{equation}
    \mathcal{S}_{\rm metric}=\underset{0\leq m \leq M_{\rm inj}}{\max}\mathcal{S}(m)\,,
\end{equation}
with $M_{\rm inj}$ the number of gates before the injection (in the Heisenberg picture). We note that since $\mathcal{S}_{\rm metric}\leq \mathcal{S}_{\rm max}$, taking the standard metric $ \mathcal{S}_{\rm max}$ instead of this metric $\mathcal{S}_{\rm metric}$ would only render the computations more difficult. We make this choice to anticipate and circumvent a simple and significant resource sparing.

Firstly, in Fig \ref{psdfig:1}a, we show the number of strings as a function of truncation parameter $\epsilon$, for the circuit on $4\times 4$ with $1$ step, when evaluating the correlation between sites $0,0$ and $1,1$. On the log-log plot, we observe a roughly linear behaviour of the number of strings at the end of the circuit $\mathcal{S}(M)$ and for the maximal value before injection $\mathcal{S}_{\rm metric}$, as a function of truncation $\epsilon$. This is in agreement with previous observations \cite{gharibyan2025practical}. We note that we do not necessarily have $\mathcal{S}_{\rm metric}\geq \mathcal{S}(M)$ because the maximum is only taken before the injection. Secondly, in Fig \ref{psdfig:1}b, we show the relative norm squared (namely $||\mathcal{O}(m)||^2/||\mathcal{O}(0)||^2$) as a function of $\mathcal{S}(M)$ and $\mathcal{S}_{\rm metric}$. We observe a generic "S curve" that is quadratic at the beginning, displays a linear behaviour for most of the values of Pauli strings, and then eventually curves down when approaching relative norm $1$. Thirdly, in Fig \ref{psdfig:1}c, we compare for a same circuit $4\times 4$ with $2$ steps, the cost in computing different correlations. We observe that the nearest correlation $(1,0)$ is the cheapest to compute, and obtaining a same relative norm as $(1,0)$ for $(2,0)$ or $(2,1)$ incurs a factor $10$ to $100$ in terms of number of Pauli strings.

\subsubsection{Relation between norm and precision of observables}
\begin{figure}
    \centering
    \includegraphics[scale=0.75]{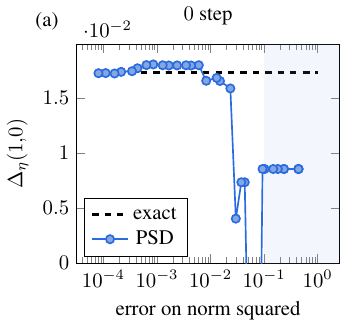}
    \includegraphics[scale=0.75]{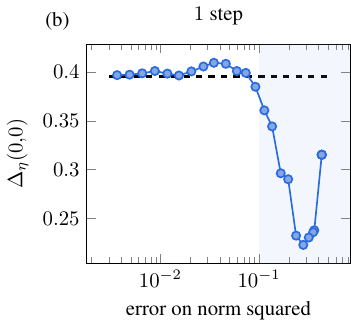}
    \includegraphics[scale=0.75]{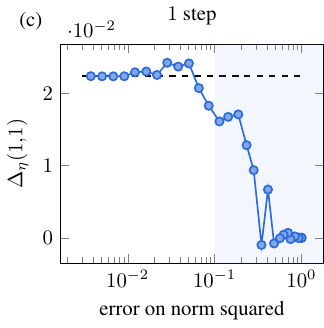}
    \includegraphics[scale=0.75]{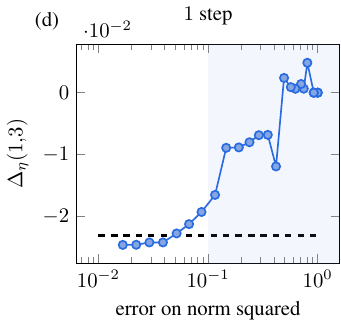}
    \includegraphics[scale=0.75]{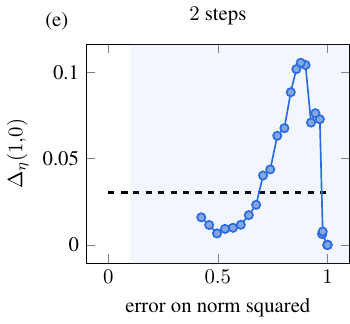}
    \includegraphics[scale=0.75]{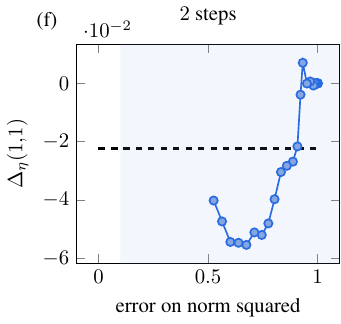}
    \includegraphics[scale=0.75]{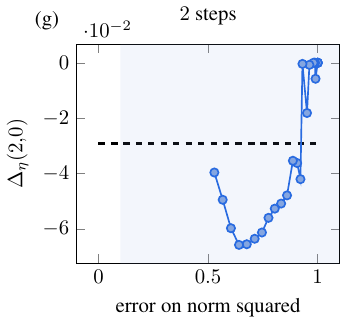}
    \includegraphics[scale=0.75]{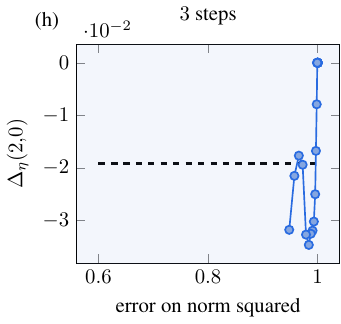}
    \caption{\textbf{Efficiency of SPD in $4\times 4$.} Expectation value of $\Delta_\eta(x,y)$ for the $4\times 4$ circuits, for different $x,y$ and different number of steps, as a function of the relative error on norm squared $1-||\mathcal{O}(m)||^2/||\mathcal{O}(0)||^2$, showing SPD (blue) and exact value (dashed black). The shaded blue region indicates that the state contains less than $90\%$ of the norm squared.}
    \label{psdfig:2}
\end{figure}
We now evaluate the relation between error on the norm and error on the expectation value of an observable. To that end, we consider the $4\times 4$ circuits that can be simulated classically with state-vector simulations when written in terms of the JW transformation, and compare several different correlations $\Delta_\eta(x,y)$ computed with SPD and with JW. We present the numerical results in Fig \ref{psdfig:2}. We observe that having a relative error on the norm squared of around say $\sim 10\%$, namely getting $90\%$ of the norm squared of the operator, is required to have an expectation value close to the exact value and seemingly starting to converge. In particular, for a norm squared smaller than $50\%$ of what it should be, we observe significant variations, preventing from performing a fit with the error on norm squared, and providing little confidence on the convergence and accuracy of the estimate obtained with SPD.

\subsubsection{Memory estimate for the circuits run on hardware}
\begin{figure}
    \centering
    \includegraphics[scale=1]{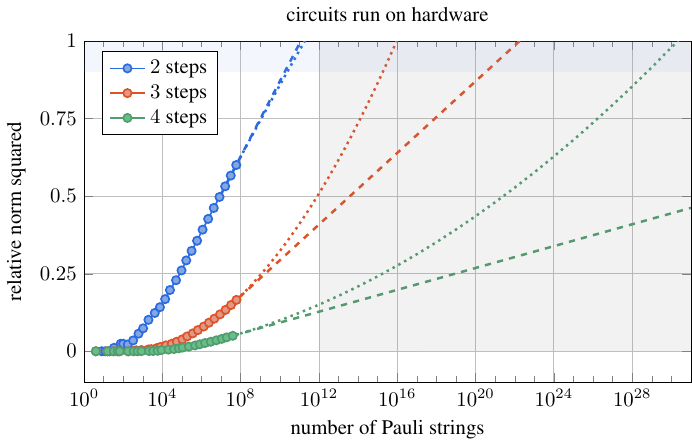}\\
    \caption{\textbf{Relative norm squared as a function of maximal number of strings before injection $\mathcal{S}_{\rm metric}$}. Data shown is for the circuits run on hardware on a $6\times 6$ lattice, performing $2$ Trotter steps after state preparation (corresponding to the light pulse), $3$ Trotter steps (one extra step on top of the light pulse), and $4$ Trotter steps (two extra steps). The dashed lines indicate quadratic extrapolations using the last $8$ points, and dotted lines linear extrapolations using the last $3$ points. The gray shaded area indicates the region that is beyond the largest SPD simulation that has been performed before the present work on a supercomputer \cite{broers2025scalable}. The blue shaded area indicates the region where $90\%$ of the norm is obtained.}
    \label{psdfig:3}
\end{figure}
We finally attempt at simulating the circuits run on hardware, corresponding to the light pulse (denoted by "$2$ steps" here), and to $1,2$ Trotter steps on top of the light pulse (denoted by "$3,4$ steps" here). The results are presented in Fig \ref{psdfig:3}. We show the relative norm squared obtained as a function of the number of Pauli strings for the easiest correlation $\Delta_\eta(1,0)$, for $j=2,3,4$ steps. We perform a linear extrapolation with the last $3$ points of each curve (denoted by dashed lines), and a quadratic extrapolation with the last $8$ points of each curve (denoted by dotted lines). Given the general "S shape" of these curves, the quadratic extrapolation necessarily overestimates the relative norm squared. The linear fit overestimates it if the linear regime is not attained yet, but underestimates it otherwise, because of the eventual curving down of the S shape. A realistic compromise can probably be found between the two dashed and dotted curves.

For $j=2$ steps, corresponding to running the light pulse, we observe that with $10^{11}$ Pauli strings, one should be able to capture a large enough amount of the norm to get a good estimate of the expectation value with good confidence. We saw that longer correlations will incur factors $10-100$ on top of this estimate. We recall that the largest SPD simulation to this date has been done with $10^{12}$ strings on the Fugaku supercomputer with $100,000$ CPUs. We can thus conclude from this numerics that multiple correlations in this circuit are probably computable with a supercomputer. However, we note the existence numerical effort that computing all the correlations represent.

For $j=3$ and $j=4$ steps, which corresponds to adding extra Trotter steps on top of the light pulse, the situation is much more difficult for the classical computers. The necessarily optimistic quadratic extrapolation yields $10^{15}$ strings to obtain $90\%$ of the norm squared, which is a factor $10^3$ above the largest SPD simulation performed to this date. More realistic extrapolations give an estimate around $10^{15}$ and $10^{23}$ strings. With $10^{12}$ strings, one should obtain around $40\%$ of the norm squared. Getting a good estimate of the expectation value with good confidence should be very challenging to do with SPD for this circuit. Finally, regarding $j=4$ with two extra steps, $10^{12}$ strings would yield less than $20\%$ of the norm squared. Extrapolating to $90\%$ gives astronomically large numbers.

\subsubsection{Runtime estimate for the circuits run on hardware}

\begin{figure*}[t]
    \centering
\includegraphics[scale=0.8]{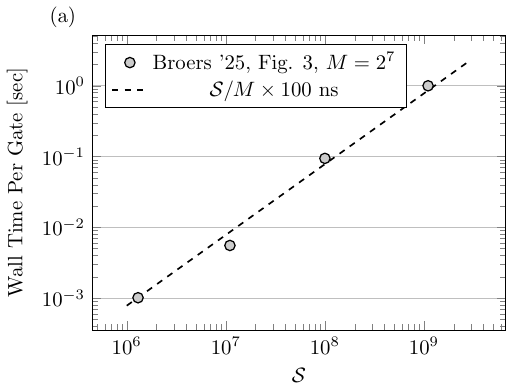}
\includegraphics[scale=0.8]{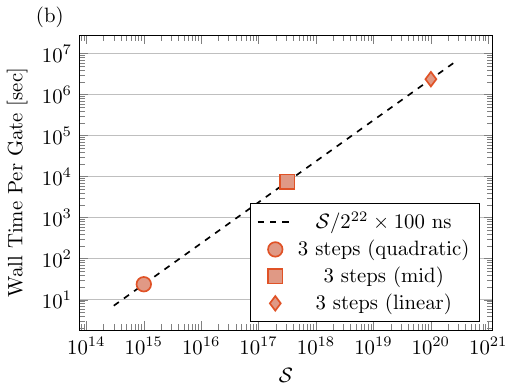}
    \caption{\textbf{Estimating the run time for Sparse Pauli Decomposition on Fugaku.} (a): Wall time per gate extracted from Fig.~3 in Broers '25 which refers to Ref.~\cite{broers2025scalable} as a function of the number of Pauli strings $\mathcal{S}$. Assuming a perfectly linear scaling of wall time per gate in $\mathcal{S}$ and perfect parallelisation, i.e. inverse linear scaling in the number of parallel processes $M$, the best fit we obtain is $\mathcal{S}/M \times 100$ nanoseconds per gate.
    (b): Wall time per gate using $M=2^{22}$ parallel processes assuming the scaling extracted in (a) for simulating the 3-step circuit in three different scenarios: A quadratic and linear scenario, as well as the geometric mean, denoted "mid". We consider the mid scenario to be the most likely of the three and $M^{22}$ to upper bound the maximum number of parallel processes achievable on Fugaku based on Appendix F of~\cite{broers2025scalable}.
\label{supplement_fig_walltime_broers}}
\end{figure*}

How do these numbers translate to actual runtimes? This question is generally difficult to answer due to differences in classical compute hardware and implementation, although in this case we can make some estimates based on the recent parallelised implementation of SPD on the Fugaku supercomputer~\cite{broers2025scalable}. In Fig.~3 of that work, wall time per gate is shown for different numbers of Pauli strings $\mathcal{S}$ and parallel processes $M$. Using plot digitisation software~\cite{WebPlotDigitizer}, we have extracted the scaling of wall time per gate $w$ as a function of $\mathcal{S}$ for a fixed number of parallel processes $M=2^7$. While we found the slope $w \sim \mathcal{S}^{1.05}$ to fit the data best, let us make the optimistic assumption that $w \propto \mathcal{S}$ scales linear, that memory is unlimited, and that furthermore the problem can be parallelised perfectly up to $M=2^{22}$ parallel processes. Given that in that work up to 12,288 nodes were used to carry out $2^{18}$ parallel processes, this larger $M$ upper bounds the number of parallel processes that could be run on all 158,976 of Fugaku's nodes. Under these assumptions, we find a best fit
\begin{equation}
    w = \frac{\mathcal{S}}{M} \times 100 \mathrm{\, nanoseconds}
\end{equation}
for the wall time per gate, as shown in Fig.~\ref{supplement_fig_walltime_broers}.

How long would these $2^{22}$ parallel processes have to work to simulate the 3-steps circuit run on the quantum hardware and obtain $90\%$ of the relative norm squared? We differentiate three scenarios shown in Fig.~\ref{psdfig:3}: A "quadratic" scenario (dotted), in which the relative norm squared grows quadratically in the logarithm of the number of Pauli strings $\mathcal{S}$, all the way up to $90\%$. As we have seen in Fig.~\ref{psdfig:1}, this is an optimistic scenario, since from $\sim 50\%$ we expect the growth of norm to be linear in the logarithm of the number of Pauli strings. In this scenario, Fig.~\ref{psdfig:3} shows that $\mathcal{S} \approx 10^{15}$ would be required to retain $90 \%$ of the relative norm squared.
There is a second "linear" scenario (dashed), in which the relative norm squared grows linearly to $90\%$. In this scenario, $\mathcal{S} \approx 10^{20}$ would be required to retain $90 \%$ of the relative norm squared. The most realistic estimate is likely somewhere between these points and we consider a third "mid" scenario, in which the required number of Pauli strings is the (geometric) mean of the two previous scenarios. We obtain wall times per gate
\begin{equation}
\begin{aligned}
        w_\mathrm{quadratic} &= \frac{10^{15}}{2^{22}} \times 100 \mathrm{\, nanoseconds} \approx 24 \mathrm{\, seconds} \\
        w_\mathrm{mid} &= \frac{10^{17.5}}{2^{22}} \times 100 \mathrm{\, nanoseconds} \approx 2 \mathrm{\, hours} \\
        w_\mathrm{linear} &= \frac{10^{20}}{2^{22}} \times 100 \mathrm{\, nanoseconds} \approx 28 \mathrm{\, days}
\end{aligned}
\end{equation}
The 3-step circuit contains 2888 two-qubit gates and 12151 gates in total. Let us assume that through rebasing of single-qubit gates, the overall gate count can be reduced to three times the number of two-qubit gates ($2888 \times 3 = 8664$) and that $50 \%$ of those gates can be done before reaching the maximum number of Pauli strings $\mathcal{S}$ and that the observable is propagated through those 4332 gates instantaneously. We obtain an overall runtime of 
\begin{equation}
\begin{aligned}
        T^\mathrm{run}_\mathrm{quadratic} &= 24 \mathrm{\, seconds} \times 4332 \approx 29 \mathrm{\, hours}\\
        T^\mathrm{run}_\mathrm{mid} &= 2 \mathrm{\, hours} \times 4332 \approx 1 \mathrm{\, year}\\
        T^\mathrm{run}_\mathrm{linear} &= 28 \mathrm{\, days} \times 4332 \approx 332 \mathrm{\, years}
\end{aligned}
\end{equation}
The most likely overall runtime estimate for obtaining a reliable output with $90 \%$ of the relative norm square using all 158,976 nodes of the Fugaku supercomputer is thus  $T^\mathrm{run}_\mathrm{mid} \approx 1$ year. The 215 shots executed on that circuit to obtain the data in Fig.~\ref{fig1}g. took approximately $215 \times 4.2$ seconds $\approx 15$ minutes of runtime on the Helios quantum computer.

\subsection{Matrix-Product Operators, Projected-Entangled-Pairs States, and Beyond}

Projected-entangled-pair state (PEPS) methods have become a cornerstone tensor-network framework for simulating quantum many-body systems~\cite{cirac2021matrix}. In contrast to MPS, PEPS employs a lattice-based architecture that more naturally and efficiently represents entanglement in higher-dimensional ($D>1$) systems. This enhanced expressiveness, however, comes at the cost of greater computational complexity and the absence of a canonical form for optimal compression. We employ a PEPS specifically adapted to the Octagon lattice geometry with periodic boundary conditions (PBC) and integrate a Belief Propagation (BP) algorithm to efficiently evolve the state~\cite{AlAr21}. 

\begin{figure}
    \centering
    \includegraphics[scale=0.75]{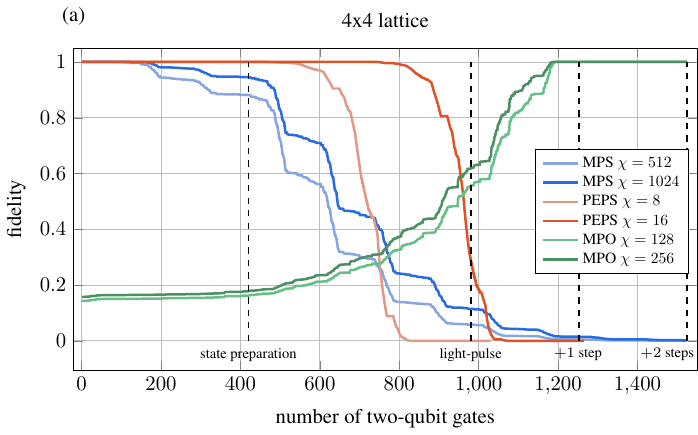}
    \includegraphics[scale=0.75]{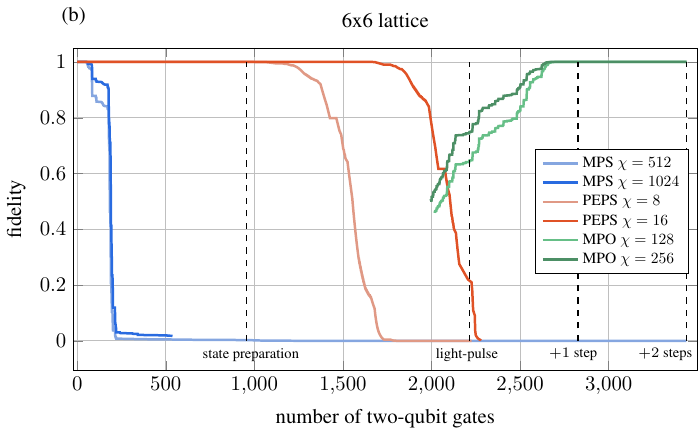}
    \caption{\textbf{Fidelity versus two-qubit gate count for simulating light-pulse dynamics (+2 extra steps) in $4\times4$ and $6\times6$ systems using MPS, PEPS, and MPO methods.}}
    \label{fig:mpspepsmpo}
\end{figure}

To evaluate the accuracy of the PEPS simulation, we report the fidelity as a proxy for the simulation precision. The fidelity is estimated using the $L2$-BP algorithm, wherein the environment tensors are approximated through product-state gauges, $F = |\langle \psi | \psi \rangle|^2$, with $\ket{\psi}$ denoting the compressed state~\cite{Tomislav:2024}. In Fig \ref{fig:mpspepsmpo}, we plot the fidelity as a function of the number of two-qubit gates (i.e., the light-pulse circuit followed by two additional Trotter relaxation steps).

For comparison, we also report the fidelities [see Eq.~(\ref{eq:fidelity})] using MPS simulations, as well as those obtained using matrix-product operator (MPO) Heisenberg evolution. Notably, MPO evolution proceeds backward in time, and its accuracy depends on both the spatial support of the initial (unevolved) operator and the light-cone structure of the circuit. The fidelity of MPO simulations drops rapidly as the light cone of the operator expands, indicating that MPO methods alone are clearly incapable of simulating the full circuit. 
Similar to both the MPS and SPD approach, linear extrapolation of $\langle O(F) \rangle$ using MPOs has also been considered to extend the limit of finite bond dimension. Empirically, fidelity-based extrapolation of expectation values has been reported to yield results consistent with exact values in the IBM quantum utility experiment~\cite{anand2023classical}. However, the expectation value as a function of fidelity is not monotonic. Even as a heuristic method, the fidelity has to be larger than a certain threshold $\mathcal{F}_\textrm{th}$ to obtain meaningful signal, either by extrapolation or claiming convergence of the method. 
Since the MPO and SPD are similar techniques---both track the evolution of observables in the Heisenberg picture and differ only in their truncation schemes---we expect this threshold to be close to $1$, which makes the stand-alone MPO approach practically infeasible and motivates the use of hybrid techniques.

From the plotted results, we observe that PEPS achieves higher fidelity than MPS with achievable bond dimensions, particularly as the system scales from $4 \times 4$ to $6 \times 6$, reflecting the benefit of the intrinsic two-dimensional connectivity of PEPS. Nevertheless, despite this advantage, the PEPS simulations remain constrained by the maximum bond dimension employed here ($\chi=16$) and cannot fully capture the deepest circuits. 

One possible direction to overcome these limitations is a hybrid approach that partially evolves the state in the Schr\"odinger picture and the operator in the Heisenberg picture, meeting in the middle --- a strategy referred to as the mixed representation. For the $4 \times 4$ system, one can imagine evolving MPS and MPO and stopping near where their fidelity curves cross to achieve better fidelity. However, given the near-vanishing MPS fidelity for the $6 \times 6$ system, one would instead need to evolve PEPS and MPO (or potentially PEPO) and meet at their crossing point to obtain reasonable fidelities. However, measuring observables (in the form of an MPO or PEPO) for PEPS with $\chi>8$ likely requires either a (potentially sign-problematic) sampling technique or a loop-series BP expansion~\cite{EvEtAl24}. It is unclear to us whether the observable can be accurately estimated in such a mixed representation with feasible computing resources, but it would be an interesting direction for further study. Finally, we note that all of these methods could also be implemented with an intrinsically fermionic PEPS and/or PEPO representation that avoids additional overhead of simulating the fermion-to-qubit encoding, generally resulting in lower bond dimension for the relevant objects. Reliably establishing the difficulty of each of these potential classical techniques for simulating these circuits requires significant further study and is left for future work.

\newpage
\section{Boundary Conditions \label{sec:boundaryconditions}}
In this section, we explain our choice of boundary conditions mentioned in section \ref{sec_supplement_fermionic_encoding}. We recall that the fermionic encoding used in this work possesses $N/2$ independent stabilisers, $N/2-2$ of which are "local", namely they involve qubits located in a definite region of the lattice, and $2$ of which "wind" around the lattice, one in the horizontal direction and one in the vertical direction. The $N/2-2$ independent local stabilisers must be initialized to $+1$ in order for the fermionic anticommutation relations to be implemented correctly. On the other hand, the value taken by the two winding stabilisers dictate the boundary conditions for the fermions in the horizontal and vertical directions. We explained in section \ref{sec_supplement_fermionic_encoding} that the fermionic encoding involves edge operators $A_{jk}$ that must obey the following equation, for every cyclic sequence of neighbouring sites $j_1,...,j_p$ that does not wind around the lattice
\begin{equation}\label{cyclicsequence}
    i^{p-1}\prod_{s=1}^{p-1}A_{j_s,j_{s+1}}=1\,.
\end{equation}
If the cyclic sequence of neighbouring sites winds around the lattice in either the horizontal or vertical directions, then the value taken by this product will depend on the boundary conditions imposed there. We recall that in our case, these edge operators are written in terms of Pauli matrices
\begin{equation}
    A_{ij}=\begin{cases}
        X_{i\uparrow} Y_{j\uparrow} X_{i\downarrow} Y_{j\downarrow} X_a\,,\qquad \text{for vertical edges oriented upwards}\\
        -X_{i\uparrow} Y_{j\uparrow} X_{i\downarrow} Y_{j\downarrow} X_a\,,\qquad \text{for vertical edges oriented downwards}\\
        X_{i\uparrow} Y_{j\uparrow} X_{i\downarrow} Y_{j\downarrow} Y_a\,,\qquad \text{for horizontal edges }\,,
    \end{cases}
\end{equation}
where $a$ denotes the ancilla adjacent to the edge $\langle i,j\rangle$. Taking the product \eqref{cyclicsequence} over a cyclic sequence that is a line or a column winding around the lattice in the horizontal or vertical direction precisely gives the two winding stabilisers $S_{\rm hor}$ and $S_{\rm vert}$ defined in section \ref{sec:stabilizers}, times a minus sign. Their value $S=\pm 1$ exactly correspond to imposing either periodic boundary conditions for $S=-1$, and antiperiodic boundary conditions for $S=1$. These two boundary conditions can be chosen independently for the vertical and horizontal directions, by setting independent values to $S_{\rm hor}$ and $S_{\rm vert}$. Imposing periodic boundary conditions in the lattice would thus correspond to prepare $S_{\rm hor}=S_{\rm vert}=-1$.

In our case, we implemented what we call \emph{mixed} boundary conditions. These correspond to a superposition of the four different combinations of possible boundary conditions, periodic or antiperiodic in the horizontal or vertical directions, with equal weight. To that end, we initialize the two winding stabilisers in the state
\begin{equation}
    \frac{1}{2}\left(|00\rangle+|01\rangle+|10\rangle +|11\rangle\right)\,.
\end{equation}
Because the winding stabilisers commute with the circuit, the circuit exactly corresponds to implementing the four different boundary conditions separately and averaging the results (as long as we do not post-select onto $S_{\rm hor/vert}=\pm 1$). 

This choice of mixed boundary conditions is justified on two grounds. Firstly, these mixed boundary conditions are less sensitive to noise than periodic boundary conditions. Indeed, for periodic boundary conditions, an error that flips one of the two winding stabilisers will modify the boundary conditions. However, for mixed boundary conditions, as long as the observables do not involve these two winding stabilisers or logical operators that can flip them, everything happens like if the winding stabilisers were already in the maximally mixed state, where all four possibilities are equally probable, and on which Pauli errors have no effects. For that reason these mixed boundary conditions are more robust to noise.

The second reason for choosing the mixed boundary conditions is that, ultimately, we are interested in the thermodynamic limit $N\to\infty$ where the boundary conditions have no effects anymore. In that respect, periodic, antiperiodic or mixed boundary conditions are only ways of dealing with the finite system, and are physically equally relevant. One can thus choose the boundary condition that will converge the fastest to the thermodynamic limit. It is known that averaging over "twisted" boundary conditions where a phase $e^{i\varphi}$ is given to fermions that wind around the lattice, with $\varphi$ the value to average over, make expectation values of observables converge faster to the thermodynamic limit than periodic boundary conditions \cite{lin2001twist}. In that point of view, the mixed boundary conditions can be seen as a rough approximation of this phase averaging, and is thus expected to give results that are closer to the thermodynamic limit than periodic boundary conditions.

\begin{figure*}[t]
    \centering
\includegraphics[width=\linewidth]{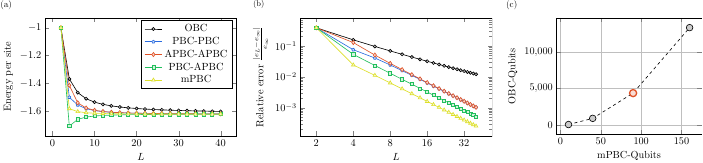}
\caption{\textbf{Effect of boundary conditions on finite size effects for the ground state energy of the non-interacting model.} (a) Both sides of the lattice can be associated either open (OBC), periodic (PBC) or antiperiodic (APBC) boundary conditions. In the thermodynamic limit, each choice converges to the thermodynamic value $e_\infty \approx -1.6211353773t$. For quantum simulation setups with a limited number of degrees of freedom, fast convergence is desired. 
(b) Relative error with respect to the thermodynamic limit value. Mixed periodic boundary conditions (mPBC), which is the defined as the average over PBC-PBC, APBC-APBC, PBC-APBC and APBC-PBC shows the best performance. All periodic setups have a scaling advantage with respect to open boundary conditions. (c) Translating the finite size error achieved with qubits modelling a periodic vs. open system, assuming the Octagon Fermion-to-Qubit encoding in each case. The experimental setting is shown in red.
\label{supplement_fig_boundary_conditions}}
\end{figure*}

As a concrete example, consider the ground state energy of the $L \times L$ square lattice of free fermions with nearest neighbour hopping 
\begin{equation}
    H = -t \sum_{\langle ij \rangle} c^\dagger_{i \sigma} c_{j \sigma}
\end{equation}
at half-filling. The ground state energy in the thermodynamic limit is given by
\begin{equation}
    e_\infty = \frac{2}{4 \pi^2} \int_{\cos(k_x) + \cos(k_y) \geq 0} -2t(\cos(k_x) + \cos(k_y)) \mathrm{d}k_x \mathrm{d}k_y \approx -1.6211353773t.
\end{equation}
In Fig.~\ref{supplement_fig_boundary_conditions}a, we show how different boundary conditions approach the thermodynamic limit. While open boundary conditions (OBC), periodic boundary conditions (PBC) and antiperiodic boundary conditions (in which every hopping term across the boundary comes with an extra minus sign, APBC) systematically \textit{overestimate} the ground state energy, boundary conditions with PBC in one direction and APBC in the other \textit{underestimate} it. Mixed periodic boundary conditions (mPBC), which we define by averaging over the 4 settings (PBC-PBC, APBC-APBC, APBC-PBC and PBC-APBC), thus yields the smallest finite-size effects. The resulting relative error with respect to the thermodynamic limit is shown in Fig.~\ref{supplement_fig_boundary_conditions}b. For the $L=6$ lattice considered in this work, mixed periodic boundary conditions already show more than 8 times smaller finite-size effects. To match the finite size effects of a $6 \times 6$ mixed periodic lattice, an open boundary conditions setup would require more than $40^2 = 1600$ sites. If such an OBC system was realised using the same fermion-to-qubit mapping as described here, this would increase the qubit count from 90 to $1600\times2 \times 5/4 = 4000$ qubits. This discrepancy increases with increasing $L$ due to the different slopes in Fig.~\ref{supplement_fig_boundary_conditions}b: While in principle there exist exact expressions for the finite size relative error $r_L=|e_{L}-e_\infty|/e_\infty$, we can obtain a simple fit using the data points from $L=32$ to $L=40$, to obtain
\begin{equation}
\begin{aligned}
r_L^\mathrm{OBC} &\approx \frac{0.555}{L^{1.037}} \\
r_L^\mathrm{mPBC} &\approx \frac{0.426}{L^{2.012}}.
\end{aligned}
\end{equation}
We note that finite-size corrections to the energy density scaling as $L^{-1}$ for open boundaries and $L^{-2}$ for periodic boundaries is a well-known and generic fact in one-dimensional critical systems, see for example \cite{pirvu2012matrix}. In the Octagon encoding, the number of qubits $Q$ is related to the linear system size as $L=\sqrt{2Q/5}$. Therefore, to obtain the same amount of finite-size effects in the two settings, one needs
\begin{equation}
    Q^\mathrm{OBC} \approx 0.704 (Q^\mathrm{mPBC})^{1.94},
\end{equation}
assuming that the ground state can prepared exactly in each case. We note that, for the free fermion case, the fermionic fast Fourier transform can prepare the ground state reasonably efficiently even without periodic boundary conditions being natively available in the hardware (though reconfigurability does confer a benefit even in this case~\cite{maskara2025fastsimulationfermionsreconfigurable}). The quantitative estimate of the advantage of (mixed) periodic qubits shown in Fig.~\ref{supplement_fig_boundary_conditions}c is thus to be seen as a representative example for the simulation of critical states.

\newpage
\section{Quantum Monte Carlo}
\label{sec_supplement_qmc}

For the Hubbard model at half-filling, the quantum Monte Carlo method can provide a numerically exact theory predictions for various physical observables for finite temperatures and the ground state.
Such predictions are often used as a metrology tool to determine the temperature of the state prepared in analog quantum simulation~\cite{mazurenko_cold-atom_2017,xu_neutral-atom_2025}. 
Here, we perform the determinant quantum Monte Carlo calculations on a $6\times 6$ lattice with an interaction strength of $U/t=8$ and all four different boundary conditions, i.e., PBC-PBC, PBC-APBC, APBC-PBC, and APBC-APBC, using the QUEST package~\cite{varney2009quantum,lee2010quest,chang2015recent}.
For the PBC-PBC configuration, we employ the default run script "test", which exploits the C\(_4\) rotational symmetry of the lattice. 
For all other boundary condition setups, simulations are carried out using the "ggeom" run script, which supports a variety of geometric and boundary configurations.
Each simulation consists of 20,000 warmup sweeps followed by 80,000 measurement sweeps with a step size $\tau=0.01$.
The mixed boundary condition results reported are the average over the simulation results of the four boundary conditions.

\begin{figure}
    \centering
\includegraphics[scale=0.85]{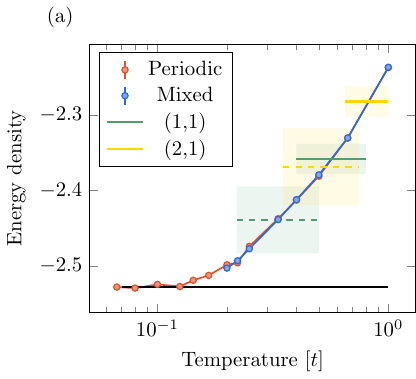}
\includegraphics[scale=0.85]{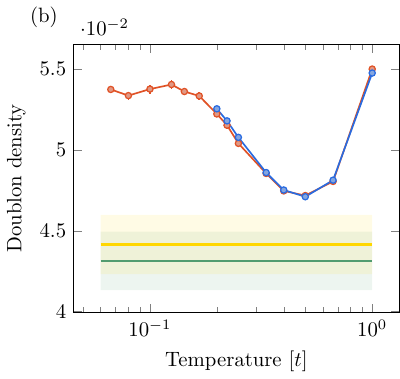}
\includegraphics[scale=0.85]{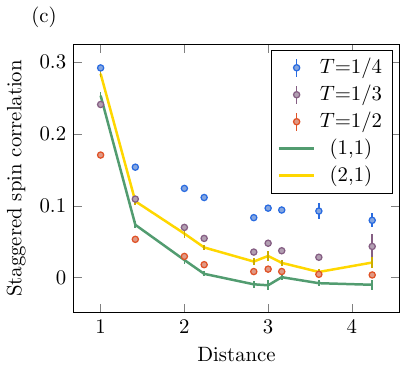}
    \caption{\textbf{Determinant Quantum Monte Carlo theory prediction.} (a): Total energy as a function of temperature, showing raw hardware data (continuous line) and mitigated hardware data (dashed line). The ground state with periodic boundaries is indicated by the black line.
    (b): double occupancy as a function of temperature with same conventions as (a).
    (c): staggered spin-spin correlation with mixed boundary conditions as a function of distance, i.e. correlations multiplied by $(-1)^{x+y}$ where $x,y$ are the distances in the horizontal and vertical directions.
    \label{supplement_fig_qmc}}
\end{figure}

\newpage
\tocless\section{Acknowledgements}
This work was made possible by a large group of
people, and the authors would like to thank the entire
Quantinuum team for their many contributions. We
are grateful for helpful discussions with Immanuel Bloch, Harry Buhrman, David Amaro and Cristina Cirstoiu. E.G. acknowledges support by the Bavarian Ministry of Economic Affairs, Regional Development and Energy (StMWi) under project Bench-QC
(DIK0425/01). At various points during the development of this work, Phasecraft's compact fermionic encoding, US patent no. 12020119 was used under license. Some of the matrix-product state simulations were run in compute resources of the National Energy Research Scientific Computing Center (NERSC), a Department of Energy Office of Science User Facility using NERSC award DDR-ERCAP0032628. All experiments were run on Quantinuum's Helios Quantum computer, powered by Honeywell.
\\
\tocless\section{Author Contributions}
E.G., S-H.L, K.H., R.N., M.I. and H.D. conceived the experimental ideas, implemented and tested the circuits and did the data analysis, with E.G. leading work on the half-filled, S-H.L. on the doped and K.H. on the bilayer model. H.D. defined the high-level experimental protocols for the three settings, developed the injection technique for state preparation and optimized circuit design for the three settings. E.G. developed the error mitigation scheme. R.N. and H.D. worked out the fermionic encodings. A.R., C.F., C.F., J.P.G., A.H., A.H., A.I., C.J.K., N.K., I.S.M., M.M., A.R.M., A.J.P., A.P.R., B.N. and J.G.B. built the experimental apparatus and took the data. J.A. developed circuit debugging tools. D.T.S. developed efficient toric code preparation circuits. R.H., P.A-M., M.F-F., and A.C.P contributed to the classical simulations and to the writing of the manuscript. H.D. drafted the initial manuscript to which all authors contributed.

\tocless\section{Data Availability}
The data supporting this study is available at Zenodo repository with DOI 10.5281/zenodo.17483197~\cite{zenodo}.

\tocless\section{Competing Interests}
H.D. is a shareholder of Quantinuum. All other authors declare no competing interests.
\end{document}